\documentclass[onecolumn,12pt]{els-mrw-arxiv} 
\usepackage{amsmath,amssymb,amsfonts,amsthm,makeidx,graphicx}
\usepackage{txfonts}
\usepackage{helvet}
\newcommand{\be}{\begin{equation}}
\newcommand{\ee}{\end{equation}}
\newcommand{\tr}{{\rm Tr}}
\usepackage{hyperref}

\begin{document}

\chapter{Quantum chaotic systems: a random-matrix approach}\label{chap1:chap1}

\author[1]{Mario Kieburg}%
\address[1]{\orgname{University of Melbourne, School of Mathematics and Statistics}, \orgaddress{813 Swanston Street, Parkville, Melbourne, VIC 3010, Australia}}

\articletag{Chapter Article tagline: April 10, 2026}

\maketitle

\begin{glossary}[Keywords]
Quantum Chaos, Random Matrix Theory, Symmetry Classification, Diagonalising Random Matrices, Unfolding, Local Spectral Statistics

\end{glossary}

\begin{abstract}[Abstract]
We review the ideas of how random matrix theory has to be properly applied to quantum physics; particularly we focus on how the spectrum has to be properly prepared and the random matrix correctly identified before the random matrix and the physical eigenvalue spectrum can be compared. We explain the ideas of the symmetry classification of symmetric matrix spaces and how that yields Dyson's threefold and Altland-Zirnbauer's tenfold way. We also outline how the joint probability density function of the eigenvalues can be calculated from a given probability density function on the matrix space. Furthermore, we dive into the subtleties of the unfolding procedure. For this purpose, we explain the ideas of the local mean level spacing, the local level spacing distribution and the $k$-point correlation functions. We outline the techniques of orthogonal polynomials, determinantal and Pfaffian point processes and their related Fredholm determinants and Pfaffians as well as the supersymmetry method. Moreover, we relate the local spectral statistics to effective Lagrangians that give the relation to non-linear $\sigma$-models. In all these discussions, we also make brief excursions to non-Hermitian random matrix theory which are useful when studying open quantum systems, for instance.
\end{abstract}

\section{Introduction}\label{chap1:sec:intro}

Random Matrix Theory (RMT) can now look back to an almost 100 years old history of applications. In 1928, Wishart~\cite{Wishart} has proposed the first random matrix model in statistics. His aim was to compute the probability distribution of the covariance matrix when the time series (samples) are normal distributed. This random matrix is nowadays called Wishart-Laguerre ensemble due to its relation to Laguerre polynomials. Born out from this work, Wishart's random matrix model was employed in statistics to quantify the impact of noise in a time series matrix as normal distributed data is nothing more than white noise, e.g., see the review~\cite{statistics}. It is rather surprising that it took almost forty years until the level density of this particular random matrix has been calculated which is nowadays known as the Mar\v{c}enko-Pastur distribution~\cite{MP-dist}. It serves as a benchmark and filtering tool in time series analysis. But it has also applications in telecommunications~\cite{telecom}.

In physics, RMT was introduced by Wigner~\cite{Wigner-semi,Wigner-surmise} in the 50's. During these difficult years, where nuclear power plants but also nuclear bombs were developed, Wigner tried to address the serious problem of predicting the statistics of excitations in nuclei. This was certainly a futile task when doing so for each single nucleus as not only heavy nuclei are tremendously complicated and, thus, chaotic many-body systems but also the amount of different nuclei including all of their possible isotopes (stable and unstable) needed to be taken care of. 

Wigner's idea was to approach this problem in a statistical way by replacing the original Hamiltonian and to consider a real symmetric random matrix model instead. From these early days two important and celebrated results were derived by Wigner. First, the Wigner semicircle~\cite{Wigner-semi} which describes the macroscopic level density of the eigenvalues of a Gaussian distributed Hermitian (can be also real symmetric or quaternion Hermitian (aka Hermitian self-dual\footnote{A matrix is called (anti-)self-dual if it satisfies $\widehat\tau_2H^T\widehat\tau_2=(-)H^T$ where $\widehat\tau_2$ is the second Pauli matrix embedded in the $2N\times2N$ matrices.})) random matrix. He also showed already in his earlier works that it does not even matter that a Gaussian has been chosen as a probability distribution as long as the matrix entries are identically and independently distributed with suitable integrability conditions and being centred. Depending on what type of convergence is wanted an existent second moment is suitable for a convergence in distribution~\cite{Wigner-semi} while a fourth moment condition is needed for uniform convergence~\cite{Fourth-moment}.

While the Wigner semi-circle does not look very physical in many situations, because spectra usually grow either polynomially for single-body Hamiltonians or exponentially in many-body systems, it is the second result by Wigner which made a big furore and had an important impact in physics. This result is the Wigner surmise~\cite{Wigner-surmise} which describes the level spacing distribution of a two-level Gaussian distributed random matrix, see~\eqref{chap1:Wigner-surmise}. It is this quantity which is often used to decide whether a system is quantum chaotic or not. Although Wigner knew that the two-level random matrices can be only an approximation to the result in the limit of large matrix dimension, he and, especially, the field of RMT was very lucky that this approximation is extremely good. Quite often the deviations to the universal result are below the statistical and systematic errors of empirical data. A non-Gaussian random matrix model with only two levels would have not shown such good agreement. Also that he had considered real eigenvalue spectra and not complex helped, because it is known~\cite{GHS} that Wigner's simple two level ansatz does not work so well for complex eigenvalues.

As a side remark, we would like to mention that this development in physics went in parallel to mathematical developments during that time, where Weyl, Harish-Chandra, Gelfand et al. developed a theory about Lie groups  in Representation Theory and Harmonic Analysis, e.g., see the books by Hua~\cite{Hua} and Helgason~\cite{Helgason} and references therein. As pointed out in~\cite{DF-Hurwitz}, this history goes even further back in time to Hurwitz' work~\cite{Hurwitz} in 1897 about creating group invariants by integration and generalisation of Euler's parametrisation~\cite{Euler} of the group ${\rm SO}(3)$. The simultaneous studies in mathematics and physics in the 50's were almost oblivious of each other. Nonetheless, these mathematical developments should have an extremely important impact in random matrix theory. One well-known result is Harish-Chandra's integral~\cite{HC-int} for compact semi-simple groups and their corresponding Lie algebras. It should not be confused with the Itzykson-Zuber integral~\cite{IZ-int}. Those two types of integrals only agree in the unitary case while they differ for the orthogonal and unitary symplectic group integrals.

Coming from many-body chaos, people first thought that the agreement with RMT comes from the `many-body part'. Yet, soon it was clear that actually quantum chaos seems to be the reason why the Wigner-Dyson statistics describe the local spectral statistics of physical Hamiltonians so well. In a seminal work~\cite{BGS-conj}, Bohigas, Giannoni and Schmit  (BGS) conjectured in the mid 80's that the spectral statistics of the Hamilton operator in a closed quantum system should agree with the Wigner-Dyson statistics if the corresponding classical system is strongly chaotic meaning in a K-system, where `K' stand for Kolmogorov. Already a few years earlier Berry and Tabor~\cite{BT-conj} conjectured a similar statement for integrable system which is, however, quite often misinterpreted. They said that in an integrable system, where the energy contours in action space are curved, one should find Poisson statistics. They do not say that when one finds Poisson statistics the underlying quantum system is integrable. Indeed, there are chaotic quantum systems that can exhibit Poisson statistics if the spectrum is not well-prepared, especially in this case they split into statistically independent sub-spectra. This has been the case for arithmetic quantum billiards at specific parameter values, e.g., see~\cite{BraunHaake}.

Starting with the early 60's, RMT entered other areas in physics. First  with Kubo's~\cite{Kubo} as well as Gor'kovs and Eliashberg's~\cite{GE-cond} works, condensed matter theory  became  aware of random matrix results and used those to describe disordered systems~\cite{Beenakker,Oxford-cond}; especially the Anderson localisation transition is well-known~\cite{Anderson} and led to the field of banded random matrices~\cite{Band}.  In the 90's, the range of applications exploded including modelling Quantum Chromodynamic (QCD) Dirac operators, first without~\cite{Chiral-threefold} and then with chemical potential~\cite{Stephanov,Osborn}, on the lattice~\cite{DSV-lat,Osborn-lat} and even with finite $\theta$-angle~\cite{KVW-theta}. Also in quantum information~\cite{Quantum-info,LZ-QC-QI} several groups used random matrix theory to model uniformly distributed quantum states~\cite{LP-qinfo,Page} or more general ensembles of density matrices~\cite{density-mat}, which was first proposed by Lubkin~\cite{Lubkin} in the late 70's. Later it was also employed to describe quantum maps~\cite{Quantum-maps}. Furthermore, telecommunications became aware of random matrix theory in the late 90's which prepared the ground for 3G and 4G telecommunications system~\cite{telecom}. This application conversely spurred new developments in RMT, 15 years ago~\cite{AKW-tele}, where products of random matrices were used to model progressive scattering. This model is actually related  to the Dorokhov-Mello-Pereyra-Kumar (DMPK) equation~\cite{DMPK-dor,DMPK-MPK} for quantum transport in disordered media, see~\cite{Ipsen-Schomerus}. Also in the 90's, first relations to 2-dimensional quantum gravity were made~\cite{2d-quantum-gravity} which was recently extended to random tensors~\cite{Tensor-gravity} and Liouville quantum gravity~\cite{Liouville-gravity}. These applications have led to a new direction in mathematics which is now called enumerative geometry~\cite{enumerative}. Another recent development, which picks up where Wigner and others left in the 60's, are new random matrix models applied to many-body systems. Those random matrix models are called embedded random matrices~\cite{Kota}. They are random matrices with a number of independent matrix entries growing polynomially in the number of particles but are embedded in an exponentially large matrix which replaces the many-body Hamiltonian, density matrix or other physical operator. Examples are the random fermionic and bosonic Gaussian states~\cite{LSZ-boson,Qinfo-review,AHHJK} and the Sachdev–Ye–Kitaev (SYK) model~\cite{SYK}. Certainly, there are many more application of RMT which we cannot list here, especially recent developments in mathematics but also in physics such as in scattering theory~\cite{Guhr-scattering, Schomerus-scattering,Oxford-scattering} are very rich.

When it comes to apply RMT to model physical or other realistic operators, it is important to prepare the empirical spectrum. If this is not done properly, one cannot expect good agreement  with the benchmarks RMT is delivering. This creates also  the difficulty to make any definitive conclusion. This review should highlight the main steps of preparation of spectra, the major spectral objects and random matrix techniques to derive the benchmarks. We would like to list them here which also gives the outline of the present work.

\begin{enumerate}

\item	First one needs to split the eigenvalue spectrum in subspectra which correspond to fixed good quantum numbers. Those can be the conserved total spin or perhaps the parity, e.g., see~\cite{unfolding}. We will not dive deeper into this as there is not much to do analytically other than splitting the spectrum.

\item	Next one needs to figure out the symmetries of the considered operator, such as the Hamiltonian or the Dirac operator, which should be modelled by a random matrix. In Section~\ref{chap1:sec:symmetries},  we review the idea of such a symmetry classification. Although, we only show how the classification schemes of symmetric matrix spaces arise, it works exactly the same for physical operators as shown in~\cite{topo-class,QCD-class,QCD-lat-class,SYK-class}. Moreover, we show in this section how random matrices are diagonalised and how the joint probability density function (jpdf) of the eigenvalues is obtained.

\item	In the last step, one needs to identify the local mean level spacing. This allows a proper unfolding of the spectrum to compare it with the universal spectral statistics of the identified random matrices. See Section~\ref{chap1:sec:unfolding}, for an outline of the main concepts of the local spectral statistics where we summarise some standard results but also show some derivations of those. For example, we sketch how the determinantal and Pfaffian point processes are obtained for the $k$-point correlation functions when starting from the jpdf. We explain the idea of gap probabilities and Fredholm determinants and Pfaffians, and we show which effective Lagrangians are hidden behind the local spectral statistics with the help of the supersymmetry method. The latter relates RMT directly to effective field theories in QCD-like systems~\cite{Oxford-QCD} and condensed matter theory~\cite[Chapter~4]{Efetov} as it delivers the low energy potentials for the non-linear $\sigma$-models.
\end{enumerate}

Although, we mainly consider Hermitian random matrices, which correspond to closed quantum systems, we also  briefly mention what the new developments for non-Hermitian random matrices and Hamiltonians are. The latter correspond to open quantum systems and many other areas and have been employed in a multitude of works, e.g., see~\cite{Stephanov,Osborn,DSV-lat,symm-open,KVZ-Wilson,time-lagged}.

\section{Symmetries of Random Matrices and their JPDF of Eigenvalues}\label{chap1:sec:symmetries}

A random matrix ensemble is essentially a set of matrices (rectangular or square) that is distributed by a probability measure. Hence, it can be in general everything from a random matrix with all of its matrix entries being non-zero random variables to diagonal matrices with random entries. Even the extreme case of only a single fixed matrix is allowed which would cover the empirical situation. The power of RMT, however, lies in simple random matrix models that are analytically feasible and give in the large matrix dimension limit $N\to\infty$ universal statistics that are also shared with realistic systems.

To decide which matrix space needs to be chosen, it is paramount to understand the symmetries of the operator that should be modelled. If the wrong symmetries are chosen one cannot expect good agreement with empirical data or numerical simulations. Hence, as a first step when applying random matrices is the symmetry analysis. This was understood quite early in physics where already Wigner~\cite{Antiunitary} and Dyson~\cite{Dyson-threefold} have drawn relations between the unitary and anti-unitary symmetries of the physical Hamiltonian and the chosen random matrix. We follow the historical development and start with the celebrated Dyson's threefold way in Subsection~\ref{chap1:sec:Dyson}. In Subsection~\ref{chap1:sec:Altland-Zirnbauer}, we will review the development of their generalisations to the tenfold way by Altland and Zirnbauer~\cite{AZ-class}. In particular, we will sketch the main steps of how to derive the jpdf of the eigenvalues with the help of the example of the Hermitian matrix space. Finally, we will go over to recent developments about non-Hermitian symmetry classes and summarise those, in Subsection~\ref{chap1:sec:non-Hermitian}.

\subsection{Dyson's threefold way}\label{chap1:sec:Dyson}

\begin{figure}[t!]
\centering
\includegraphics[width=.5\textwidth]{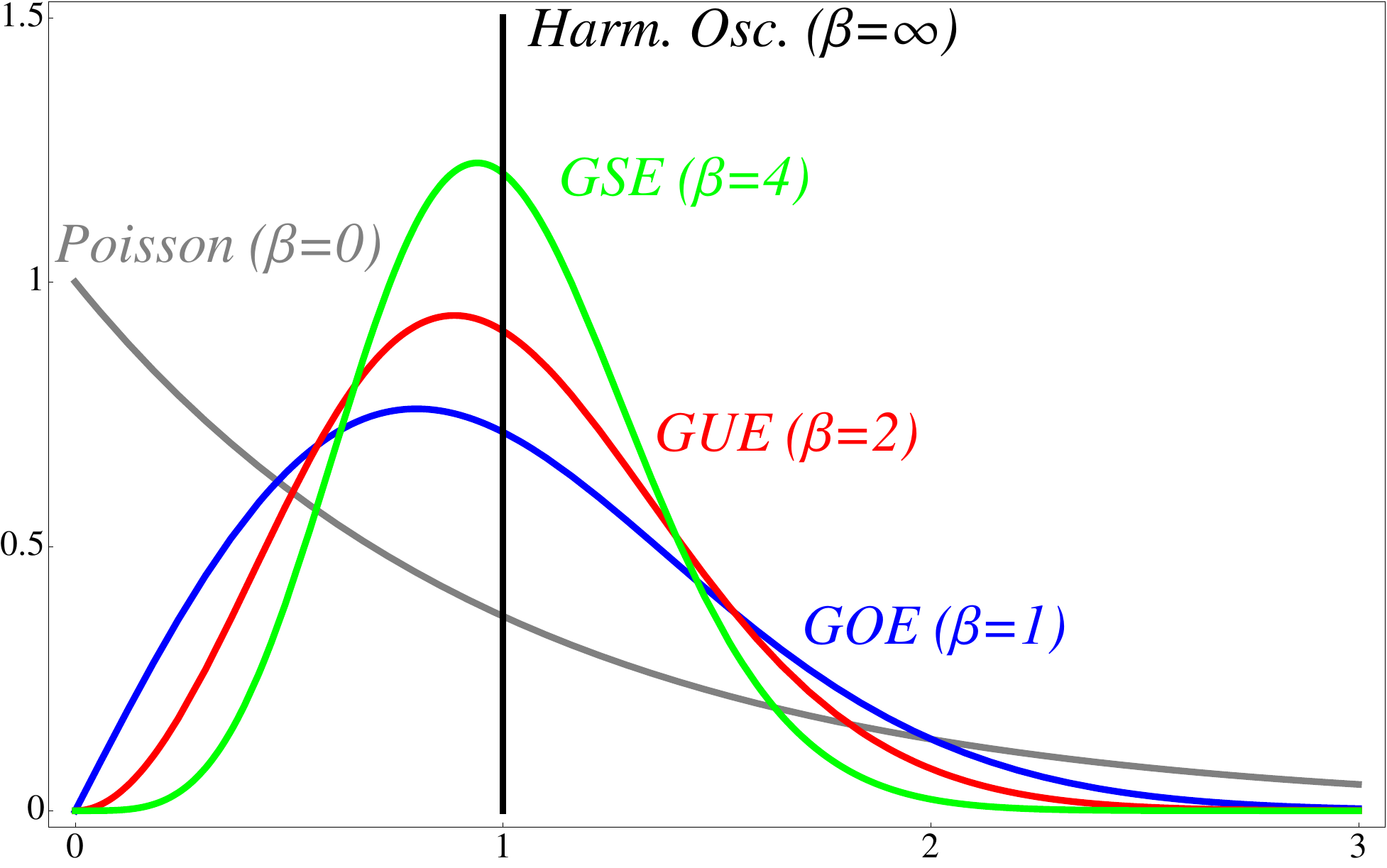}
\caption{Level spacing distribution of independently randomly distributed eigenvalues ($\beta=0$, Poisson statistics, grey curve), eigenvalues of the quantum harmonic oscillator ($\beta=\infty$, picket fence statistics, vertical black line illustrating the Dirac delta function $\delta(1-s)$), and the Wigner surmise~\eqref{chap1:Wigner-surmise} for $\beta=1$ (GOE, blue curve),  $\beta=2$ (GUE, red curve) and  $\beta=4$ (GSE, green curve). All distributions are normalised in such a way that the first moment is equal to unity. The case of the Poisson distribution $e^{-s}$, which is denoted by $\beta=0$, is not covered by Wigner's surmise~\eqref{chap1:Wigner-surmise}. Indeed, the agreement to the 2-dimensional Coulomb gas result for  $N\to\infty$ becomes worse for $\beta<1$.}
\label{chap1:fig1:level-spacing}
\end{figure}

In the 50's, Wigner proposed the Gaussian ensembles~\cite{Wigner-surmise} on the set of $2\times 2$ real symmetric matrices, i.e.,
\begin{equation}
S=\left[\begin{array}{cc} S_{11} & S_{12} \\ S_{12} & S_{22}\end{array}\right]\in{\rm Sym}(2)\qquad {\rm with}\quad P(S)=\sqrt{\frac{2}{\pi^3}}\exp[-{\rm tr} S^2].
\end{equation}
 For this ensemble, he computed the level spacing distribution between the two real eigenvalues which is
 \begin{equation}\label{chap1:Wigner-surmise-GOE}
 p_{\rm GOE}(s)=\frac{\pi}{2}s\exp\left[-\frac{\pi}{4}s^2\right]
 \end{equation}
 after rescaling to unit first moment.
 It is a surprisingly good approximation when comparing it to the $N\to\infty$ limit which is given in terms of solutions of Painlev\'e equations~\cite[Chapters 8.3 and 8.4]{log-gas} with deviations of only a few percent. The agreement becomes even better for a $2\times2$ Hermitian Gaussian random matrix and the best for $4\times 4$ quaternion Hermitian where the deviation is only a tiny fraction of a percent from the large $N$ limit~\cite{HaakeDietz}. This agreement led to call the level spacing distribution
 \begin{equation}\label{chap1:Wigner-surmise}
 p_{\rm G\beta E}(s)=2\frac{(\Gamma[(\beta+2)/2])^{\beta+1}}{(\Gamma[(\beta+1)/2])^{\beta+2}}s^\beta\exp\left[-\left(\frac{\Gamma[(\beta+2)/2]}{\Gamma[(\beta+1)/2]}\right)^2s^2\right],\qquad\text{with the Gamma function $\Gamma(z)$,}
 \end{equation}
the Wigner surmise. The Dyson index $\beta$ can takes a positive value. However, only for $\beta=1$ (real symmetric), $\beta=2$ (Hermitian), and $\beta=4$  (quaternion Hermitian) they correspond to the classical Gaussian matrix ensembles (also known as Wigner-Dyson ensembles), though a tridiagonal matrix representation also exists~\cite{beta-ens} for general $\beta\geq0$. The distribution~\eqref{chap1:Wigner-surmise} is illustrated in Fig.~\ref{chap1:fig1:level-spacing}.

The Wigner-Dyson ensembles are the Gaussian Orthogonal Ensemble (GOE) for real symmetric matrices, the Gaussian Unitary Ensemble (GUE) for Hermitian matrices, and the Gaussian Symplectic Ensemble (GSE) for quaternion Hermitian matrices. Their name reflects the invariance group and not the matrix space chosen. They are all distributed by
 \begin{equation}\label{chap1:Gaussian.Wigner}
 P_{\rm G\beta E}(H)=\frac{1}{\mathcal{Z}_{\rm G\beta E}}\exp\left[-\frac{1}{2\sigma_\beta^2}{\rm tr} H^2\right],
 \end{equation}
 where $H$ is in one of the three classes, $\sigma_\beta>0$ sets the standard deviation and $\mathcal{Z}_{\rm G\beta E}>0$ is the normalisation constant. Usually one chooses $\sigma_4=\sqrt{2}\sigma_1=\sqrt{2}\sigma_2$  because quaternion Hermitian matrices have Kramers degenerate eigenvalues.
 
 The reason for these three  types of classical ensembles is born out of the observation that Hamiltonians may satisfy anti-unitary symmetries. In a Hilbert space $\mathcal{H}$ with a scalar product $\langle\psi|\phi\rangle$ an anti-unitary operator $\mathfrak{A}$ satisfies
 \begin{equation}
 \langle\mathfrak{A}\psi|\mathfrak{A}\phi\rangle=\langle\phi|\psi\rangle\qquad {\rm for\ all}\quad |\psi\rangle,|\phi\rangle\in\mathcal{H},
 \end{equation}
 unlike a unitary operator $\mathfrak{U}$, whose property is
 \begin{equation}
 \langle\mathfrak{U}\psi|\mathfrak{U}\phi\rangle=\langle\psi|\phi\rangle\qquad {\rm for\ all}\quad |\psi\rangle,|\phi\rangle\in\mathcal{H},
 \end{equation}
keeping the order of the vectors the same. Such anti-unitary operators can be the time reversal operator $\mathfrak{T}$ in physics. Already Wigner found out~\cite{Antiunitary} that any anti-unitary operator $\mathfrak{A}$ can be represented by a unitary operator $\mathfrak{U}$ and the complex conjugation operator $\mathfrak{K}$, i.e., $\mathfrak{A}=\mathfrak{K}\mathfrak{U}$. Especially, anti-unitary operators that lead to involutions when conjugating other operators with them, meaning they satisfy $\mathfrak{A}^2X\mathfrak{A}^{-2}=X$ for all complex linear operators $X:\mathcal{H}\to\mathcal{H}$, are of particular importance in physics. Schur's lemma implies for such operators that $\mathfrak{A}^2=\lambda\mathbf{1}$ must be proportional to the identity. On the other hand, it is also $\mathfrak{A}^2=\mathfrak{U}^*\mathfrak{U}$ with $\mathfrak{U}=\mathfrak{K}\mathfrak{A}$ the unitary operator relating to $\mathfrak{A}$ and $\mathfrak{U}^*$ its complex conjugate. Conjugating $\mathfrak{U}^*\mathfrak{U}=\lambda\mathbf{1}$ with $\mathfrak{U}$  we see that $\lambda$ must be real and as $\mathfrak{U}^*\mathfrak{U}$ is also unitary it must be a complex phase, too. Thus, we find that the corresponding anti-unitary operators with the involutive property above satisfy
 \begin{equation}\label{chap1:anti-unitary-square}
\mathfrak{A}^2=\pm \mathbf{1}.
 \end{equation}
 
Equation~\eqref{chap1:anti-unitary-square} is the origin of Dyson's threefold way~\cite{Dyson-threefold}. A Hamiltonian which commutes with an anti-unitary operator $\mathfrak{A}$, such as the time reversion operator $\mathfrak{T}$, implies that when $|\psi\rangle$ is an eigenvector then $\mathfrak{A}|\psi\rangle$ is one, too. Out of this knowledge one can create particular bases. For instance the Hamiltonian has a real eigenbasis when $\mathfrak{A}^2=+\mathbf{1}$ or a quaternion eigenbasis when $\mathfrak{A}^2=-\mathbf{1}$. In the case that there is no anti-unitary operator which commutes with the Hamiltonian, the eigenbasis is generically complex.
This analysis of anti-unitary symmetries is, thus, as important as those of ordinary symmetries as they have a crucial impact on the spectral statistics (eigenvalue- as well as eigenvector-wise).

\subsection{Altland-Zirnbauer's tenfold Way}\label{chap1:sec:Altland-Zirnbauer}

\subsubsection{  Cartan Classification}

Altland and Zirnbauer, while studying normal-superconducting hybrid structures~\cite{AZ-class}, extended Dyson's classification of Hamiltonians. The existence of particle-hole symmetries (charge conjugation) and chirality symmetries opened new possibilities which actually create a particular symmetry point in the spectrum (known as Dirac point in condensed matter theory as the band structure looks locally like a Dirac cone). Around such particular points (which are usually set to be equal to the origin), it becomes immediate that more than only three local statistics may arise despite Dyson's threefold way, e.g., see~\cite{Ivanov-micro}. The symmetry classification is based on Cartan decompositions~\cite[Chapter III.7]{Cartan}.

The idea is to start with a Lie-Algebra $\mathfrak{g}$ and consider a Cartan involution $\iota$, which satisfies
\begin{equation}\label{chap1:Cartan-properties}
\iota(\iota(X))=X \qquad{\rm and}\qquad \iota([X,Y])=[\iota(X),\iota(Y)]\qquad{\rm for\ all}\ X,Y\in\mathfrak{g}
\end{equation}
with $[X,Y]$ the Lie bracket and a positivity condition regarding the Killing form. Then, the eigenspaces $\mathfrak{h}=\{\iota(X)=X\in \mathfrak{g}\}$ and  $\mathfrak{p}=\{\iota(X)=-X\in \mathfrak{g}\}$ of fixed points and anti-fixed points of $\iota$, respectively, constitute symmetric matrix spaces. The matrix spaces of the Wigner-Dyson ensembles are then results of the identification
\begin{equation}
\begin{aligned}
&\mathfrak{g}=\mathbb{R}^{N\times N},\ \iota(X)=-X^T&\qquad&\Rightarrow&\qquad &\mathfrak{p}={\rm Sym}(N)&\quad&\text{real symmetric matrices},\\
&\mathfrak{g}=\mathbb{C}^{N\times N},\ \iota(X)=-X^\dagger&\qquad&\Rightarrow&\qquad &\mathfrak{p}={\rm Herm}(N)&\quad&\text{Hermitian matrices},\\
&\mathfrak{g}=\mathbb{H}^{N\times N}\subset\mathbb{C}^{2N\times 2N},\ \iota(X)=-\widehat{\tau}_2X^T\widehat{\tau}_2&\qquad&\Rightarrow&\qquad &\mathfrak{p}={\rm Self}(N)&\quad&\text{quaternion/self-dual Hermitian matrices},
\end{aligned}
\end{equation}
where $\widehat{\tau}_2=\tau_2\otimes \mathbf{1}_N$ is the embedding of the second Pauli matrix in the $2N\times2N$ matrices, reflecting the complex representation of the quaternion numbers $\mathbb{H}$. The eigenspace of fixed points $\mathfrak{h}$ is actually a sub-Lie algebra which anew can be subjected to another Cartan involution.

In physical terms, the Cartan involutions can be the conjugation of operators with time reversion operation $\iota(X)=\mathfrak{T}X\mathfrak{T}^{-1}$, the charge conjugation operator $\iota(X)=-\mathfrak{C}X\mathfrak{C}^{-1}$ (the minus sign guarantees the properties~\eqref{chap1:Cartan-properties} as $\mathfrak{C}$ reverses the order of operators) and chirality operator $\iota(X)=\chi X\chi^{-1}$. Depending on whether a symmetry with respect to one or the other involution is present Altland and Zirnbauer~\cite{AZ-class} found ten symmetric matrix spaces which is nowadays known as Altland-Zirnbauer's tenfold way. These ten classes in terms of matrices are listed in Table~\ref{chap1:tab:RMT-Symmetry}. 3 classes are up to an imaginary unit equal to the classical Lie algebras of compact groups, 5 classes are sets of chiral matrices and the other 5 classes have no chiral symmetry. Actually, the tenfold way only consists of the symmetric matrix spaces $\mathfrak{p}$ where $\mathfrak{h}$ is the Lie algebra of a compact classical group. One can also consider the Lie algebras themselves though they are also found as a matrix $\mathfrak{p}$ when multiplying an imaginary unit to it, so that it is sufficient to concentrate on  $\mathfrak{p}$ alone.

\begin{table}[t!]
\centering
\rotatebox{0}{\small
\begin{tabular}{|c|c|c|c|}
  \hline
 symmetric matrix space $\mathfrak{p}$
                 & \hspace*{-0.2cm}$\begin{array}{c} \text{abbreviation} \\ \text{for\ Gaussian}\end{array}$\hspace*{-0.2cm} & \hspace*{-0.2cm}$\begin{array}{c} {\rm Cartan} \\ {\rm class}\end{array}$\hspace*{-0.2cm} & random matrix $H\in\mathfrak{p}$  \\
  \hline\hline
  $\begin{array}{c} \text{Hermitian matrices} \\ \text{$i$ times Lie algebra of }{\rm U}(N) \end{array}$		&	GUE$(N)$	&	A	&	$H=H^\dagger\in\mathbb{C}^{N\times N},\ N\in\mathbb{N}$  \\ \hline
  real symmetric matrices		&	GOE$(N)$	&	AI	&	$\overset{\ }{H=H^T=H^\dagger\in\mathbb{R}^{N\times N}},\ N\in\mathbb{N}$  \\ \hline
   $\begin{array}{c} \text{Hermitian self-dual (quaternion) matrices} \end{array}$		&	GSE$(N)$	&	AII	&	$\overset{\ }{\begin{array}{c} H=\hat{\tau}_2H^T\hat{\tau}_2=H^*\in\mathbb{C}^{2N\times 2N},\ N\in\mathbb{N}\end{array}}$  \\ \hline\hline
  $\begin{array}{c} \text{imaginary antisymmetric matrices} \\ \text{$i$ times Lie algebra of }{\rm O}(N) \end{array}$	&	GAOE$_\nu(N)$	&	B/D	&	$\overset{\ }{\begin{array}{c} H=-H^T=H^\dagger\in i\mathbb{R}^{(2N+\nu)\times(2N+\nu)},\  N\in\mathbb{N},\ \nu=0,1\end{array}}$  \\ \hline
  $\begin{array}{c} \text{Hermitian anti-self-dual matrices} \\ \text{$i$ times Lie algebra of }{\rm USp}(2N) \end{array}$	&	GASE$(N)$	&		C &	$\overset{\ }{\begin{array}{c} H=-\hat{\tau}_2H^T\hat{\tau}_2=H^*\in\mathbb{C}^{2N\times2N},\ N\in\mathbb{N}\end{array}}$  \\ \hline
  $\begin{array}{c} \text{chiral Hermitian matrices} \end{array}$		&	$\chi$GUE$_\nu(N)$	&	AIII	&	$\overset{\ }{\begin{array}{c} H=\left[\begin{array}{cc} 0 & W \\ W^\dagger & 0 \end{array}\right],\ W\in\mathbb{C}^{N\times(N+\nu)},\ N\in\mathbb{N},\ \nu\in\mathbb{N}_0\end{array}}$  \\ \hline
  $\begin{array}{c} \text{chiral real symmetric matrices} \end{array}$		&	$\chi$GOE$_\nu(N)$	&	B/DI	&	$\overset{\ }{\begin{array}{c} H=\left[\begin{array}{cc} 0 & W \\ W^\dagger & 0 \end{array}\right],\ W=W^*\in\mathbb{R}^{N\times(N+\nu)},\ N\in\mathbb{N},\ \nu\in\mathbb{N}_0\end{array}}$  \\ \hline
   $\begin{array}{c} \text{chiral Hermitian self-dual matrices} \end{array}$		&	$\chi$GSE$_\nu(N)$	&	CII	&	\hspace*{-0.3cm}$\overset{\ }{\begin{array}{c} H=\left[\begin{array}{cc} 0 & W \\ W^\dagger & 0 \end{array}\right],\ W=\hat{\tau}_2W^*\hat{\tau}_2\in\mathbb{C}^{2N\times2(N+\nu)},\ N\in\mathbb{N},\ \nu\in\mathbb{N}_0\end{array}}$\hspace*{-0.3cm}  \\ \hline
  $\begin{array}{c} \text{symmetric Bogoliubov-de Gennes matrices} \end{array}$	&	GBOE$(N)$	&	CI	&	$\overset{\ }{\begin{array}{c} H=\left[\begin{array}{cc} 0 & W \\ W^\dagger & 0 \end{array}\right],\ W=W^T\in\mathbb{C}^{N\times N},\ N\in\mathbb{N}\end{array}}$  \\ \hline
   	$\begin{array}{c} \text{antisymmetric Bogoliubov-de Gennes matrices} \end{array}$	&	GBSE$_\nu(N)$	&	B/DIII	&	\hspace*{-0.3cm}$\overset{\ }{\begin{array}{c} H=\left[\begin{array}{cc} 0 & W \\ W^\dagger & 0 \end{array}\right],\ W=-W^T\in\mathbb{C}^{(2N+\nu)\times (2N+\nu)},\ N\in\mathbb{N},\ \nu=0,1\end{array}}$\hspace*{-0.3cm}	  \\ \hline
\end{tabular}}
\caption{
The ten Hermitian symmetric (flat) matrix spaces~\cite{AZ-class}. In the second column the abbreviations for the invariant Gaussian random matrix ensembles are listed. The third column shows the Cartan symbol that is often used in condensed matter theory and it is reminiscent to the Cartan classification of the classical Lie algebras. The first three classes correspond to Dyson's threefold way.
}\label{chap1:tab:RMT-Symmetry}
\end{table}

Why should we care about this classification scheme? First of all, the ten classes have helped in the classification of topological insulators~\cite{topo-class} ten years later. Not only there they have had an impact. Actually, the Cartan involutions identified by Altland and Zirnbauer are not the only ones which exist in physics. Already in the early 90's, Verbaarschot~\cite{Chiral-threefold} has introduced the chirality operation to derive the threefold classification of QCD-like field theories in 4 dimensions. They helped to explain the strong oscillations of the level density of Dirac operators in the infrared limit which were observed in lattice simulations. 20 years later, DeJonghe, Frey, and Imbo~\cite{QCD-class} classified those for all dimensions which highlighted a Bott periodicity in the dimension. Only a few years later, we extended this to lattice discretisations of the QCD-like theories~\cite{QCD-lat-class} where we observed the Bott periodicity not only in the dimension but also the number of lattice directions where a parity (split of even and odd) of lattice sites can be defined. During the same year, also Hermitian SYK models have been classified in~\cite{SYK-class}. In all these classification schemes of physical systems, it became apparent that all ten symmetry classes can be realised and none seems to be purely of mathematical interest.

It is exactly the application in lattice QCD which underscores a second important point of the symmetry analysis. When discretising a quantum field theory like QCD it can happen that symmetries change, see~\cite{QCD-lat-class}. It is still unsolved that in these situation the limit to the continuum theory actually exists as the wrong symmetry is rather persistent~\cite{BKPW-staggered}. RMT approaches tried to help to unravel this problem, see~\cite{Osborn-staggered}, but unfortunately with only limited success as the random matrix models reflecting the transition become easily analytically intractable. Nevertheless, RMT is helping to explore such limits which are numerically hard to come by due to a rapid growth of the matrix dimension of the physical operator in the system size.

\subsubsection{  Diagonalising random matrices -- The case of Hermitian matices}

The difference of the spectral statistics of the various classes becomes immediate if one considers the jpdf of the eigenvalues. Thus, we would like to sketch how to obtain the jpdf in the case of the Hermitian matrices ${\rm Herm}(N)$.

First of all, we need to create the setting. Say we consider a Hermitian random matrix $H\in{\rm Herm}(N)$ with a probability density $P(H)$ which should be invariant under conjugation of any unitary matrix $U\in{\rm U}(N)$, i.e., $P(H)=P(UHU^{-1})$. Then, we can diagonalise $H=UEU^{-1}$ with $E={\rm diag}(E_1,\ldots,E_N)$ comprising the eigenvalues. The unitary matrix $U$ encodes the eigenvectors as its column vectors. It is actually not drawn fully from the whole unitary group but from the subset ${\rm U}(N)/[{\rm U}^N(1)\times\mathbb{S}_N]$ where ${\rm U}^N(1)$ describes the diagonal unitary matrices that commute with $E$ and $\mathbb{S}_N$ is the symmetric group which permutes the eigenvalues but still gives a diagonal matrix. In mathematical terms, $\mathcal{N}={\rm U}^N(1)\times\mathbb{S}_N\subset{\rm U}(N)$ is the normaliser corresponding to the maximal Abelian subalgebra of the diagonal Hermitian matrices $\mathfrak{a}=\{{\rm diag}(E_1,\ldots,E_N)|E_1,\ldots, E_N\in\mathbb{R}\}\subset{\rm Herm}(N)$.

We need to take the coset ${\rm U}(N)/[{\rm U}^N(1)\times\mathbb{S}_N]$ instead of the whole unitary group ${\rm U}(N)$ as we would have otherwise an overcounting of the real degrees of freedom which is $N^2$ for ${\rm Herm}(N)$. It also shows when we compute the Jacobian of the change of variables from $H$ to $U$ and $E$. This can be effectively done with the help of the Riemannian metric as it is used in General Relativity Theory, namely
\begin{equation}\label{chap1:Riem.metric}
ds^2={\rm tr}\,dH^2=\sum_{a,b}dx^ag_{ab}dx^b={\rm tr}\,dE^2+{\rm tr}[U^{-1}dU,E]^2=\sum_{j=1}^NdE^2+2\sum_{1\leq a<b\leq N}(E_a-E_b)^2|\{U^{-1}dU\}_{ab}|^2
\end{equation}
where we used
\begin{equation}
dH=U(dE+[U^{-1}dU,E])U^{-1} \qquad{\rm and}\qquad [E,dE]=0.
\end{equation}
We identify the coordinates $dx^a$ with $dE_j$ and $\{U^{-1}dU\}_{ab}$ for $a\neq b$. We recall that $(U^{-1}dU)^\dagger=-U^{-1}dU$ is anti-Hermitian, meaning its image as a matrix differential $1$-form is in the Lie algebra of ${\rm U}(N)$. It is at this position apparent that the diagonal unitary matrices ${\rm U}^N(1)$ must be taken out as otherwise the Riemannian metric $g$ would be degenerate, meaning it would have zero eigenvalues. Furthermore, it highlights a connection to representation theory because the factors $(E_a-E_b)^2$ are the squared roots, meaning the eigenvalues of the linear map $U^{-1}dU\mapsto [[U^{-1}dU,E],E]$.

The Riemannian volume element is then equal to
\begin{equation}
\begin{split}
d[H]=&\left(\prod_{j=1}^NdH_{jj}\right)\left(\prod_{1\leq a<b\leq N}^Nd{\rm Re}(H_{ab})d{\rm Im}(H_{ab})\right)\\
=&2^{-N(N-1)/2}\sqrt{g}\prod_{a}dx^a=2^{-N(N-1)/2}\left(\prod_{1\leq a<b\leq N}2(E_a-E_b)^2\right)\left(\prod_{j=1}^NdE_j\right)d\mu(U)=\Delta_N^2(E)d[E]d\mu(U).
\end{split}
\end{equation}
This result can be also computed via wedge products as exercised by Hua~\cite[Chapter III]{Hua}.
The factor
\begin{equation}\label{chap1:Vandermonde}
\Delta_N(E)=\prod_{1\leq a<b\leq N}(E_b-E_a)=\det\left[E_{a}^{b-1}\right]_{a,b=1,\ldots,N}
\end{equation}
is the Vandermonde determinant, which has got its name as it is the determinant of the Vandermonde matrix $\{E_{a}^{b-1}\}_{a,b=1,\ldots,N}$ which plays an important role in signal analysis and, thus, in telecommunications.
We use the notation of square brackets such as $d[H]$ and $d[E]$ if it is only the product of real independent differentials. The first factor $2^{-N(N-1)/2}$ comes from the identification of $d[H]$ with the Riemannian volume element resulting from the length element~\eqref{chap1:Riem.metric}. The measure $d\mu(U)$ is the induced Haar measure of the coset ${\rm U}(N)/[{\rm U}^N(1)\times\mathbb{S}_N]$. It can be explicitly constructed with the help of wedge products, i.e.,
\begin{equation}
d\mu(U)=\bigwedge_{1\leq a<b\leq N}{\rm Re}(\{U^{-1}dU\}_{ab})\wedge{\rm Im}(\{U^{-1}dU\}_{ab}).
\end{equation}
We will not dive further into this part as the probability density $P(H)$ only depends on the eigenvalues $E$ and we can integrate over $U$ yielding the volume of the coset ${\rm U}(N)/[{\rm U}^N(1)\times\mathbb{S}_N]$. The volume can be computed by making use of the fact that the constant is independent of the chosen probability density so that the Gaussian choice fixes it,
\begin{equation}
\int_{{\rm U}(N)/[{\rm U}^N(1)\times\mathbb{S}_N]}d\mu(U)=\frac{\int_{{\rm Herm}(N)}\exp[-\tr H^2/2]d[H]}{\int_{\mathbb{R}^N}\exp[-\tr E^2/2]\Delta_N^2(E)d[E]}.
\end{equation}
The integral in the numerator factorises into univariate Gaussians and gives $2^{N/2}\pi^{N^2/2}$. The denominator can be computed in various ways. One way is via orthogonal polynomials~\cite{Oxford-orthogonal}, which are in the current case the Hermite polynomials, and the Andr\'eief integral identity~\cite{Andreief}. We will come to that approach in Subsection~\ref{chap1:sec:universal}. Another one is via Selberg integrals~\cite[Chapter 4]{log-gas}. In the present case, this would yield
\begin{equation}
\int_{{\rm U}(N)/[{\rm U}^N(1)\times\mathbb{S}_N]}d\mu(U)=\prod_{j=1}^{N}\frac{\pi^{ j-1}}{ j!}=\frac{1}{(2\pi)^NN!}\prod_{j=1}^{N}\frac{2\pi^{ j}}{ j!}.
\end{equation}
Writing this product in terms of the second expression allows us to identify the volumes of the individual components of the coset ${\rm U}(N)/[{\rm U}^N(1)\times\mathbb{S}_N]$. The factor $N!$ is the number of elements of the symmetric group $\mathbb{S}_N$ and $(2\pi)^N$ the volume of the subgroup  ${\rm U}^N(1)$ while $2\pi^{ j}/j!$ is the volume of a $2j-1$ dimensional hypersphere $S^{2j-1}$ which is related to ${\rm U}(N)$ by the factorisation ${\rm U}(N)\simeq S^1\times S^3\times\ldots S^{2N-1}$. One can interpret $S^{2N-1}$ as the first column of a unitary matrix which is uniformly distributed on a sphere. $S^{2N-3}$ is then the second column which is not only normalised but also conditioned to be orthogonal to the first column. Since we deal with complex vectors such an orthogonality condition gives two linearly independent linear equations reducing the real dimension of the remaining vector space by $2$. This is then carried on for the other columns of a unitary matrix leading to this factorisation.

Summarising, the jpdf of the eigenvalues of a random matrix $H\in{\rm Herm}(N)$ is given by
\begin{equation}\label{chap1:GUE-jpdf}
p_{\rm Dyson}^{(2)}(E)=\prod_{j=1}^{N}\frac{\pi^{ j-1}}{ j!}P(E)\Delta_N^2(E).
\end{equation}
When choosing $P(H)=P_{\rm G\beta E}(H)$, see~\eqref{chap1:Gaussian.Wigner}, we obtain the classical Gaussian result.

\subsubsection{  Generalisation to all ten classes}

This procedure can be generalised to any of the ten symmetric matrix spaces, see Table~\ref{chap1:tab:RMT-Symmetry}. There are, however, some subtleties. For instance, the matrices cannot always be diagonalised with their corresponding group. This is for instance for the imaginary antisymmetric matrices the case whose eigenvectors are inherently complex and, hence, cannot be the column vectors of a real orthogonal matrix, which constitute the corresponding invariance group. Thus, the concept of diagonalisation is generalised to  quasi-diagonalisation and maximal Abelian subalgebras, which is for the antisymmetric matrices the $2\times2$ antisymmetric block-diagonal matrices. The other steps are essentially the same. Certainly, the normaliser group $\mathcal{N}$ we have to divide out as well as the squared roots, meaning the eigenvalues of the linear map $U^{-1}dU\mapsto [[U^{-1}dU,E],E]$, would change with the respective symmetric matrix space. The Jacobian is then a product of these squared roots.

\begin{table}[t!]
\centering
\rotatebox{0}{\small
\begin{tabular}{|c|c|c|c|c|c|}
  \hline
  RMT
                 & $\begin{array}{c} \text{Abbreviation for} \\ \text{Gaussian ensemble}\end{array}$ & \hspace*{-0.2cm} Cartan class \hspace*{-0.2cm} & $\alpha$ & $\beta$  & $\begin{array}{c} \text{Number of generic} \\ \text{zero eigenvalues}\end{array}$ \\
  \hline\hline
  $\begin{array}{c} \text{imaginary antisymmetric matrices } \end{array}$	&	$\begin{array}{c}  {\rm GAOE}_\nu(2N+\nu);\ \nu=0,1 \end{array}$	&	B/D	&	$2\nu$ & $2$ & $\nu$ \\ \hline
  $\begin{array}{c} \text{Hermitian anti-self-dual matrices} \end{array}$	&	GASE$(N)$	&		C &	$2$ & $2$ & $0$ \\ \hline
  $\begin{array}{c} \text{chiral Hermitian matrices} \end{array}$		&	$\begin{array}{c} \chi{\rm GUE}_\nu(N);\ \nu\in\mathbb{N}_0 \end{array}$	&	AIII	&	$2\nu+1$ & $2$ & $\nu$  \\ \hline
  $\begin{array}{c} \text{chiral real symmetric matrices} \end{array}$		&	$\begin{array}{c} \chi{\rm GOE}_\nu(N);\ \nu\in\mathbb{N}_0 \end{array}$	&	B/DI	&	$\nu$ & $1$ & $\nu$  \\ \hline
   $\begin{array}{c} \text{chiral Hermitian self-dual matrices} \end{array}$		&	$\begin{array}{c} \chi{\rm GSE}_\nu(N);\ \nu\in\mathbb{N}_0 \end{array}$	&	CII	&	$4\nu+3$ & $4$ & $2\nu$  \\ \hline
  $\begin{array}{c} \text{symmetric Bogoliubov-de Gennes matrices} \end{array}$	&	GBOE$(N)$	&	CI	&	$1$ & $1$  & $0$ \\ \hline
   	$\begin{array}{c} \text{antisymmetric Bogoliubov-de Gennes matrices} \end{array}$	&	$\begin{array}{c} {\rm GBSE}_\nu(N);\ \nu=0,1 \end{array}$	&	DIII	&	$4\nu+1$ & $4$ & $2\nu$	  \\ \hline
\end{tabular}}
\caption{
Parameters $\alpha$ and $\beta$ in~\eqref{chap1:jpdf-non-Dyson} of the seven Hermitian symmetric matrix spaces which do not belong to Dyson's threefold way.
}\label{chap1:tab:RMT-param}
\end{table}

When following this procedure, the jpdf of the eigenvalues in one of the three Dyson classes (first three ensembles in Table~\ref{chap1:tab:RMT-Symmetry}) is given by
\begin{equation}\label{chap1:jpdf-Dyson}
p_{\rm Dyson}^{(\beta)}(E)=\prod_{j=1}^N\frac{\pi^{\beta (j-1)/2}\Gamma[1+\beta /2]}{\Gamma[1+\beta j/2]}P(E)|\Delta_N(E)|^\beta,
\end{equation}
where $\beta=1,2,4$ is the Dyson index.
The jpdf of the remaining seven symmetry classes $\mathcal{M}$ is equal to~\cite{Ivanov-micro}
\begin{equation}\label{chap1:jpdf-non-Dyson}
p^{(\alpha,\beta)}(E)=C\prod_{j=0}^{N-1}\frac{\Gamma[1+\beta /2]}{\Gamma[1+\beta (j+1)/2]\Gamma[(\alpha+1+\beta j)/2]}P(E)|\Delta(E^2)|^\beta\ \prod_{j=1}^N|E_j|^{\alpha}\ ,
\end{equation}
where the constant is a Gaussian integral
\begin{equation}
C=\left\{\begin{array}{cl} \int_{\mathcal{M}} \exp[-\tr H^2/4]d[H], & \beta=1,2,\\  \int_{\mathcal{M}} \exp[-\tr H^2/8]d[H], & \beta=4 \end{array}\right.
\end{equation}
with $\mathcal{M}$ any of the ten matrix spaces. The constant $C$ is only a product of powers of $2$ and $\pi$. The factor $1/2$ difference in the exponential is due to Kramers degeneracy of the eigenvalues in the quaternion case.

The Dyson index $\beta$ is responsible for the level repulsion between the eigenvalues. This means in practical terms that the larger $\beta$ is the stiffer the spectrum becomes. This shows in the level spacing distribution that it is more concentrated around $1$, cf., Fig.~\ref{chap1:fig1:level-spacing}, as well as in stronger oscillations of the microscopic level densities or other correlation functions, cf., Figs.~\ref{chap1:fig8:hard-mic} and~\ref{chap1:fig9:soft-mic}.  In contrast, the index $\alpha$ creates a level repulsion from the origin. This parameter is related to the number of generic zero eigenvalues which a matrix of the considered the symmetric matrix space has. For instance, $\nu$ of $\chi$G$\beta$E is the rectangularity (modulus difference of the matrix dimensions) of the off-diagonal blocks, see Table~\ref{chap1:tab:RMT-param}.

As can be seen after closer inspection of~\eqref{chap1:jpdf-non-Dyson} with Table~\ref{chap1:tab:RMT-Symmetry}, the odd-dimensional imaginary antisymmetric matrices (class B, $\nu=1$) and the Hermitian self-dual matrices (class C) give the very same jpdf. A similar agreement can be found for the class B/DI with $\nu=1$ and CI. The reason is that the root system of these different matrix spaces is almost the same. Only the length of the roots is different which cancel with the normalisation of the probabilistic measures. Therefore, the difference in their statistics is not encoded in the eigenvalues, but the eigenvector statistics will be different as one deals with real vectors in one setting while they are quaternion vectors in the other setting.

From the jpdf~\eqref{chap1:jpdf-Dyson} and~\eqref{chap1:jpdf-non-Dyson} we can read off that those eigenvalues in the bulk  which are far away from the origin, meaning $E_j=E_0+\delta E_j$ with $|E_0|\gg|\delta E_j|$, see an effective jpdf which is approximately
\begin{equation}\label{chap1:bulk.approximation}
P(E)|\Delta(E^2)|^\beta\ \prod_{j=1}^N|E_j|^{\alpha}\approx {\rm const.}\, |\Delta(\delta E_j)|^\beta.
\end{equation}
This is the reason why for a long time only three Dyson classes were studied. The level repulsion for very small spacing looks like $s^\beta$ which is reflected in Wigner's surmise~\eqref{chap1:Wigner-surmise}. However, when zooming into the spectrum around the origin, in particular $E_0=0$, the spectral statistics drastically change and almost all ten symmetry classes show a very distinct behaviour for their own class~\cite{Ivanov-micro}.

\subsection{Non-Hermitian Classification}\label{chap1:sec:non-Hermitian}

Inspired by Altland and Zirnbauer's work~\cite{AZ-class}, Bernard and LeClair~\cite[version 1 of 2001]{BC-class} have tried to extend the classification scheme to non-Hermitian matrix spaces. They have used the Cartan decomposition approach explained at the beginning of Subsection~\ref{chap1:sec:Altland-Zirnbauer} and found 43 symmetry classes. Magnea~\cite{Magnea-class} retried a few years later to get explicit matrix representations of those matrices and found two new classes raising the number to 45. About ten years after Magnea's article, in a work by Kawabata et al.~\cite{KSUS-class}, when they studied topological properties of these non-Hermitian matrix spaces when introducing energy point and line gaps,  this number again changed to 38. After this work, in a new version of their original work also Bernard and LeClair agreed to this new number~\cite[version 1 of 2020]{BC-class}.

Where is this disagreement coming from? Firstly, the Cartan decomposition procedure is not very complicated but it is tedious of going through all kinds of combinations of possible involutions. Moreover, it is challenging to show that the various matrix spaces found are actually inequivalent and not a different representation of another one. This makes this procedure vulnerable for over-counting, and it is the reason why a rigorous proof of this classification is still missing, yet.

Also this classification scheme for the non-Hermitian matrix spaces found applications in the physical systems. 19 of the 38 classes have been found for dissipative many-body quantum chaotic SYK models~\cite{SYK-nonHerm-class}. 10 symmetry classes for steady states and again 19 for non-steady states were found for many-body Lindbladians in open quantum systems, see~\cite{Lindblad-class}. In~\cite{stochastic-class}, the authors considered the symmetry classification of Markov generators of classical stochastic processes which fall into 15 classes.

Already far earlier than Bernard and LeClair's work, Ginibre~\cite{Ginibre} has proposed a non-Hermitian counterpart of Dyson's ensembles. He considered the Lie algebras of the general linear groups as matrix spaces, meaning $\mathbb{R}^{N\times N}$ ($\beta=1$), $\mathbb{C}^{N\times N}$ ($\beta=2$), and $\mathbb{H}^{N\times N}\subset \mathbb{C}^{2N\times 2N}$ ($\beta=4$), with a Gaussian distribution of the form
\begin{equation}\label{chap1:Gin.dist}
P_{\rm Gin\beta E}(X)=\frac{1}{\mathcal{Z}_{\rm Gin\beta E}}\exp\left[-\frac{1}{2\sigma_\beta^2}{\rm tr} XX^\dagger\right].
\end{equation} 
Those ensembles are nowadays known as Ginibre ensembles, and they found applications in the study of open chaotic quantum system, which ranges back to a work by Grobe, Haake and Sommers~\cite{GHS}. Yet, also in many-body quantum chaos Ginibre statistics were recently observed~\cite{SLHC-Gin}. There are, however, works such as~\cite{NKL-repulsion} which caution that a similar conjecture such as the BGS conjecture may not hold in the non-Hermitian case.

Other non-Hermitian random matrix ensembles found also applications. To name  a few, the chiral counterparts of Ginibre's three ensembles were introduced and studied in QCD at finite baryon chemical potential~\cite{Stephanov,Osborn} and time-lagged time series in statistics~\cite{time-lagged}. A pseudo-Hermitian random matrix was introduced to study the Wilson lattice discretisation of the Dirac operator~\cite{DSV-lat,KVZ-Wilson}. The complex symmetric matrices found applications as scattering matrices for open chaotic quantum systems~\cite{symm-open}.

Ginbre was actually able to compute the expressions for the jpdf of the eigenvalues of Ginibre matrices which can be now complex. He used essentially the same method as demonstrated with the help of Hermitian matrices in Subsection~\ref{chap1:sec:Altland-Zirnbauer}, only that he went via wedge products instead of the Riemannian metric. This approach is equivalent so that we stick with the Riemannian metric here. There are, however, several significant differences to be considered.

Firstly, the Riemmannian metric must be group invariant and, hence, psudo-Riemannian which forces us to employ
\begin{equation}\label{chap1:pseudo-Riemann}
ds^2_{\rm Haar}={\rm Re}({\rm tr}\, dX^2),
\end{equation}
which is related to the Haar measure of the corresponding group, and not the positive Riemannian metric
\begin{equation}
ds^2_{\rm Euclidean}={\rm tr}\, dXdX^\dagger,
\end{equation}
which breaks the invariance under the conjugation $X\to VXV^{-1}$ of a group element $V$. For Hermitian matrices, the metric~\eqref{chap1:pseudo-Riemann} simplifies to the original one~\eqref{chap1:Riem.metric}.

Secondly, the diagonalisation $X=UZU^{-1}$ with $Z={\rm diag}(z_1,\ldots,z_N)$ and $S$ an invertible matrix (or quasi-diagonalisation in the general case) does not work for all matrices in a non-Hermitian symmetric matrix space. Indeed, the Jordan normal form may have Jordan blocks that are larger than only one-dimensional. As for those matrices the eigenvalues must be degenerate, the sets of those matrices are usually of measure zero. Thence, this difference may be not so severe when one does not consider probability weights which concentrate on these sets via Dirac delta functions.

More problematic is that the diagonalising matrix $S$ will be drawn from a coset $\mathcal{C}$ of non-compact groups whose group invariant measure $d\mu(S)$ cannot be normalised. Hence, the probability density cannot be group invariant as in the Hermitian case. This leads to coset integrals of the form $\int P(SXS^{-1})d\mu(S)$ which is in the Gaussian case of the form
\begin{equation}\label{chap1:non-Herm-int}
I(Z)=\int_{\mathcal{C}} \exp\left[-\frac{1}{2\sigma_\beta^2}{\rm tr} (S^\dagger S)Z(S^\dagger S)^{-1}Z^\dagger \right]d\mu(S).
\end{equation}
These integrals were computed by Ginibre~\cite{Ginibre} for the cosets $\mathcal{C}={\rm Gl}_{\mathbb{C}}(N)/[{\rm Gl}_{\mathbb{C}}^N(1)\times\mathbb{S}_N]$ and  $\mathcal{C}={\rm Gl}_{\mathbb{H}}(N)/[{\rm Gl}_{\mathbb{H}}^N(1)\times\mathbb{S}_N]\subset\mathbb{C}^{2N\times2N}$ which correspond to the complex ($\beta=2$) and quaternion ($\beta=4$) Ginibre ensemble, leading to the jpdf
\begin{equation}\label{chap1:GINUE}
p_{\rm GinUE}(Z)=\frac{1}{(2\sigma_2^2)^{N^2}N!\prod_{j=1}^{N}2\pi(j-1)!}|\Delta_N(Z)|^2\exp\left[-\frac{1}{2\sigma_2^2}\sum_{j=1}^N|z_j|^2\right].
\end{equation}
for the complex Ginibre ensemble ($\beta=2$) and
\begin{equation}\label{chap1:GINSE}
p_{\rm GinSE}(Z)=\frac{1}{\sigma_4^{4N^2}N!\prod_{j=1}^{N}4\pi(2j-1)!}\Delta_{2N}(Z,Z^*)\prod_{j=1}^N(z_j-z_j^*)\exp\left[-\frac{1}{\sigma_4^2}|z_j|^2\right].
\end{equation}
for the quaternion Ginibre ensemble ($\beta=4$). The jpdf for the quaternion case looks at first glance to be complex but one can readily check by writing out the Vandermonde determinant, see~\eqref{chap1:Vandermonde} that it is real and non-negative.

Ginibre partially failed in computing~\eqref{chap1:non-Herm-int} for the real case. This is also due to the second big problem with non-Hermitian random matrices. It might be that there are inequivalent maximal Abelian subalgebras $\mathfrak{a}_l$ of a non-Hermitian symmetric matrix space. This happens for the real square matrices without any symmetries for which the eigenvalues can come either in complex conjugate pairs or are real. The number $l$ of complex conjugate pairs  can be used to label these inequivalent maximal Abelian subalgebras $\mathfrak{a}_l$. This has also implication for the integral~\eqref{chap1:non-Herm-int}, as the coset changes with $l$ which is $\mathcal{C}={\rm Gl}_{\mathbb{R}}(N)/[{\rm Gl}_{\mathbb{R}}^{N-l}(1)\times{\rm O}^l(2)\times\mathbb{S}_{N-2l}\times\mathbb{S}_{l}]$. Ginbre could do the calculation for $l=0$ which interestingly reduces to the GOE result apart from an overall normalisation, see~\eqref{chap1:jpdf-Dyson} for $\beta=1$ and $P(X)$ a Gaussian.

It took almost 40 years until a closed form expression for the real Ginibre ensemble was derived by Lehmann and Sommers~\cite{Real-Gin}. They even extended the ensemble to the elliptic ensemble where a parameter $\tau$ introduces the strength of the non-Hermitian part in the random matrix $X$, with the motivation to model asymmetric spin flip interactions in neuronal networks. Sommers and Wieczorek~\cite{Gen-real-Gin} have proposed to write all the different sectors in terms of one jpdf with the help of Dirac delta functions, in particular they found for even $N$
\begin{equation}\label{chap1:GINOE}
p_{\rm GinOE}(Z)=\frac{1}{(2\sigma_1^2)^{N^2/2}N!\prod_{j=1}^{N}2^{j/2}\Gamma(j/2)}\Delta_{N}(Z){\rm Pf}\left[F(z_a,z_b)\right]_{a,b=1,\ldots,N},
\end{equation}
with
\begin{equation}
F(z_a,z_b)=\exp\left[-\frac{z_a^2+z_b^2}{2\sigma_1^2}\right]\biggl[2i\delta^{(2)}(z_a-z_b^*)\biggl(\Theta[{\rm Im}(z_a)]{\rm erfc}\left[\frac{\sqrt{2}{\rm Im}(z_a)}{\sigma_1}\right]-\Theta[{\rm Im}(z_b)]{\rm erfc}\left[\frac{\sqrt{2}{\rm Im}(z_b)}{\sigma_1}\right]\biggl)+{\rm sign}[{\rm Re}(z_a-z_b)]\delta[{\rm Im}(z_a)]\delta[{\rm Im}(z_b)]\biggl].
\end{equation}
The function ${\rm erfc}(x)$ is the complementary error function and ${\rm Pf}$ is the Pfaffian determinant which is essentially the exact square root of a determinant of the antisymmetric matrix inside the Pfaffian. The complex Dirac delta function is denoted by $\delta^2(z)=\delta[{\rm Re}(z))\delta[{\rm Im}(z))$ and ${\rm sign}(x)$ is the sign of a real number $x$. In this formula, we have also employed the Heaviside step function $\Theta(x)$ which is $1$ when $x>0$ and vanishes otherwise.

Interestingly, when going away from the real axis and zooming into the bulk of the eigenvalues, meaning $z_j=z^{(0)}+\delta z_j$ with $|{\rm Re}(z^{(0)})|\gg|{\rm Re}(\delta z_j)|$, one can show that the jpdfs for all three Ginibre ensembles behave like 
\begin{equation}\label{chap1:GinUE.approx}
p_{\rm Gin\beta E}(Z)\approx {\rm const.}\, |\Delta_N(\delta z)|^2,
\end{equation}
meaning they are independent of $\beta$ apart from the constant. When comparing this with the Hermitian counterpart~\eqref{chap1:bulk.approximation}, one would conclude that for the bulk statistics one would find only a single type of local statistics unlike Dyson's threefold way. This is, however, not the case despite that the three Ginibre ensembles are indeed in the same universality class of the local spectral statistics in the bulk of eigenvalues~\cite{Bulk-3Gin}, which was extended to more general Wigner matrix ensembles in these matrix spaces in~\cite{TaoVu-non-Herm}, see Subsection~\ref{chap1:sec:unfolding} for more about this.

Similar results as those in~\eqref{chap1:GINUE},~\eqref{chap1:GINSE} and~\eqref{chap1:GINOE} have been derived in~\cite{Gin-prod-complex} for the chiral complex Ginibre ensembles (also called induced Ginibre ensembles). The jpdf of the eigenvalues of the Gaussian ensemble for those six ensembles, the original Ginbre and their chiral counterpart, and for the pseudo-Hermitian matrices~\cite{DSV-lat,KVZ-Wilson}, which satisfy $\gamma_5 X\gamma_5=X^\dagger$ with $\gamma_5={\rm diag}(\mathbf{1}_p,-\mathbf{1}_{N-p})$, are the only one of 38 classes that are known. Despite that the probability density on the matrix side, which always has the form $P(X)\propto \exp[-{\rm tr}\,XX^\dagger/(2\sigma^2)]$, looks extremely simple, the jpdfs for the Gaussian ensembles of the 31 other ensembles are, unfortunately, unknown. Mostly, this is due to coset integrals of the form~\eqref{chap1:non-Herm-int}.

We refer to a comprehensive overview of Ginibre and Ginibre-like ensembles given in the recent monograph~\cite{Byun-Forrester}.

It is also those group integrals which have prevented these computations far beyond the Gaussian ensembles. Two of those where it worked to compute the jpdf have been the truncated orthogonal, unitary and unitary symplectic matrices, see~\cite{real-trunc,complex-trunc,quaternion-trunc}, which are equal in distribution to the blocks of the three circular Dyson ensembles. The unitary case has found applications in modelling blocks of scattering matrices~\cite{Oxford-scattering,M-transport}, which is known as the Mexico approach~\cite{Mexico-model}. The second non-Gaussian class are the spherical ensembles~\cite{real-spheric,complex-spheric,quaternion-spheric} which are essentially the product random matrices $X_1^{-1}X_2$ where both matrices are Ginibre matrices. It was actually via the products where a whole set of new non-Gaussian random matrices have been studied~\cite{AkemannIpsen}. The motivation was driven by progressive scattering in telecommunication~\cite{AKW-tele} which has also a quantum chaotic counterpart~\cite{Ipsen-Kieburg} and is indeed related to quantum transport~\cite{Ipsen-Schomerus}.

In the complex case, even those could be extended to what is nowadays called P\'olya ensembles~\cite{KK-prod,KKF-prod}. Sadly, not much is known about the jpdfs of the eigenvalues beyond these ensembles except of matrix models of  normal matrices\footnote{A normal matrix $H$ is a matrix which commutes with its own Hermitian adjoint, i.e., $[H,H^\dagger]=0$.}.

\section{Local Spectral Statistics}\label{chap1:sec:unfolding}

Experience has taught us that universality of random matrix results sets in when we send the matrix dimension to infinity ($N\to\infty$). The double scaling limit, where one zooms onto the scale of the local mean level spacing,  is especially of physical interest. This endeavour requires the preparation of the eigenvalue spectrum of the empirical operator, such as the Hamiltonian or Dirac operator, as well as of the random matrix. It is as  important as the selection of the correct symmetry class of the matrices before any comparison with RMT can be done. 

In Subsection~\ref{chap1:sec:density}, we discuss the general concept of the level density, and recall its relation to the Green-function and the log-gas picture. We continue with the idea of the local mean level spacing and the level spacing distribution, in Subsection~\ref{chap1:sec:spacing}.  Therein, we explain their intimate relation to the level density and gap probabilities and derive the Poisson statistics for a broad class of probability distributions for illustration. We also make a small excursion to what the situation is for complex eigenvalue spectra. Subsection~\ref{chap1:sec:unfolding} is devoted to the unfolding procedure which prepares the ground for the local spectral spectral statistics in Subsection~\ref{chap1:sec:universal}, especially the $k$-point correlation functions. Here, we also review the orthogonal polynomial method, determinantal and Pfaffian point processes and the idea of Fredholm determinants and Pfaffians to compute gap probabilities. In Subsection~\ref{chap1:secLagrangians}, we finish with a brief discussion of how the spectral statistics are related to the potentials of the effective Lagrangians as they appear in non-linear $\sigma$-models. For this purpose, we go through the main steps of the supersymmetry method.

We would like to briefly mention that there is another approach by considering spacing ratios, which we do not cover here. In 2007, Oganesyan and Huse have proposed~\cite{OH-spacing} to consider ratios of the form
\begin{equation}
r_j=\frac{\min\{s_j,s_{j-1}\}}{\max\{s_j,s_{j-1}\}}\qquad {\rm with}\quad s_n=\lambda_{j+1}-\lambda_j
\end{equation}
with  ordered eigenvalues $\lambda_j$. The advantage is that it can be readily experimentally or numerically measured and that this approach avoids an unfolding if one considers only eigenvalues in the bulk of the spectrum. This is true because the eigenvalues are very close to each other so that the macroscopic level density, see Subsection~\ref{chap1:sec:density}, looks at most like a non-vanishing linear function. At the edges, where the level density may vanish or diverge, this idea breaks down which is one disadvantage. Another is that they are harder analytically tractable though also here Wigner-like surmises exist with low dimensional matrices~\cite{ABGVV}. There are also higher order spacing ratios possible~\cite{B-spacing}  which replace the $k$-point correlation functions. Furthermore, analogues for complex eigenvalue spectra have been tested and applied~\cite{SRP-spacing}.

\subsection{Level Density}\label{chap1:sec:density}

\subsubsection{  Definition and idea}

We consider the eigenvalues $E={\rm diag}(E_1,\ldots,E_N)$ of a matrix $H$. The empirical level density is then given by a sum of Dirac delta functions
\begin{equation}\label{chap1:emp.density}
\rho(\lambda)=\frac{1}{N}\sum_{j=1}^N\delta(E_j-\lambda).
\end{equation}
This level density is called empirical as this construction also holds for fixed matrices and operators and, therefore, experimental situations. In more practical terms, it means for an observable $F(\lambda)$ that its average over the eigenvalue spectrum is equal to
\begin{equation}
\int F(\lambda)\rho(\lambda)d\lambda=\frac{1}{N}\sum_{j=1}^NF(E_j),
\end{equation}
which is known as linear spectral statistics.
This construction works for both, real and complex eigenvalue spectra.

 The interpretation is very natural as an integral $\int_{\mathcal{I}}\rho(\lambda)d\lambda$ over a set $\mathcal{I}$ gives the fraction of eigenvalues in this set. In the case of a real eigenvalue spectrum, one can choose the family of intervals $\mathcal{I}=(-\infty,\mu]$ which is related to the level counting function
\begin{equation}\label{chap1:number-count}
N\,\Omega(\mu)=\int_{-\infty}^\mu \rho(\lambda)d\lambda=\sum_{E_j\leq\mu} 1
\end{equation}
with $\Omega(\mu)$ the empirical cumulative density function.

The tricky part with empirical eigenvalue spectra is to split it into a part $\rho_{\rm fl}(\lambda)$, which results from the statistical fluctuation of the realisation and should average out when one would have an ensemble of matrices or eigenvalue spectra, and a smooth part $\overline{\rho}(\lambda)$ which should be the ensemble average, i.e.,
\begin{equation}\label{chap1:av.density}
\overline{\rho}(\lambda)=\langle\rho(\lambda)\rangle=\left\langle\frac{1}{N}\sum_{j=1}^N\delta(E_j-\lambda)\right\rangle=\left\langle\frac{1}{N}{\rm tr}\,\delta(H-\lambda\mathbf{1}_N)\right\rangle,
\end{equation}
which is called the averaged level density (or in short: level density). We denote with $\langle.\rangle$ the ensemble average. The interpretation of an integral over a set $\mathcal{I}$ against the average level density is  richer than its empirical counterpart because it can be also understood as the probability to find an eigenvalue in the set. We underline that it does not specify how many we may find therein though the average value of eigenvalues therein will be $N\int_{\mathcal{I}}\overline\rho(\lambda)d\lambda$.

In experimental and numerical situations an ensemble average can be only done in approximation and sometime even not that if one has only a single Hamiltonian, for instance. In such a situation, the average over spectral windows has to be done, where one cuts the spectrum in several intervals which should be large enough to have enough eigenvalues inside but small enough so that the curvature of the density of the eigenvalue spectrum does not become a problem and prevents comparability.

\subsubsection{ Log-gas approach}

For random matrices the situation is much more straightforward since the ensemble average is the average over the matrix ensemble with the given probability density. If the jpdf of the eigenvalues $p(E)$ is known one integrates out all eigenvalues but one,
\begin{equation}\label{chap1:jpdf.density}
\overline{\rho}(\lambda)=\int p(\lambda,E_2,\ldots,E_N)dE_2\cdots dE_N.
\end{equation}
As we have seen in Subsection~\ref{chap1:sec:non-Hermitian}, such a jpdf is sometimes hard to come by, though. 

In the case of a jpdf of the form as in~\eqref{chap1:jpdf-Dyson},~\eqref{chap1:jpdf-non-Dyson},~\eqref{chap1:GINUE} and~\eqref{chap1:GINSE} and under the assumption that the probability weight can be written as $P(E)\propto \exp[-\beta N\,{\rm tr}\, V(E)]$, one can map the problem to a variational one for a static 2-dimensional Coulomb gas~\cite[Chapter 1]{log-gas}, which was already proposed by Dyson~\cite{Dyson-Brownian} for a harmonic potential on the real line and vanishing potential on the circle in the early 60's. For instance in the case of the Dyson classes, the logarithm of the jpdf is equal to
\begin{equation}
{\rm ln}[p_{\rm \beta}(E)]=-\beta N\sum_{j=1}^NV(E_j)+\beta\sum_{1\leq a<b\leq N}{\rm ln}|E_b-E_a|+{\rm const.}\approx -\beta N^2\left(\int_{-\infty}^\infty V(\lambda)\overline{\rho}(\lambda)d\lambda-\frac{1}{2}\int_{\mathbb{R}^2 }{\rm ln}|\lambda_1-\lambda_2|\overline{\rho}(\lambda_1)\overline{\rho}(\lambda_2)d\lambda_1d\lambda_2\right)+o(N^2).
\end{equation}
The idea is to minimise the energy functional in the leading term in the density $\overline{\rho}(\lambda)$, which gives the equation
\begin{equation}\label{chap1:variation}
0= V'(\lambda)-\int_{-\infty }^\infty\frac{\overline{\rho}(\widetilde{\lambda})d\widetilde{\lambda}}{\lambda-\widetilde{\lambda}}\qquad\text{under the conditions}\quad \overline{\rho}(\lambda)\geq0\quad{\rm and}\quad \int_{-\infty}^\infty \overline{\rho}(\lambda)d\lambda=1
\end{equation}
after varying in $\overline{\rho}(\lambda)$ and then differentiating in $\lambda$. The integral has to be understood in terms of a Cauchy principal value integral. This integral equation is a particular form of the linear Volterra equation of the first kind with the integral kernel $1/(\lambda-\widetilde{\lambda})$.

 In the case of a one cut solution, meaning the spectral density has a single interval $[a,b]$ as a support, and super-logarithmically growing potential $V(x)$ one can derive a closed expression which is called Tricomi's formula~\cite[Chapter 4]{Tricomi}
\begin{equation}\label{chap1:Tricomi}
\begin{split}
			\overline{\rho}(\lambda)=\frac{1}{\pi^2}\int_a^b\frac{V'(\mu)-V'(\lambda)}{\mu-\lambda}\frac{\sqrt{(b-\lambda)(\lambda-a)}}{\sqrt{(b-\mu)(\mu-a)}}d\mu\qquad{\rm with}\quad\int_a^b\frac{V'(\mu)}{\sqrt{(b-\mu)(\mu-a)}}d\mu=0\quad\text{and}\quad\int_a^b\frac{V'(\mu)\mu}{\sqrt{(b-\mu)(\mu-a)}}\frac{d\mu}{\pi}=1.
\end{split}
\end{equation}
The last two equations are the defining equations for the interval terminals $a$ and $b$ and result from the fact that the integral in~\eqref{chap1:variation} must asymptotically behave like $1/\lambda$ for any complex $|\lambda|\to\infty$, meaning it encodes the normalisation condition of $\overline{\rho}(\lambda)$.

The log-gas approach can be also employed for complex eigenvalue spectra. For those one can use the electrostatic solution $\overline\rho(z)=\Delta V(z)/\pi$ where $\Delta=\partial_z\partial_{z^*}$ is the 2-dimensional Laplacian. It seems that the problem is much simpler in this case. However, this is a wrong conclusion. The tricky part is to find the correct support of the level density $\overline\rho(z)$ which becomes a highly non-trivial potential theoretical question. We refer the reader to~\cite{Oxford-complex} for a deeper discussion about this problem.

\subsubsection{  Green function approach}

There is another way to obtain the level density via the Green function (also known as Cauchy, Hilbert or Stieltjes transform in other fields). For real eigenvalue spectra, it satisfies the following bijective relation to the level density
\begin{equation}\label{chap1:real.Green}
G(z)=\int_{-\infty}^\infty \frac{\overline{\rho}(\lambda)d\lambda}{z-\lambda}=\left\langle\frac{1}{N}{\rm tr}(z\mathbf{1}_N-H)^{-1}\right\rangle \qquad\Longleftrightarrow\qquad \overline{\rho}(E)=\frac{1}{\pi}\lim_{\varepsilon\searrow0}{\rm Im} G(E-i\varepsilon)\qquad{\rm\bf (for\ real\ spectra)},
\end{equation}
which is a result of the Sokhotski–Plemelj theorem. There is an analogue for complex eigenvalue spectra  where one uses a limit for the two-dimensional Dirac delta function which is the distributional relation $\partial_{z^*}(z-\lambda)^{-1}=\pi\delta^{(2)}(z-\lambda)$ so that
\begin{equation}\label{chap1:complex.Green}
G(z)=\int_{\mathbb{C}} \frac{\overline{\rho}(\lambda)d{\rm Re}(\lambda)d{\rm Im}(\lambda)}{z-\lambda}=\left\langle\frac{1}{N}{\rm tr}(z\mathbf{1}_N-H)^{-1}\right\rangle \qquad\Longleftrightarrow\qquad \overline{\rho}(z)=\frac{1}{\pi}\partial_{z^*}G(z)\qquad{\rm\bf (for\ complex\ spectra)}.
\end{equation}
This relation is based on the fact that the derivative with respect to $\partial_{z^*}$ vanishes wherever the function is holomorphic, and the Green function looses its holomorphicity only where the level density has its support and nowhere else. 

There are several ways to compute the Green function. One works via a moment expansion (given by a Neumann series) for asymptotically large $z$,
\begin{equation}
{\rm tr}(z\mathbf{1}_N-H)^{-1}=\frac{1}{z}+\frac{1}{z}\sum_{j=1}^\infty z^{-j}H^j \qquad\Longrightarrow\qquad G(z)=\frac{1}{z}+\frac{1}{z}\sum_{j=1}^\infty z^{-j}\left\langle\frac{1}{N}{\rm tr}H^j\right\rangle,
\end{equation}
which is also the reason why $zG(z)-1$ is called moment generating function. Wigner~\cite{Wigner-semi} went along  this idea to compute the famous semi-circle law. The problem with this method is that all moments need to exist which is not necessarily the case for all random matrices and applications one may have in mind. The benefit of this method is that it relates in the large $N$-limit to Free Probability~\cite{Free-prob-book} which is based on the planar diagram approximation. Voiculescu~\cite{Voiculescu} has developed free probability techniques to create short-cuts of this approximation and developed a whole theory about it in the mid 80's.

A second method of computing the Green function is via loop equations, e.g., see~\cite{Loop-equations}. For this method one starts with the expectation value of a total derivative with respect to a matrix entry, meaning
\begin{equation}
0=\left\langle\frac{1}{P(H)}\partial_{H_{ab}}[P(H)F_{c_1,\ldots,c_L}\{(z\mathbf{1}_N-H)^{-1}\}_{de}]\right\rangle,
\end{equation}
where $P(H)$ is the probability density of the random matrix $H$ and $F_{c_1,\ldots,c_L}$ is some tensor of order $L$. Carrying out the derivative and contracting the indices in various ways gives equations that relate different expectation values. It is the contraction of indices and its diagrammatic depiction which gives this method its name. For the Gaussian ensembles, one chooses $L=2$ and $F_{c_1,c_2}=H_{c_1c_2}$. Loop equations give a hierarchy of equations which in the large $N$-limit usually simplify and reduce to algebraic equations. For instance, this is the case of the level densities resulting in Wigner's semi-circle law~\cite{Wigner-semi} and the Mar\v{c}enko-Pastur distribution~\cite{MP-dist} where the equations are quadratic.

Another method is via ratios of characteristic polynomials and supersymmetry, e.g., see~\cite{Oxford-SUSY,Oxford-char-poly} and Subsection~\ref{chap1:secLagrangians}. Here one uses the identities
\begin{equation}
\begin{aligned}
G(z)&=\frac{1}{N}\lim_{J\to0}\partial_{J}\left\langle\frac{\det[(z+J)\mathbf{1}_N-H]}{\det[z\mathbf{1}_N-H]}\right\rangle&\qquad&\text{for Hermitian}\ H,\\
G(z)&=\frac{1}{N}\lim_{J,\varepsilon\to0}\partial_{J}\left\langle\frac{\det[(z+J)\mathbf{1}_N-H]\det[z^*\mathbf{1}_N-H^\dagger]}{\det\left[\begin{array}{cc} \varepsilon\mathbf{1}_N & z\mathbf{1}_N-H\\ z^*\mathbf{1}_N-H^\dagger &  \varepsilon\mathbf{1}_N \end{array}\right]}\right\rangle&\qquad&\text{for non-Hermitian}\ H.
\end{aligned}
\end{equation}
The more complicated expression for the non-Hermitian case is needed to guarantee the integrability when rewriting the determinants in terms of Gaussian supervectors. This approach is known as Girko's Hermitisation trick~\cite{Girko}.

There are even more techniques to compute the level density and/or the Green functions, yet, we would like to go on with another quantity which is important in the spectral statistics, namely the level spacing distribution.

\subsection{Local Level Spacing Distribution}\label{chap1:sec:spacing}

\subsubsection{  Definition of the local mean level spacing and the local level spacing distribution}

Let us first start with a real eigenvalue spectrum where the unordered eigenvalues $E={\rm diag}(E_1,\ldots,E_N)$ shall be ordered in the new variables $\lambda_1\leq\lambda_2\leq\ldots\leq\lambda_N$. In a heuristic approach, we can define the empirical local mean level spacing inside an interval $[a,b]$ as follows
\begin{equation}
\overline{s}_{\rm emp}([a,b])=\frac{\sum_{j=1,\ldots,N; \lambda_j,\lambda_{j+1}\in[a,b]}(\lambda_{j+1}-\lambda_j)}{\sum_{j=1,\ldots,N; \lambda_j,\lambda_{j+1}\in[a,b]}1}.
\end{equation}
The sum in the numerator is actually telescopic and measures the distance between the smallest and largest eigenvalue in $[a,b]$ while the denominator counts the number of pairs. When it comes to an ensemble average one actually defines it as
\begin{equation}\label{chap1:mean-level-spacing}
\overline{s}([a,b])=\frac{\left\langle\sum_{j=1,\ldots,N; \lambda_j,\lambda_{j+1}\in[a,b]}(\lambda_{j+1}-\lambda_j)\right\rangle}{\left\langle\sum_{j=1,\ldots,N; \lambda_j,\lambda_{j+1}\in[a,b]}1\right\rangle}.
\end{equation}
In the limit of large $N\to\infty$, the denominator becomes
\begin{equation}
\left\langle\sum_{j=1,\ldots,N; \lambda_j,\lambda_{j+1}\in[a,b]}1\right\rangle\overset{N\gg1}{\sim} N\int_{a}^b\overline{\rho}(\lambda)d\lambda.
\end{equation}
while the numerator is the distance between $\lambda_{\max}([a,b])=\max_{\lambda_j\in[a,b]}\{\lambda_j\}$ and $\lambda_{\min}([a,b])=\min_{\lambda_j\in[a,b]}\{\lambda_j\}$. When $[a,b]$ lies fully inside the support of $\overline{\rho}(\lambda)$ it simplifies to the length of the interval, so that we have
\begin{equation}\label{chap1:spacing.interv}
\overline{s}([a,b])\overset{N\gg1}{\sim}\frac{b-a}{N\int_{a}^b\overline{\rho}(\lambda)d\lambda}.
\end{equation}
When choosing $a=\lambda_0-\varepsilon$ and $b=\lambda_0+\varepsilon$ for $\varepsilon\to0$, this becomes the well-known relation between the local mean level spacing and the the level density at the point where one zooms in
\begin{equation}\label{chap1:mean-level-spacing.limit}
\overline{s}(\lambda_0)\overset{N\gg1}{\sim}\frac{1}{N\overline{\rho}(\lambda_0)}.
\end{equation}
This derivation shows that there are deviations to be expected when a  spectral edge or another type of critical point is approached. Then, the level density may vanish or diverge. Those situations need a non-linear unfolding which is also the reason why  the approach of spacing ratio will break down at such points, see more in Subsection~\ref{chap1:sec:unfolding}.

The level spacing distribution can be defined in a very similar way. The problem is, however, that one wants to have a comparable quantity across various eigenvalue spectra, e.g., see the comparison of level statistics in the seminal lecture notes~\cite{Bohigas-Giannoni} or the extended collection in~\cite{Hayes}. Those spectra can vary over several magnitudes and may have even different physical units. But also inside a single eigenvalue spectrum it is very likely that their local mean level spacing changes drastically. To guarantee this comparability, one normalises the local level spacing distribution so that its first moment is one, especially its definition over the spectral window $[a,b]$ can be written as follows
\begin{equation}\label{chap1:level-spacing}
p_{\rm sp}(s,[a,b])=\frac{\left\langle \sum_{j=1,\ldots,N; \lambda_j,\lambda_{j+1}\in[a,b]}\delta(s- (\lambda_{j+1}-\lambda_j)/\overline{s}([a,b]))\right\rangle}{\left\langle\sum_{j=1,\ldots,N; \lambda_j,\lambda_{j+1}\in[a,b]}1\right\rangle}.
\end{equation}
As can be readily checked the distribution is indeed properly nomalised, meaning $\int_{0}^\infty p_{\rm sp}(s,[a,b])ds=\int_{0}^\infty sp_{\rm sp}(s,[a,b])ds=1$.

Let us go through three standard choices to see what we get. For the $2$-level GOE, GUE and GSE we choose $[a,b]=\mathbb{R}$ for which the denominator in~\eqref{chap1:mean-level-spacing} and~\eqref{chap1:level-spacing} is simply unity as there is only one pair of eigenvalues. The remaining calculation is straightforward when plugging in the jpdf $p_{\rm  G\beta E}\propto |\lambda_2-\lambda_1|^\beta \exp[-(\lambda_1^2+\lambda_2^2)/(2\sigma_\beta^2)]$, and we find Wigner's surmise~\eqref{chap1:Wigner-surmise}. Note that the standard deviation $\sigma_\beta$ automatically drops out as it should because we normalise the distribution with respect to the mean level spacing which is proportional to it.

The next simple example is the quantum harmonic oscillator (or picket fence spectrum) where the eigenvalues $\lambda_j=\hbar\omega( j+1/2)$ with $j=1,2,3,\ldots$ are equidistant, with $\hbar$ Planck's constant divided by $2\pi$. Here, $\omega>0$ is the frequency and determines the level spacing $\bar{s}=\hbar\omega$. It actually drops out, regardless how we set $[a,b]$ as long as $b-a\geq \hbar\omega$ and $a\geq \hbar\omega/2$. Then, the level spacing distribution is simply the Dirac delta function
	\begin{equation}\label{chap1:PF}
	p_{\rm sp,\ PF}(s)=\delta(s-1).
	\end{equation}
It appears as a vertical line in Fig.~\ref{chap1:fig1:level-spacing} and can be obtained as the limit $\beta\to\infty$ of Wigner's surmise.

\subsubsection{Poisson statistics}\label{chap1:sec:poisson}
	
As a third standard example, we choose $N$ independently distributed eigenvalues $E_1,\ldots,E_N$ drawn from a probability density $\omega(x)$ which should be continuous almost everywhere, and the interval $[a,b]$ is assumed to be fully in its support, though it can be also at its edge. With the help of this example we would like to show how such a calculation works and some subtleties that may occur.

The jpdf in the ordered eigenvalues $\lambda_1\leq\ldots\leq \lambda_N$ is given by
\begin{equation}
			P(\lambda)=N!\omega(\lambda_N)\prod_{j=1}^{N-1} \Theta(\lambda_{j+1}-\lambda_j)\omega(\lambda_j).
\end{equation}
The factor $N!$ reflects the number of possible orderings the original eigenvalues $E_1,\ldots,E_N$, and the Heaviside step function enforces the ordering for $\lambda_j$. In the first step of computing the local level spacing distribution in an interval $[a,b]$, we derive a closed expression for the marginal probability density of a consecutive eigenvalue pair $(\lambda_l,\lambda_{l+1})$ which is
\begin{equation}\label{chap1:int.poisson}
			\begin{split}
			p_l(\lambda_l,\lambda_{l+1})=&N!\Theta(\lambda_{l+1}-\lambda_l)\int_{-\infty< \lambda_1\leq\ldots\leq \lambda_{l-1}\leq \lambda_l}d\lambda_1\cdots d\lambda_{l-1}\int_{\lambda_{l+1}\leq\lambda_{l+2}\leq\ldots\leq\lambda_N<\infty}d\lambda_{l+2}\cdots d\lambda_{N}\prod_{j=1}^{N}\omega(\lambda_j)\\
			=&\Theta(\lambda_{l+1}-\lambda_l)\omega(\lambda_{l})\omega(\lambda_{l+1})\ \frac{N!}{(l-1)!(N-l-1)!} \left(\int_{-\infty}^{\lambda_{l}}\omega(\lambda) d\lambda\right)^{l-1}\left(\int_{\lambda_{l+1}}^{\infty}\omega(\lambda) d\lambda\right)^{N-l-1}.
			\end{split}
\end{equation}
Here, we have undone the ordering of the integration variables smaller than  $\lambda_l$ as well as those that are larger than $\lambda_{l+1}$ which gives factorising integrals, though we need to pay the price in the form of the factorials in the denominator. We already see in this expression that the cumulative density function,
\begin{equation}
\Omega(\lambda)=\int_{-\infty}^\lambda \omega(x)dx\in[0,1],
\end{equation}
 plays an important role.
 
 The number of consecutive eigenvalues in the interval $[a,b]$ can be expressed in the density~\eqref{chap1:int.poisson},
\begin{equation}
\begin{split}
			\left\langle\sum_{j=1,\ldots,N; \lambda_j,\lambda_{j+1}\in[a,b]}1\right\rangle=&\sum_{l=1}^N\int_{a\leq \lambda_l\leq\lambda_{l+1}\leq b}p_l(\lambda_l,\lambda_{l+1})d\lambda_ld\lambda_{l+1}=N(N-1)\int_{a\leq \mu_1\leq\mu_2\leq b}\left[1+\Omega(\mu_1)-\Omega(\mu_2)\right]^{N-2}d\Omega(\mu_1)d\Omega(\mu_2)\\
			=&N\left[\Omega(b)-\Omega(a)\right]+[1-(\Omega(b)-\Omega(a))]^N-1,
\end{split}
\end{equation}
where  we used that the sum is a binomial sum.  In this expression, it becomes apparent that the number of pairs grows whenever $\Omega(b)-\Omega(a)=\int_a^b\omega(x)dx\gg1/N$   which is exactly what we want to have good statistics. Then it means also that the terms $[1-(\Omega(b)-\Omega(a))]^N-1$ can be neglected as the first term is exponentially small while the second term is constant.

In a similar way, we can compute
\begin{equation}
			\begin{split}
			\left\langle\sum_{j=1,\ldots,N; \lambda_j,\lambda_{j+1}\in[a,b]}(\lambda_{j+1}-\lambda_j)\right\rangle=&N(N-1)\int_{a\leq \mu_1\leq\mu_{2}\leq b}(\mu_{2}-\mu_{1})\left[1+\Omega(\mu_1)-\Omega(\mu_2)\right]^{N-2}d\Omega(\mu_1)d\Omega(\mu_2).
			\end{split}
\end{equation}
Shifting and rescaling $\mu_2\to\mu_1+\mu_2/[N\omega(\mu_1)]$, this becomes asymptotically
\begin{equation}
			\begin{split}
			\left\langle\sum_{j=1,\ldots,N; \lambda_j,\lambda_{j+1}\in[a,b]}(\lambda_{j+1}-\lambda_j)\right\rangle=&\frac{N-1}{N}\int_a^bd\mu_1\int_0^{N\omega(\mu_1)(b-\mu_1)}d\mu_{2}\frac{\omega(\mu_1+\mu_2/[N\omega(\mu_1)])}{\omega(\mu_1)}\left[1+\Omega(\mu_1)-\Omega\left(\mu_1+\frac{\mu_2}{N\omega(\mu_1)}\right)\right]^{N-2}\mu_2\overset{N\gg1}{\sim}b-a,
			\end{split}
\end{equation}
as expected when $[a,b]$ lies in the support of $\omega(x)$; if this is not the case we would have found the length of the intersection of the support and the interval. Thence, the local mean level spacing is as given in~\eqref{chap1:spacing.interv}, in particular it is $\overline{s}([a,b])\ll b-a$. When the interval shrinks to a single point $\lambda_0$, the mean level spacing is equal to $\overline{s}(\lambda_0)=1/[N\omega(\lambda_0)]$.

For the last term to be computed, we have	
\begin{equation}
			\begin{split}
			\left\langle\sum_{j=1,\ldots,N; \lambda_j,\lambda_{j+1}\in[a,b]}\delta\left(s-\frac{\lambda_{j+1}-\lambda_j}{\overline{s}([a,b])}\right)\right\rangle=&N(N-1)\int_{a\leq \mu_1\leq\mu_{2}\leq b}\delta\left(s-\frac{\mu_{2}-\mu_1}{\overline{s}([a,b])}\right)\left[1+\Omega(\mu_1)-\Omega(\mu_2)\right]^{N-2}d\Omega(\mu_1)d\Omega(\mu_2)\\
			&\hspace*{-2cm}=N(N-1)\overline{s}([a,b])\int_a^bd\mu_1\Theta(b-\mu_1-\overline{s}([a,b])s)\omega(\mu_1)\omega(\mu_1+\overline{s}([a,b])s)\left[1+\Omega(\mu_1)-\Omega\left(\mu_1+\overline{s}([a,b])s\right)\right]^{N-2},
			\end{split}
\end{equation}	
where we evaluated the Dirac delta function. The difference
$b-\mu_1$ as well as $\mu_1$ itself is almost surely of the same order as the length of the interval $b-a$ which is much larger than $\overline{s}(\lambda_0)$ so that we can Taylor expand in it, especially we use
\begin{equation}
\Omega\left(\mu_1+\overline{s}([a,b])s\right)-\Omega(\mu_1)\overset{N\gg1}{\sim}\frac{(b-a)\omega(\mu_1)}{N[\Omega(b)-\Omega(a)]}\,s
\end{equation}
for almost all $\mu_1\in[a,b]$. Then, the limit for the level spacing distribution~\eqref{chap1:level-spacing} is in this setting
\begin{equation}\label{chap1:result.Poisson}
p_{\rm sp}(s,[a,b])\overset{N\gg1}{\sim}\frac{b-a}{[\Omega(b)-\Omega(a)]^2}\int_a^bd\mu_1\omega^2(\mu_1)\exp\left[-\frac{b-a}{\Omega(b)-\Omega(a)}\omega(\mu_1)s\right],
\end{equation}
which becomes a Poisson distribution when taking the limit of the interval $[a,b]\to\{\lambda_0\}$ to a single point, namely
\begin{equation}\label{chap1:Poisson}
p_{\rm sp,\ 1D-Poisson}(s,\lambda_0)=e^{-s}.
\end{equation}
This result is the reason why this kind of statistics is called Poisson statistics.

As one can see, the result~\eqref{chap1:Poisson} follows from a very large class of functions. The continuity of $\omega(x)$, which should be almost everywhere,  should guarantee that its cumulative density $\Omega(x)$ is almost everywhere differentiable. Furthermore, the condition that $[a,b]$ must be fully in the support of $\omega(x)$ is needed so that $1/\omega(\mu_1)$ is defined almost everywhere in the integration domain. We note that the interval $[a,b]$ can touch the edges of the support, highlighting that also there one finds Poisson statistics.

When studying this calculation in detail, we notice that the order of limits is important. One needs to take first a finite interval  before taking the limit of large matrix dimension. It is required that the number of consecutive pairs of eigenvalues grows otherwise one cannot expect that universal statistics will appear. Once the limit $N\to\infty$ is done, one can perform the limit of the interval to a single point. Taking the reverse order of the limits would be wrong. It also reflects nicely what experimentalists know for a long time, that the spectral window $[a,b]$ must be chosen such that $1\ll N\int_a^b\overline{\rho}(\lambda)d\lambda\ll N$.

Let us mention that quite often the Poisson statistics is associated with integrability. This can be, however, a fallacy. Poisson statistics always appear when statistically independent spectra are overlapping. This is the reason why the Berry-Tabor conjecture~\cite{BT-conj} was stated only in one way, namely that when an integrable system, where the energy eigenvalues depend on more than one quantum number and they are  incommensurable in the energy, then the statistics will follow the Poisson statistics. They never stated the conjecture in the converse way because it would be wrong as can be seen for taking a direct sum of infinitely many statistically independent GUE ensembles. This was also observed in arithmetic billiards~\cite{BraunHaake} where the underlying dynamics is always chaotic but only at specific points of the parameter space.

\subsubsection{  Gap probabilities}

There is a more elegant way to calculated the level spacing distribution via gap probabilities. The probability that a certain set $\mathcal{I}$ is void is given by
\begin{equation}\label{chap1:gap.prob}
\mathbb{P}(\mathcal{I})=\left\langle\prod_{j=1}^N[1-\chi_{\mathcal{I}}(E_j)]\right\rangle,
\end{equation}
where $\chi_{\mathcal{I}}(E_j)$ is the indicator function which vanishes when $E_j\notin\mathcal{I}$ and is unity if $E_j\in\mathcal{I}$. In this way, the distribution of the largest and smallest eigenvalues for real eigenvalue spectra are, for instance, expressed as first derivatives of
\begin{equation}\label{chap1:extreme.fredholm}
p_{\max}(\lambda)=\partial_\lambda\mathbb{P}([\lambda,+\infty)) \qquad{\rm and}\qquad p_{\min}(\lambda)=-\partial_\lambda\mathbb{P}((-\infty,\lambda]).
\end{equation}
The derivative sets one of the eigenvalues at $\lambda$ while the others must be either smaller or larger than this value, respectively.

For the local level spacing distribution in an interval $[a,b]$, we rewrite
\begin{equation}
\begin{split}
p_{\rm sp}(s,[a,b])=&\frac{1}{\left\langle\sum_{j=1,\ldots,N; \lambda_j,\lambda_{j+1}\in[a,b]}1\right\rangle}\int_{a\leq \mu_1\leq\mu_2\leq b}\delta\left(s-\frac{\mu_2-\mu_1}{\overline{s}([a,b])}\right)\sum_{j=1}^Np_j(\mu_1,\mu_2)d\mu_1d\mu_2\\
=&-\frac{1}{\left\langle\sum_{j=1,\ldots,N; \lambda_j,\lambda_{j+1}\in[a,b]}1\right\rangle}\int_{a\leq \mu_1\leq\mu_2\leq b}\delta\left(s-\frac{\mu_2-\mu_1}{\overline{s}([a,b])}\right)\partial_{\mu_1}\partial_{\mu_2}\mathbb{P}([\mu_1,\mu_2])d\mu_1d\mu_2
\end{split}
\end{equation}
with $p_j(\mu_1,\mu_2)$ the distribution of the $j$th consecutive eigenvalue pair. When going over to relative $\delta\mu=(\mu_2-\mu_1)/2$ and center of mass coordinates $\mu=(\mu_2+\mu_1)/2$ and carrying out the Dirac delta function for the relative coordinate we find
\begin{equation}
\begin{split}
p_{\rm sp}(s,[a,b])=&\frac{\overline{s}([a,b])}{4\left\langle\sum_{j=1,\ldots,N; \lambda_j,\lambda_{j+1}\in[a,b]}1\right\rangle}\int_a^b  \left(\frac{4}{\overline{s}^2([a,b])}\partial_{s}^2-\partial_\mu^2\right)\mathbb{P}\left(\left[\mu-\frac{\overline{s}([a,b])}{2}s,\mu+\frac{\overline{s}([a,b])}{2}s\right]\right)d\mu\\
\overset{N\gg1}{\sim}&\frac{1}{b-a}\int_a^b  \partial_{s}^2\mathbb{P}\left(\left[\mu-\frac{\overline{s}([a,b])}{2}s,\mu+\frac{\overline{s}([a,b])}{2}s\right]\right)d\mu,
\end{split}
\end{equation}
where we assumed that $[a,b]$ is completely comprised in the support of the level density of $\overline{\rho}(\lambda)$ and assumed that $N\int_a^b\overline{\rho}(\lambda)d\lambda\gg1$. Taking the limit of the interval $[a,b]$ to a single point $\lambda_0$ yields the well-known relation~\cite[Chapter 6.1]{Mehta}
\begin{equation}\label{chap1:spacing.fredholm}
\begin{split}
p_{\rm sp}(s,\lambda_0)=&\partial_{s}^2\mathbb{P}\left(\left[\lambda_0-\frac{\overline{s}([a,b])}{2}s,\lambda_0+\frac{\overline{s}([a,b])}{2}s\right]\right).
\end{split}
\end{equation}

\begin{figure}[t!]
\centering
\includegraphics[width=.45\textwidth]{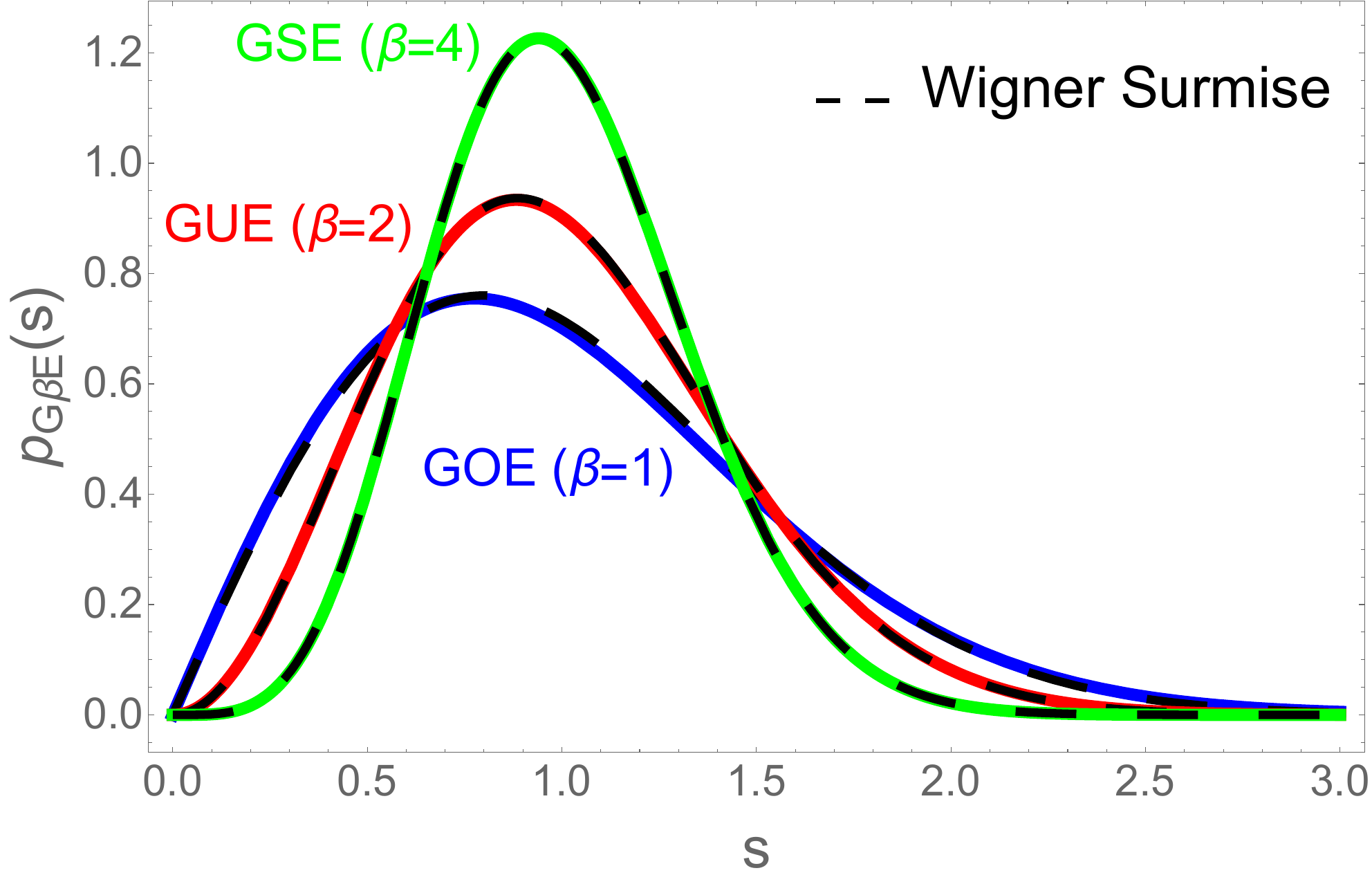}\qquad\includegraphics[width=.48\textwidth]{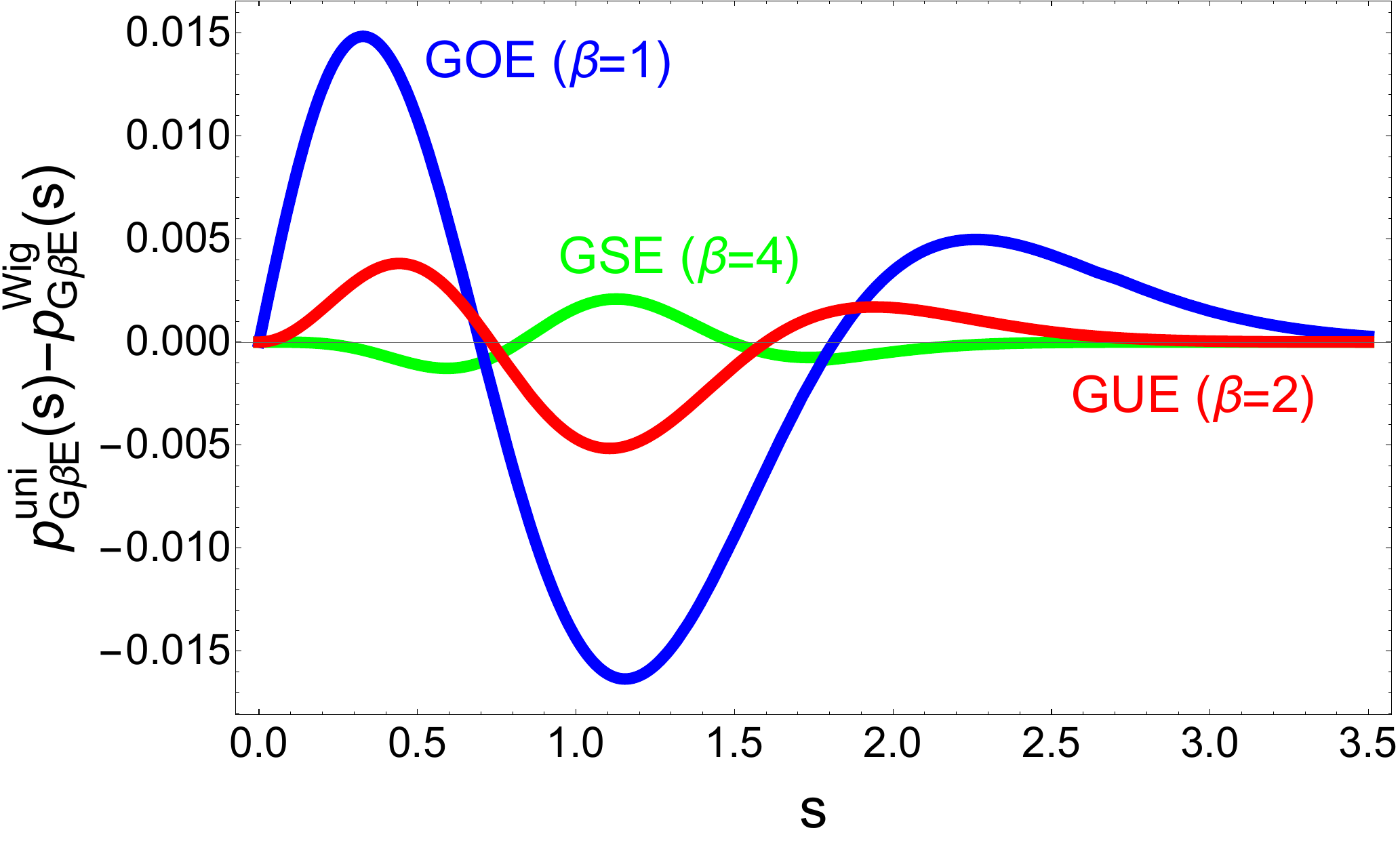}
\caption{{\bf Left plot:} level spacing distribution of the Wigner surmise (black dashed curves) for the G$\beta$E $2$-level system, see~\eqref{chap1:Wigner-surmise}, and the exact  bulk statistics when $N\to\infty$ (coloured solid curves). {\bf Right plot:}  difference of the exact bulk statistics and the Wigner surmise which is mostly less than one percent. Both plots were generated by the data and method described in~\cite{HaakeDietz}, where the Taylor expansion about the origin is interpolated to the asymptotic behaviour by a Pad\'e expansion.}
\label{chap1:fig2:Wigner-surmise-deviation}
\end{figure}

There are several techniques to compute the gap probability. The most successful is when the jpdf exhibits a determinantal or Pfaffian point process, see Subsection~\ref{chap1:sec:unfolding}. Then, one can approach the problem with the help of Fredholm determinants and Pfaffians~\cite[Chapter 9]{log-gas}, see also~\cite{Forrester-Fredholm}. Those let to the famous expressions in terms of solutions of Painlev\'e equations, such as the Tracy-Widom distribution for the largest eigenvalue at a soft-edge~\cite{TW-dist}. For the level spacing distribution, one finds that the Wigner surmise~\eqref{chap1:Wigner-surmise} is an extremely good approximation, see Fig.~\ref{chap1:fig2:Wigner-surmise-deviation} for the GUE case. Quite often the deviation is below the statistical noise in the empirical data such that the exact expression, which is tremendously more complicated, is not really needed.

\subsubsection{Complex eigenvalue spectra}

The gap probability approach is ideal when going over to complex eigenvalue spectra. Then, the interval of the level spacing distribution is usually replaced by a disc. In this way, closed expression of the local level spacing distribution are given for the 2-dimensional Poisson statistics  (independent eigenvalues),
\begin{equation}
p_{\rm sp,\ 2D-Poisson}(s)=\frac{\pi}{2}s\exp\left[-\frac{\pi}{4}s^2\right],
\end{equation}
and for the complex Ginibre ensemble~\cite{GHS},
\begin{equation}
p_{\rm sp,\ GinUE}(s)=\hat{s}\left(\prod_{j=1}^\infty \frac{\Gamma[1+j,\hat{s}^2s^2]}{\Gamma[1+j]}\right)\sum_{l=1}^\infty\frac{2(\hat{s}s)^{2l+1}e^{-\hat{s}^2s^2}}{\Gamma[1+l,\hat{s}^2s^2]}\qquad{\rm with}\quad \hat{s}\approx1.143,
\end{equation}
where $\Gamma[1+j,s^2]$ is the incomplete Gamma function.

 Comparison of the level spacing distribution of the 2-dimensional Poisson statistics and the one of Wigner's surmise~\eqref{chap1:Wigner-surmise-GOE} for the GOE shows that they are identical. The difference, however, is that for Wigner's surmise it is an approximation for the limit $N\to\infty$, while it is exact for the Poisson statistics. Moreover, Grobe, Haake and Sommers~\cite{GHS} have already mentioned that the $N=2$ surmises for complex eigenvalue spectra of the GinUE are a very bad approximation for the large $N$ limit in contrast to the statistics on the real line. However, the $N=4$ approximation is already in a regime where the deviations are only of a few percent, and $N=6$ is almost on top.

What is with the other non-Hermitian matrix spaces? Grobe and Haake gave a general argument in~\cite{Grobe-Haake} that in the bulk of eigenvalues away from any symmetry point and axes must have a level repulsion like $s^3$. This would agree with the observation for the jpdf of the three Ginibre ensembles~\eqref{chap1:GinUE.approx}, where the modulus square of the Vandermonde determinant gives a factor $s^2$ and an additional factor $s$ comes from the polar decomposition of the complex difference between the next neighbouring eigenvalues. Yet, Kawabata et al.~\cite{HKKU-univ} have seen numerically that some non-Hermitian symmetric matrices have a different spacing distribution despite that the level repulsion for very small spacing is like $s^3$. Those two matrix spaces that do not follow the Ginibre statistics are the complex symmetric matrices, $X=X^T\in{\rm Sym}_{\mathbb{C}}(N)$, (denoted by the Cartan symbol AI$^\dagger$) and the complex self-dual matrices, $X=\widehat{\tau}_2X^T\widehat{\tau}_2\in{\rm Self}_{\mathbb{C}}(2N)$, (denoted by the Cartan symbol AII$^\dagger$).  This was confirmed in another numerical study~\cite{AMP-spacing} by Akemann at al., where they matched the spacing distribution with a 2-dimensional Coulomb gas with Dyson indices $\beta=1.4$ for the class AI$^\dagger$ and $\beta=2.6$ for the class AII$^\dagger$. These observations gave rise to two conjectures:
\begin{enumerate}
\item	The classes AI$^\dagger$ (complex symmetric matrices) and AII$^\dagger$ (complex self-dual matrices) have a local bulk statistics of the complex eigenvalues different to the complex Ginibre ensemble (GinUE, class A).

\item	There are only three universal bulk statistics of the complex eigenvalues like in the real case of Dyson's threefold way with the classes $A$, AI$^\dagger$ and AII$^\dagger$ as their representatives
\end{enumerate}
The conjectures, actually, had been extended as it was observed for the three Ginibre ensembles that the universality even holds at the edges as long as one stays away from the symmetry axis~\cite{TaoVu-non-Herm,Edge-3Gin}, which was the real axis in the real and quaternion case.

\begin{figure}[t!]
\centering
\includegraphics[width=.5\textwidth]{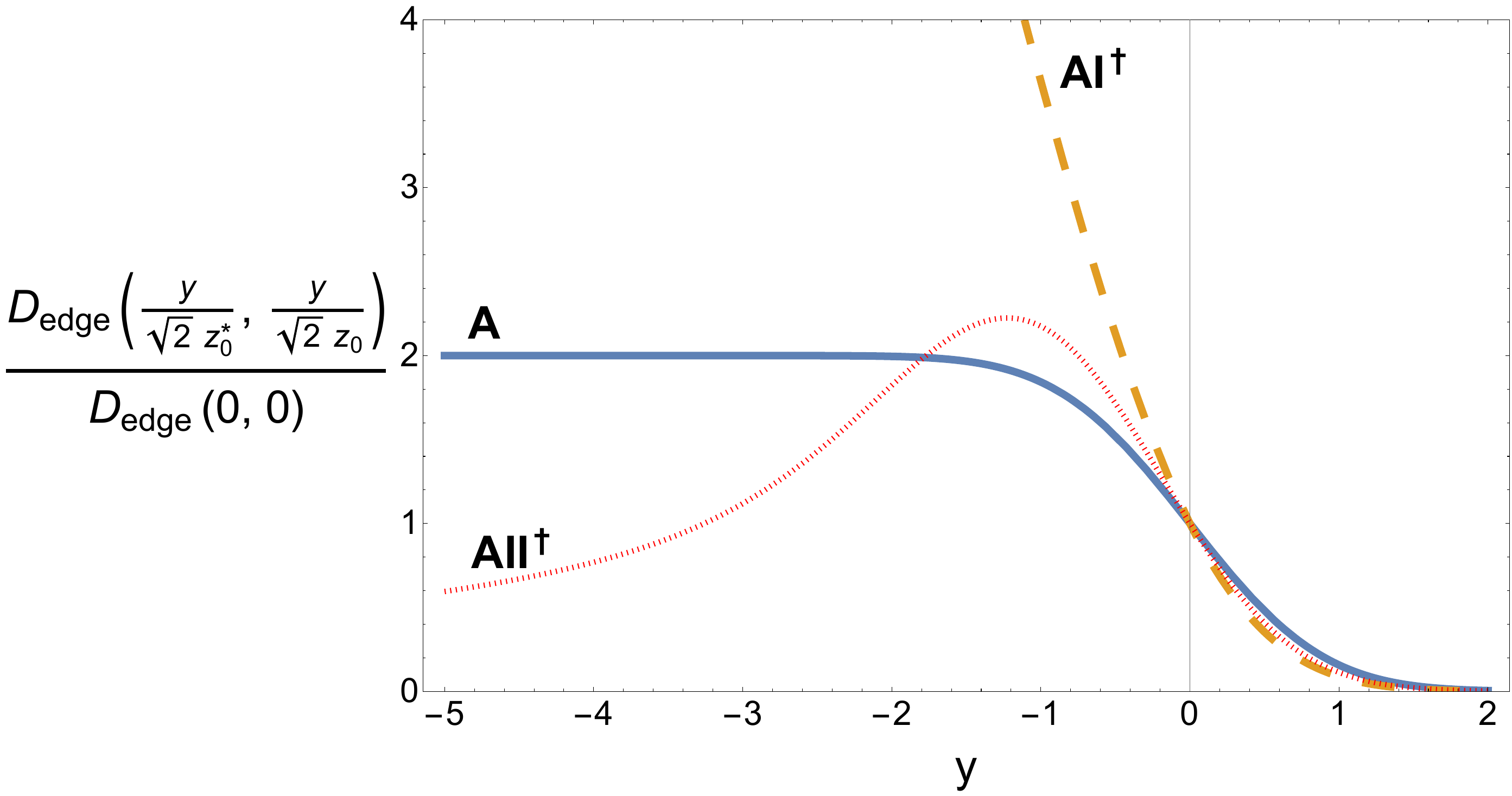}
\caption{Edge behaviour of the average of two characteristic polynomials~\eqref{chap1:def-char-pol} for the three symmetry classes A, AI$^\dagger$ and AII$^\dagger$. The results~\eqref{chap1:Dedge} are normalised at the origin for better comparability. The image is taken from~\cite{AAKP-determinants}.}
\label{chap1:fig3:edge-difference}
\end{figure}

Conjecture (1) has been very recently analytically shown in a series of works~\cite{LZ-duality,AAKP-determinants,KKR-spacing,Forrester-determinants,AAAMPRW-spacing} from various groups. In all of these groups, they independently looked at the same quantity namely averages of products of characteristic polynomials, such as of only two characteristic polynomials
\begin{equation}
D_N(z,w^*):=\langle \det[z\,\mathbf{1}-X]\det[w^*\,\mathbf{1}-X^\dagger]\rangle. 
\label{chap1:def-char-pol}
\end{equation}
For instance in~\cite{AAKP-determinants}, it has been derived that at an edge\footnote{There is no symmetry axis in the classes A, AI$^\dagger$ and AII$^\dagger$.} point $z_0\in\mathbb{C}$ with $|z_0|=1$ the limits are given by
\begin{equation}\label{chap1:Dedge}
\begin{aligned}
D_{\rm edge}^{\rm A}(\chi,\eta^*)=\frac{1}{2}{\rm erfc}(y),\quad D_{\rm edge}^{\rm AI^\dag}(\chi,\eta^*)=\frac{1}{\sqrt{2\pi}}e^{-y^2}-\frac{y}{\sqrt{2}}{\rm erfc}(y),\quad
D_{\rm edge}^{\rm AII^\dag}(\chi,\eta^*)=
\left(-\frac{\sqrt{\pi}}{4}{\rm erfc}\left(y\right){\rm erfi}(y/\sqrt{2})-i\sqrt{\pi}\,T\left(\sqrt{2}y,i/\sqrt{2}\right)\right) e^{-y^2/2}
\end{aligned}
\end{equation}
with $y=(z_0^*\chi+z_0\eta^*)/\sqrt{2}$.
The function ${\rm erfi}$ is imaginary error function and $T$ is Owen's T-function~\cite{Owen}. These results are shown in Fig.~\ref{chap1:fig3:edge-difference}. There are also other indicators such as effective Lagrangians that show a different behaviour, see Subsection~\ref{chap1:sec:unfolding}.

Unfortunately, there exist no explicit formulas for the level spacing distribution of the classes AI$^\dagger$ and AII$^\dagger$, yet. It will be certainly a challenge to get those as an explicit jpdf is lacking, as well, due to unknown coset integrals of the form~\eqref{chap1:non-Herm-int}. Hence, what is left with comparing physical systems and the random matrix models are either Monte-Carlo simulations or the numerical evaluation of 2-dimensional Coulomb gases with Dyson index $\beta=1.4$ and $\beta=2.6$, see~\cite{AMP-spacing}.

\subsection{Unfolding Spectra}\label{chap1:sec:unfolding}

\subsubsection{  Unfolding procedure}

The example of the Poisson statistics in Subsection~\ref{chap1:sec:poisson} has shown something surprising and, yet, very common. The result~\eqref{chap1:result.Poisson} gives an average of Poisson distributions with different rates, while zooming in at any point yields the Poisson distribution~\eqref{chap1:Poisson}. The reason is that the level density, here equal to the weight function $\omega(x)$, varies over the interval. Hence, the eigenvalue spectrum gets mixed up, especially this means that we have overlapping eigenvalue spectra that live on different scales and are therefore not comparable. Actually, if we had bijectively mapped the spectrum to the variables $0\leq\Omega(\lambda_1)\leq\Omega(\lambda_2)\leq\ldots\leq \Omega(\lambda_N)\leq1$ first, meaning the image of the cumulative density function, and computed the level spacing distribution in these new variables, we would have found for the whole spectrum the Poisson distribution without any distortion like in~\eqref{chap1:result.Poisson}.
This approach is called unfolding of a spectrum.

In general, unfolding of a real eigenvalue spectrum is easier than those of a complex spectrum or any 2-dimensional point process, see for instance~\cite{ACKMOP-spacing} where complex eigenvalue spectra and empirical data of bird observations have been unfolded. Thus, we concentrate on real spectra, only, at the moment.

Say we want to zoom in around the point $\lambda_0\in\mathbb{R}$ which should lie somewhere in the support of the level density $\overline{\rho}(\lambda)$, including its edges. One then starts with a number counting function similar to~\eqref{chap1:number-count} only centred at $\lambda_0$, namely
\begin{equation}
\Omega_{\lambda_0}(\lambda)=\int_{\lambda_0}^\lambda \overline{\rho}(\lambda)d\lambda.
\end{equation}
This quantity can be negative or positive since $\lambda_0$ can sit in the middle of the spectrum. $N\Omega_{\lambda_0}(\lambda)$ is then the average number of eigenvalues in the interval $[\lambda_0,\lambda]$ if $\lambda>\lambda_0$ and minus the average number in the interval $[\lambda,\lambda_0]$ when $\lambda<\lambda_0$. The local spectral scale is given when that number is of order one. Thus, it is natural to go over to the new variables
\begin{equation}\label{chap1:unfolding}
\mu_j=N\Omega_{\lambda_0}(\lambda_j)\qquad\Rightarrow\qquad d\mu_j=N\overline{\rho}(\lambda_j)d\lambda_j
\end{equation}
which are called unfolded eigenvalues. As $\overline{\rho}(\lambda)>0$ in the bulk of eigenvalues, one uses usually the macroscopic large $N$ limit  of the level density in this unfolding. But sometimes it is necessary to switch to some mesoscopic level density as it has been case for an infinite product of random matrices at the soft edge~\cite{ABK-product}. This is also the case when going to one of the edges where the macroscopic level density $\overline{\rho}(\lambda)$ may vanish or diverge at $\lambda_0$, e.g., see~\cite{AGK-hard-soft} where a transition from the hard to the soft edge has been studied.

\begin{figure}[t!]
\centering
\includegraphics[width=.9\textwidth]{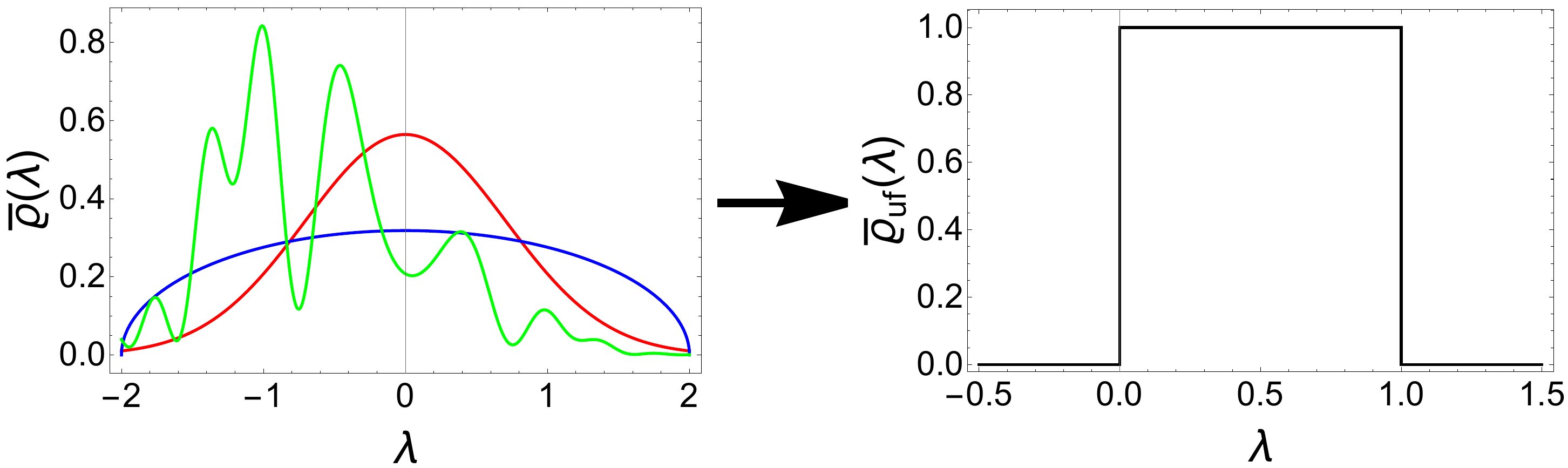}
\caption{Schematic picture of the unfolding. Various eigenvalue spectra (left plot) look in the unfolded eigenvalues flat (right plot). This increases the comparability of local spectral statistics on the scale of the local mean level spacing~\eqref{chap1:mean-level-spacing.limit} at any point $\lambda_0$ inside the support of the macroscopic level density $\overline{\rho}(\lambda_0)$.}
\label{chap1:fig4:unfolding}
\end{figure}

What is actually happening? The new unfolded eigenvalues $\mu_j$ have now a constant level density equal to the height $1/N$, see~\eqref{chap1:unfolding}. For the choice $\mu_j/N=\Omega_{-\infty}(\lambda_j)$ it has the form as in the schematic picture illustrated in Fig.~\ref{chap1:fig4:unfolding}. All densities are put on the same footing in this way, regardless where in the spectrum one zooms in. This has the advantage of comparability and gives a conclusion whether a system is chaotic or not better credibility. Moreover, in the limit $N\to\infty$ the local mean level spacing is in these new variables automatically equal to $1$, which is another indicator that the unfolded eigenvalues are preferable. This automatic rescaling of the level spacing distribution is due to the relation~\eqref{chap1:mean-level-spacing.limit}.

\subsubsection{  An example: unfolded spectra of integrable Hamiltonians}

Let us see what the unfolding does. For that reason, we consider a classical integrable bounded system. Following the Hamilton-Jacobi formalism and going over to action-angle coordinates, we know that the Hamiltonian $H=H(J_1,\ldots,J_M)$ becomes a differentiable function in terms of the action variables $J_1,\ldots, J_M$ and is independent of the angles canonical to $J_j$. The Bohr-Sommerfeld quantisation tells us that we can replace those by the quantum numbers $\hbar n_1,\ldots,\hbar n_M$. Thus, the energies are given by $E_{n_1,\ldots,n_M}=H(\hbar n_1,\ldots,\hbar n_M)$.

We fix all but one quantum number $n_L$ which corresponds to splitting the spectrum. Furthermore, we assume that $H(J_1,\ldots,J_M)$ is strictly increasing or decreasing in $J_L$. Then, $\widetilde{H}(J_L)=H(\hbar n_1,\ldots,\hbar n_{L-1},J_L,\hbar n_{L+1},\ldots,\hbar n_M)$ is a diffeomorphism due to the differentiability of the Hamiltonian in the action variable $J_L$. Consequently, the macroscopic level density of the eigenvalues $\widetilde{E}(n_L)=E_{n_1,\ldots,n_M}$ is given by $\overline{\rho}(\widetilde{E}(J_L/\hbar))=1/[\hbar|\partial_{J_L}\widetilde{H}(J_L)|]$ which follows from the empirical level density $\rho(\lambda)=\sum_{n_L}\delta(\lambda-\widetilde{E}(n_L))$ and the transformation properties of the Dirac delta functions under changing variables. The unfolded eigenvalues can be computed by the differential relation
\begin{equation}
d\mu=\overline{\rho}(\widetilde{E})d\widetilde{E}=\frac{d\widetilde{E}(J_L)}{\hbar|\partial_{J_L}\widetilde{H}(J_L)|} \qquad\Rightarrow\qquad \mu_j=n_{L,j}+{\rm const.}
\end{equation}
where the constant depends on the chosen $\lambda_0$ where we zoom in. 
This means that the unfolded eigenvalues are equidistant with distance $1$. 
Assuming that $n_L$ takes at least two integer values so that at least one consecutive pair exists, the unfolded level spacing distribution is equal to the one of a quantum Harmonic oscillator, see~\eqref{chap1:PF}.

One may wonder whether this argument would work for any empirical Hamiltonian, and, hence, empirical real eigenvalue spectrum, as every Hermitian operator can be diagonalised. The difference in a chaotic and an integrable system is that there is no differentiable invertible map $J_L\to\widetilde{H}(J_L)$ from the action angle $J_L$ to the energy which we used here.

The Berry-Tabor conjecture~\cite{BT-conj}, yielding the Poisson statistics~\eqref{chap1:Poisson}, enters here when averaging over all quantum numbers meaning one does not fix all apart from one. If the quantised action variables lead to curved energy contours in action space, the Poisson statistics should be visible. A standard example is the rectangular quantum billiard with incommensurate lengths, where the energies are given by $E_{j,l}=a j^2+ b l^2$ with $a/b\notin\mathbb{Q}$. From this insight, it becomes clear that one needs a collection of energies in at least two independent quantum numbers in an integrable system to find the Poisson statistics. Fixing all up to one will always return us to the picket fence statistics of the 1-dimensional quantum harmonic oscillator after unfolding.

\subsection{Universal Local Spectral Statistics}\label{chap1:sec:universal}

\subsubsection{  $k$-point correlation functions and point processes}

What is with other spectral quantities? For this purpose we consider the $k$-point correlation function. Like the empirical level density~\eqref{chap1:emp.density}, the empirical $k$-point correlation function can be defined in terms of a sum of products of Dirac delta-functions,
\begin{equation}
R_k(\lambda_1,\ldots,\lambda_k)=\sum_{\substack{1\leq j_1,\ldots,j_k\leq N \\ j_l\neq j_m\ {\rm for}\ l\neq m}}\delta (E_{j_1}-\lambda_1)\cdots\delta (E_{j_k}-\lambda_k),
\end{equation}
where we use  the normalisation $\int_{\mathbb{R}^k}R_k(E_1,\ldots,E_k)dE_1\cdots dE_k=N!/(N-k)!$. A probabilistic normalisation so that it can be seen as a marginal density is employed in the literature, as well. Integrating over the Cartesian  product of sets $\mathcal{I}_1\times\ldots\times\mathcal{I}_k$ gives the number of all $k$-tuples of eigenvalues with at least one eigenvalues in each of the sets $\mathcal{I}_j$.

For $k=1$, it is the empirical level density $\rho(\lambda)=R_1(\lambda)/N$, while $k=N$ yields the empirical jpdf of the eigenvalues $R_N(\lambda_1,\ldots,\lambda_N)/N!$ which can be written in the form of a permanent $R_N(\lambda_1,\ldots,\lambda_N)={\rm Perm}[\delta(E_a-\lambda_b)]_{a,b=1,\ldots,N}$.
When going over to expectation values we can also write the $k$-point correlation function in terms of integrals over the jpdf of the eigenvalues, namely
\begin{equation}\label{chap1:k-point-int}
\overline{R}_k(\lambda_1,\ldots,\lambda_k)=\frac{N!}{(N-k)!}\int dE_{k+1}\ldots dE_{N}p(\lambda_1,\ldots,\lambda_k,E_{k+1},\ldots,E_N).
\end{equation}
Like for the averaged level density, the interpretation gets richer when we integrate over $\mathcal{I}_1\times\ldots\times\mathcal{I}_k$. After properly normalising with $(N-k)!/N!$, this gives the probability to find at least one $k$-tuple of eigenvalues with at least one eigenvalues in each of the sets $\mathcal{I}_j$.

Particularly the $2$-point correlation function plays an important role in computing the covariance of two linear statistics, in particular for the observables $F(\lambda)$ and $G(\lambda)$ it is
\begin{equation}\label{chap1:covariance}
\left\langle \left[\sum_{j=1}^N F(E_j)-\left\langle\sum_{l=1}^N F(E_l)\right\rangle\right]\left[\sum_{j=1}^N G(E_j)-\left\langle\sum_{l=1}^N G(E_l)\right\rangle\right] \right\rangle=\int F(\lambda_1)G(\lambda_2)[\overline{R}_2(\lambda_1,\lambda_2)-\overline{R}_1(\lambda_1)\overline{R}_1(\lambda_2)]d\lambda_1d\lambda_2+\int F(\lambda)G(\lambda)\overline{R}_1(\lambda)d\lambda.
\end{equation}
The combination $C(\lambda_1,\lambda_2)=\overline{R}_2(\lambda_1,\lambda_2)-\overline{R}_1(\lambda_1)\overline{R}_1(\lambda_2)$ is also known as $2$-point cluster function~\cite[Chapter 6.1]{Mehta}.
In the case of $G(\lambda)=F(\lambda)=\Theta[(b-\lambda)(\lambda-a)]$ and real spectra  one would obtain the level number variance $\Sigma^2(L)$ inside the interval $[a,b]$ of length $L=b-a$, see~\cite{Dyson-Mehta-number-variance}. It is used quite often as a test of the present of quantum chaos, e.g., see the review~\cite{GMW-review} and references therein. Another choice would be $G(\lambda)=e^{2\pi ik \lambda}$ and $F(\lambda)=e^{-2\pi ik \lambda}$ which gives the spectral form-factor $S^{(\beta)}(k)$ which was introduced by Berry~\cite{Berry-spectral-form-factor} in the 80's. It is connected to quantum chaotic systems especially periodic orbits of those via the Gutzwiller trace formula~\cite{WG-periodic-orbits,T-Hamiltonian} as it is applied in the study of various systems such as the quantum kicked top~\cite{PB-kicked-top}, quantum graphs~\cite{BG-graphs} and quantum billiards~\cite{Dietz-billiards}. Recent applications of the spectral form factor can be found in the study of the SYK model~\cite{Jia-Verbaarschot-spectral-form}, the related problem of black hole quantum physics~\cite{Okuyamaa-Sakai-spectral-form}, out-of-time-correlations (OTOC)~\cite{KHH-spectral-form}, topological phases~\cite{SPAD-spectral-form} and various other fields. Those references are only a tiny fraction of examples and are by far exhaustive as vast as the applications of this observable are.

All these definitions and relations hold true for complex eigenvalue spectra, as well, and is nothing particular for real spectra. This is  also the case for the following discussion.

Considering the jpdfs~\eqref{chap1:jpdf-Dyson} and~\eqref{chap1:jpdf-non-Dyson} for the Gaussian ensembles, meaning $P(E)\propto \prod_{j=1}^N\exp[-E_j^2/2]$ for a specific choice of variance, or the jpdfs of the three Ginibre ensembles~\eqref{chap1:GINUE},~\eqref{chap1:GINSE}, and~\eqref{chap1:GINOE}, it becomes apparent that they satisfy a specific algebraic structure due to the Vandermonde determinant~\eqref{chap1:Vandermonde}. All of them can be written in one of the two functional forms
\begin{equation}\label{chap1:jpdf-det-Pf}
p(E)=\frac{1}{\mathcal{Z}_N}\det[f_a(E_b)]_{a,b=1,\ldots,N}{\rm det}\left[ h_a(E_b)\right]_{a,b=1,\ldots,N}\qquad{\rm or}\qquad p(E)=\frac{1}{\mathcal{Z}_N}\det[f_a(E_b)]_{a,b=1,\ldots,N}{\rm Pf}\left[g(E_a,E_b) \right]_{a,b=1,\ldots,N}
\end{equation}
with some suitably integrable functions or distributions (including Dirac delta functions) $f_a(E_b)$, $h_a(E_b)$ and $g(E_a,E_b) =-g(E_b,E_a) $, see the tables in~\cite{Kieburg-Guhr-1,Kieburg-Guhr-2} for various examples. 
For the Pfaffian case we assumed that there is only an even number $N$ of eigenvalues otherwise one would get an additional column in the Pfaffian, e.g., see~\cite[Chapter 6.3]{log-gas}. To comprise also those cases and unify the approach of both forms~\eqref{chap1:jpdf-det-Pf} one can instead consider the general form
\begin{equation}\label{chap1:jpdf-general}
p(E)=\frac{1}{\mathcal{Z}_N}\det[f_a(E_b)]_{a,b=1,\ldots,N}{\rm Pf}\left[\begin{array}{cc} g(E_a,E_c) & h_d(E_a)\\ -h_b(E_c) & 0\end{array}\right]_{\substack{a,c=1,\ldots,N\\ b,d=1\ldots,L}}.
\end{equation}
The Pfaffian case in~\eqref{chap1:jpdf-det-Pf} with an even number of eigenvalues is obtained when setting $L=0$ and for the odd case when $L=1$. The determinantal form is obtained when setting $g(E_a,E_c)=0$ as well as $L=N$ and using the identity
\begin{equation}
{\rm Pf}\left[\begin{array}{cc} 0& h_d(E_a)\\ -h_b(E_c) & 0\end{array}\right]_{\substack{a,c=1,\ldots,N\\ b,d=1\ldots,N}}={\rm det}\left[ h_a(E_b)\right]_{a,b=1,\ldots,N}.
\end{equation}
Actually, the general form~\eqref{chap1:jpdf-general} with arbitrary $L$ found indeed an application~\cite{Kieburg-mixing} when studying the eigenvalues of pseudo-Hermitian matrices~\cite{DSV-lat,KVZ-Wilson} and their Hermitised counterpart~\cite{Akemann-Nagao}  which were employed as random matrix models of the Wilson Dirac operator, a particular form of discretisation of the Dirac operator in QCD.

Usually the functions $f_a(E_b)$ are the monomials $E_b^{a-1}$ reflecting the Vandermonde determinant. However, there are cases such as in multi-matrix models, e.g., see~\cite{Akemann-Strahov}, where the Vandermonde determinant cancels out. Then, $f_a(E_b)$ can be some other set of functions. The functions $h_a(E_b)$ and $g(E_a,E_b)$ usually comprise the statistical weights. They also may contain monomials, such as those of the second Vandermonde determinant in the GUE jpdf~\eqref{chap1:GUE-jpdf}. 

What is the benefit of these algebraic structures? When computing the integrals~\eqref{chap1:k-point-int} these structures help. For instance, in the determinantal case~\eqref{chap1:jpdf-general}. Due to the properties of the determinant we are free to choose $f_a(x)$ and $h_b(x)$ to be biorthonormal, i.e.,
\begin{equation}
\int f_a(x)h_b(x)dx=\delta_{ab}
\end{equation}
which leads to a normalisation constant $\mathcal{Z}_N=N!$.
Then, the generalised Andr\'eief identity~\cite[Appendix C.1]{Kieburg-Guhr-1} (generalising~\cite{Andreief}) gives us
\begin{equation}\label{chap1:det.proc}
\begin{split}
\overline{R}_k(\lambda_1,\ldots,\lambda_k)=&\frac{1}{(N-k)!}\int dE_{k+1}\ldots dE_{N}\det[f_a(\lambda_b),\ f_a(E_c)]_{\substack{a=1,\ldots,N\\b=1,\ldots,k\\c=k+1,\ldots,N}}{\rm det}\left[ h_a(\lambda_b),\ h_a(E_c)\right]_{\substack{a=1,\ldots,N\\b=1,\ldots,k\\c=k+1,\ldots,N}}\\
=&(-1)^k\det\left[\begin{array}{cc} 0 & f_d(\lambda_a) \\ h_b(\lambda_c) & \int f_a(x)h_b(x)dx \end{array}\right]_{\substack{a,c=1,\ldots,N\\b,d=1,\ldots,k}}=\det[K_N(\lambda_a,\lambda_b)]_{a,b=1,\ldots,k}
\end{split}
\end{equation}
with the kernel
\begin{equation}\label{chap1:kernel-det}
K_N(\lambda_a,\lambda_b)=\sum_{j=1}^{N} f_j(\lambda_a)h_j(\lambda_b).
\end{equation}
In the last step, we employed the Schur complement identity for the determinant
\begin{equation}
\det\left[\begin{array}{cc} A & B\\ C& D\end{array}\right]=\det[A-BD^{-1}C]\det[D]
\end{equation}
and the biorthonormality between $f_a(x)$ and $h_b(x)$. The result~\eqref{chap1:det.proc} holds true for all $k=1,\ldots,N$, and eigenvalue statistics that satisfy this algebraic structure are called determinantal point processes~\cite{Oxford-determinant}. The level density at finite matrix dimension $N$ takes then the form
\begin{equation}\label{chap1:level-kernel}
\overline{\rho}(\lambda)=\frac{1}{N}\overline{R}_1(\lambda)=K_N(\lambda,\lambda).
\end{equation}
There is a ${\rm Gl}_{\mathbb{C}}(1)$ gauge invariance of the kernel which allows a rescaling of the form $K_N(\lambda_a,\lambda_b)\to e^{\xi(\lambda_a)-\xi(\lambda_b)}K_N(\lambda_a,\lambda_b)$ as those terms cancel when pulling them out of the determinant, which is also called cocycle. This is the reason why kernels may differ from one literature to the other despite they describe the same statistics. This gauge invariance is especially useful when taking asymptotic limits as very often the kernel has only a limit after a proper $N$-dependent choice of the function $\xi(\lambda)$. A similar ambiguity exists for the two sets of functions $f_a(x)$ and $h_b(x)$. We have the full freedom of an arbitrary basis, say of $h_1(x),\ldots,h_N(x)$. Once such a choice has been done the functions $f_1(x),\ldots, f_N(x)$ are uniquely determined by the biorthornormality conditions.

An example of determinantal point processes are the GUE with $f_a(x)=h_a(x)$ being the quantum harmonic oscillator wave functions, meaning essentially the Hermite polynomials with the Gaussian weight. Another classical example is the complex Wishart-Laguerre ensemble with $f_a(x)=h_a(x)$ being the generalised Laguerre polynomials with the weight $x^{\nu/2}e^{-x}$ and support $x\geq0$. Actually, the various classes of P\'olya ensembles~\cite{KKF-prod} exhibit all determinantal point structures. They are quite large and comprise  the complex Jacobi ensemble (truncated unitary matrices), four of the ten Gaussian ensembles in the Altland-Zirnbauer classification, see Subsection~\ref{chap1:sec:Altland-Zirnbauer}, as well as solutions of the Dyson-Brownian motion on the real line~\cite{Dyson-Brownian} and the circle~\cite{Polya-cyclic} and the solution of the DMPK equation~\cite{Ipsen-Schomerus} describing quantum transport in disorded quantum channels. What is less known is that also the picket fence statistic, meaning the statistic of the 1-dimensional quantum Harmonic oscillator and any other integrable quantum system with only one quantum number, follows a determinantal point process, see~\cite{ABK-product}. The kernel in this system is equal to 
\begin{equation}\label{chap1:Pf-kernel}
K_{{\rm PF}}(\lambda_a,\lambda_b)=\sum_{j=1}^N\frac{\sin[\pi(j-\lambda_a)]}{\pi(j-\lambda_a)}\delta(j-\lambda_b)
\end{equation}
for an operator of $N$ equidistant unfolded eigenvalues. The combination of the determinant and the {\it sinus cardinalis} function in front of the Dirac delta function selects with its zeros only those terms such that at most a single eigenvalue can be at any position $1,\ldots,N$.

For the Pfaffian form in~\eqref{chap1:jpdf-det-Pf}, we can find a similar result
\begin{equation}\label{chap1:Pf.proc}
\overline{R}_k(\lambda_1,\ldots,\lambda_k)={\rm Pf}\left[\begin{array}{cc} D(\lambda_a,\lambda_c) & K(\lambda_a,\lambda_d) \\ -K(\lambda_c,\lambda_b) & J(\lambda_b,\lambda_d) \end{array}\right]_{a,b,c,d=1,\ldots,k}
\end{equation}
 with the three kernel functions
\begin{equation}\label{chap1:kernel-Pf}
\begin{split}
D(\lambda_a,\lambda_c)=&\sum_{j=1}^{N/2}[f_{2j-1}(\lambda_a)f_{2j}(\lambda_c)-f_{2j}(\lambda_a)f_{2j-1}(\lambda_c)]=-D(\lambda_c,\lambda_a),\\
K(\lambda_a,\lambda_d)=&\sum_{j=1}^{N/2}\int[f_{2j-1}(\lambda_a)f_{2j}(x)-f_{2j}(\lambda_a)f_{2j-1}(x)]g(x,\lambda_b)dx,\\
J(\lambda_b,\lambda_d)=&g(\lambda_b,\lambda_d)-\sum_{j=1}^{N/2}\int[f_{2j-1}(x)f_{2j}(y)-f_{2j}(x)f_{2j-1}(y)]g(\lambda_b,x)g(y,\lambda_d)dxdy=-J(\lambda_d,\lambda_b),
\end{split}
\end{equation}
when applying the generalised de Bruijn identity~\cite[Appendix C.2]{Kieburg-Guhr-1} which generalises the original identity in~\cite{deBruijn}. Anew, this holds for all $k=1,\ldots,N$, and ensembles satisfying this structure are called Pfaffian point process~\cite{Kargin-Pfaffian}.
To obtain this result, we need to go over to a basis $f_a(x)$ that is skew-orthonormal which replaces the biorthonormality relations, meaning the following identities are now satisfied
\begin{equation}
\int f_{2a}(x)g(x,y)f_{2b}(y)dxdy=\int f_{2a-1}(x)g(x,y)f_{2b-1}(y)dxdy=0\qquad{\rm and}\qquad \int f_{2a-1}(x)g(x,y)f_{2b}(y)dxdy=\delta_{ab}
\end{equation}
for all $a,b=1,\ldots, N/2$. The skew-orthonormality does not fix the functions $f_{a}(x)$ uniquely as for instance we can shift every even index function as $f_{2a}(x)\to f_{2a}(x)+c_af_{2a-1}(x)$ with any constants $c_a\in\mathbb{C}$. This replaces the ambiguous choice of the functions $h_a(x)$ in the determinantal case. Furthermore, there is again a gauge invariance which is now with respect to the larger group ${\rm Sl}_{\mathbb{C}}(2)$, due to the Pfaffian factorisation identity ${\rm Pf}[BAB^T]=\det[B]{\rm Pf}[A]$. This group would mix up the three kernels and can be anew used to manipulate those such that they have a limit when $N\to\infty$.

The Pfaffian structure~\eqref{chap1:Pf.proc}  can be also achieved for the general form~\eqref{chap1:jpdf-general}. Then, however, the kernels~\eqref{chap1:kernel-Pf} will be slightly changed as a mixing of the orthonormality condition and the skew-orthonormality takes place, see~\cite{Kieburg-mixing}.

\subsubsection{  Local spectral statistics}

\begin{figure}[t!]
\centering
\includegraphics[width=.8\textwidth]{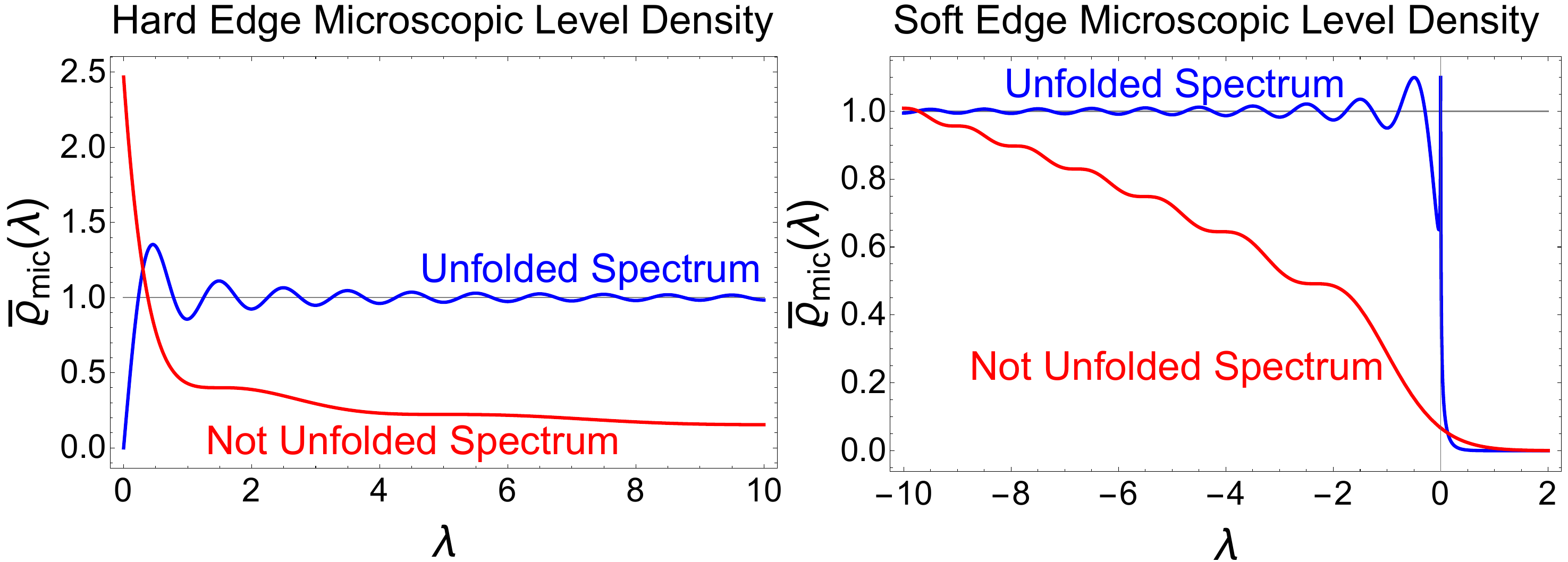}
\caption{Microscopic level densities of complex Wishart-Laguerre random matrix $H=WW^\dagger$ at the two edges. The left plot is the hard edge meaning the origin where the Bessel kernel statistics show up. The right plot is at the soft edge where the largest eigenvalue can be seen and the Airy kernel statistics manifest itself. The red curves are obtained when the spectrum is not properly unfolded, while the blue curves are those when the unfolding with the mesoscopic level density function has been performed. The divergence of the unfolded level density at the origin is a relict because of the non-analyticity of the macroscopic level density of the non-unfolded spectrum there as it vanishes like a square root.}
\label{chap1:fig5:unfolding-mic}
\end{figure}

We once again restrict ourselves to real eigenvalue spectra. Combining the unfolding with the $k$-point correlation functions, we can derive the $k$-point correlation functions for the unfolded spectral statistics around the eigenvalue $\lambda_0$ as follows
\begin{equation}
\widehat{R}_k(\mu_1,\ldots,\mu_k)d\mu_1\cdots d\mu_k=\overline{R}_k(\lambda_1,\ldots,\lambda_k)d\lambda_1\cdots d\lambda_k=\frac{\overline{R}_k(\Omega^{-1}_{\lambda_0}(\mu_1/N),\ldots,\Omega^{-1}_{\lambda_0}(\mu_k/N))}{\prod_{j=1}^NN\overline{\rho}(\Omega^{-1}_{\lambda_0}(\mu_j/N))}d\mu_1\cdots d\mu_k.
\end{equation}
Here, we want to emphasise that $\overline{\rho}(\lambda)$ shall be understood as the limiting level density for $N\to\infty$ in a mesoscopic scale while $R_1(\lambda)/N$ is the one for finite $N$. This latter point is crucial as the microscopic or local level density
\begin{equation}\label{chap1:micro.dens}
\overline{\rho}_{\rm mic,\lambda_0}(\mu)=\lim_{N\to\infty}\frac{\overline{R}_k(\Omega^{-1}_{\lambda_0}(\mu/N))}{N\overline{\rho}(\Omega^{-1}_{\lambda_0}(\mu/N))}
\end{equation}
is not always trivially equal to $1$. Particularly at the spectral edges, non-trivial behaviour manifests. We note that we take the mesoscopic spectral scale, which lies between the scale of the local mean level spacing $\overline{s}(\lambda_0)$, see~\eqref{chap1:spacing.interv}, and the scale where the whole spectrum is seen. We want to avoid the macroscopic scale of the level density, as it is usually found in the literature, because that approach does not work well at the edges where the macroscopic level density vanishes. This approach was particularly helpful for the double scaling limit of an infinite product of random matrices~\cite{ABK-product}, where at the soft edge, meaning the edge where the largest eigenvalue can be seen, a transition from the Airy to the picket fence statistics happens. It was similarly useful in the study of the transition of the hard to the soft edge at  the lower boundary for the complex Wishart-Laguerre ensemble~\cite{AGK-hard-soft}. An illustration of the effect of a proper unfolding is shown in Fig.~\ref{chap1:fig5:unfolding-mic} where the Wishart-Laguerre matrix $H=WW^\dagger$ with Gaussian distributed $W\in\mathbb{C}^{N\times N}$ has been employed. It becomes visible that the mean positions of the unfolded eigenvalues lie approximately on half integers when choosing $\lambda_0=0$ (hard edge) or when  choosing a position $\lambda_0$ where the macroscopic level density vanishes (soft edge).

There are various universal local spectral statistics not only because of the rich number of symmetry classes but  also because of the different regions a spectrum can have. The following list gives an overview:
\begin{itemize}
\item	{\bf Bulk Statistics:} We zoom in at a fixed ($N$-independent) point $\lambda_0\in\mathbb{R}$ of the macroscopic level density $\overline{\rho}(\lambda)$ with support $\mathcal{I}$ such that there is an open set $U\subset\mathcal{I}$ with $\lambda_0\in\mathcal{U}$ and any point $\lambda\in\mathcal{U}$ yields the same local spectral statistics. The scale on which the mean level spacing of two adjacent eigenvalues lives is  $\overline{s}(\lambda_0)\propto 1/N$ if the averaged level density $\overline{\rho}(\lambda_0)>0$ is finite and positive. This rather mathematical description only means that the local spectral statistics must be translation invariant. Indeed this is the case as we will see below.

\item	{\bf Soft Edge Statistics:} In this limit, we consider the local spectral statistics at a fixed boundary point $\lambda_0$ of the support $\mathcal{I}$ where each distribution of an individual eigenvalue on the scale $\overline{s}(\lambda_0)$ has a part inside the support as well as outside the support of the macroscopic level density $\overline{\rho}(\lambda)$.

\item	{\bf Hard Edge Statistics:} It is either a fixed inner or a fixed boundary point $\lambda_0$ of the support $\mathcal{I}$ where the distribution of each individual eigenvalue on the scale $\overline{s}(\lambda_0)$ has only a support on one side (either left or right) of $\lambda_0$. Hard edge points arise from Linear Algebra considerations and not probabilistic ones. Quite often the hard edge is at the origin, though for truncated orthogonal, unitary or unitary symplectic matrices (Jacobi ensembles) it can happen that there is also a hard edge at $\lambda=1$. This becomes relevant when studying quantum transport~\cite{Beenakker-transport}, scattering~\cite{Schomerus-scattering,Oxford-scattering,M-transport}, fibre optics~\cite{KMV-optics} and quantum information~\cite{complex-trunc,real-trunc,LZ-QC-QI} problems.

\item	{\bf Inner Multi-Critical Point (Merging Cuts) Statistics:} The fixed point $\lambda_0$ is a point where two soft edges start to merge with each other, meaning it is neither a bulk nor a hard edge point inside the support. Those points usually appear in multi-matrix models~\cite{GZ-two-matrix} or matrix models with external sources~\cite{BH-source}.

\item	{\bf Tail Statistics:} It is the region which after inverting the spectrum $\lambda\to1/\lambda$ becomes a hard edge. Those statistics can be found in applications of the spherical ensembles~\cite{real-spheric,complex-spheric,quaternion-spheric}. Those were recently encountered in the statistics of topological properties of random matrix fields~\cite{HKGG-topo-1,HKGG-topo-2} which should model topological properties of Hamiltonian quantum systems.

\item	{\bf Outliers:} These are singular points of the spectrum which do not lie in the support of the macroscopic level density $\overline{\rho}$ and are given by Dirac delta distributions on the macroscopic scale comprising a finite number of eigenvalues. They usually appear in the study of time series, e.g., see~\cite{statistics,LCBP-finance,VGS-time}, and give the trend of the whole or parts of the system. Those outliers hint at strongly correlated subsystems.
\end{itemize}

In quantum chaos studies quite often the bulk statistics are completely sufficient to determine whether a system is chaotic or not. Though one does not fully see in which of the ten universality classes the Hamiltonian lies in. Nonetheless, we will mostly concentrate our summary on the bulk statistics and will be briefer when it comes to the hard and soft edge statistics.

\begin{figure}[t!]
\centering
\centerline{\includegraphics[width=.4\textwidth]{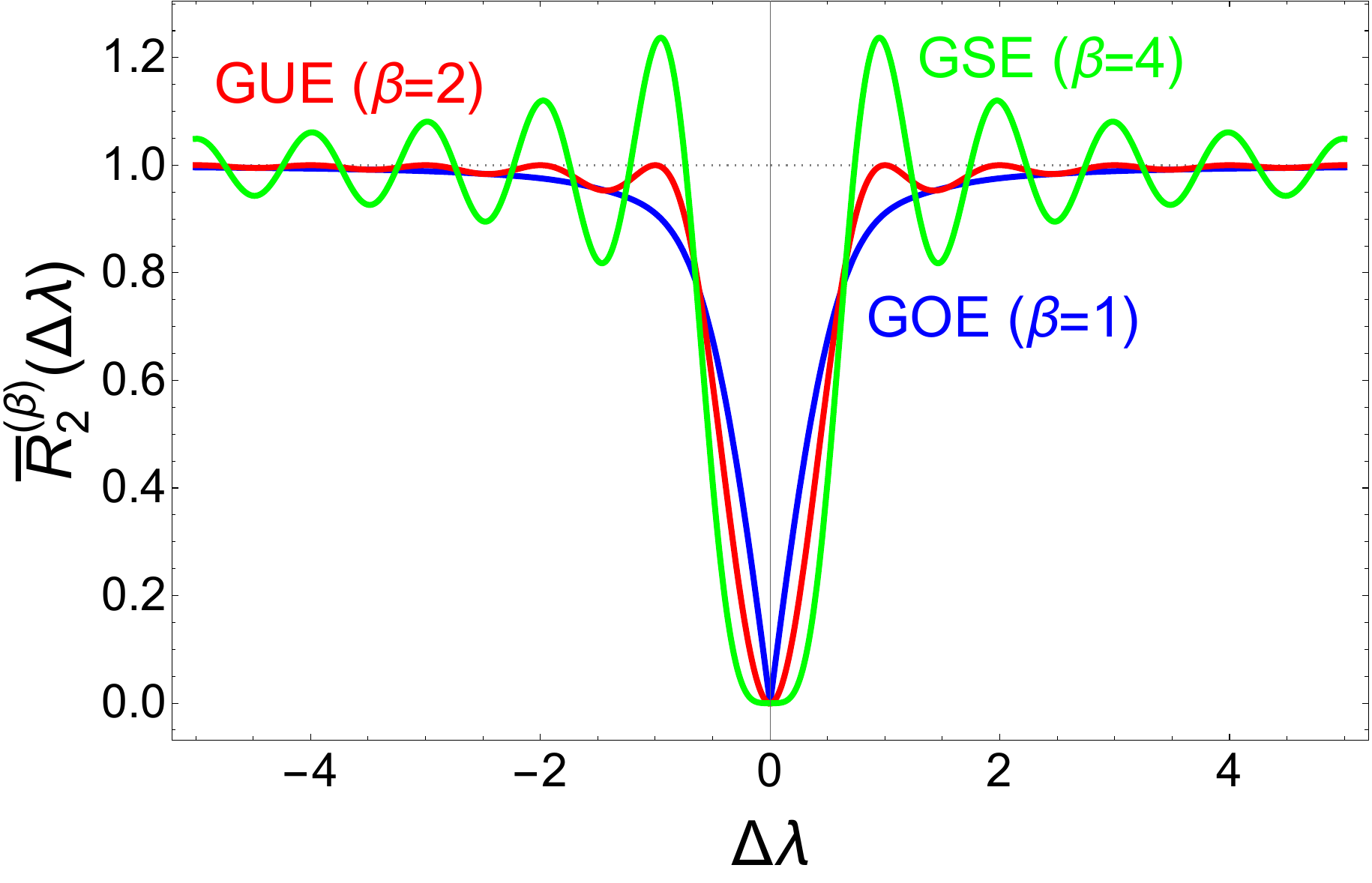}\qquad\includegraphics[width=.4\textwidth]{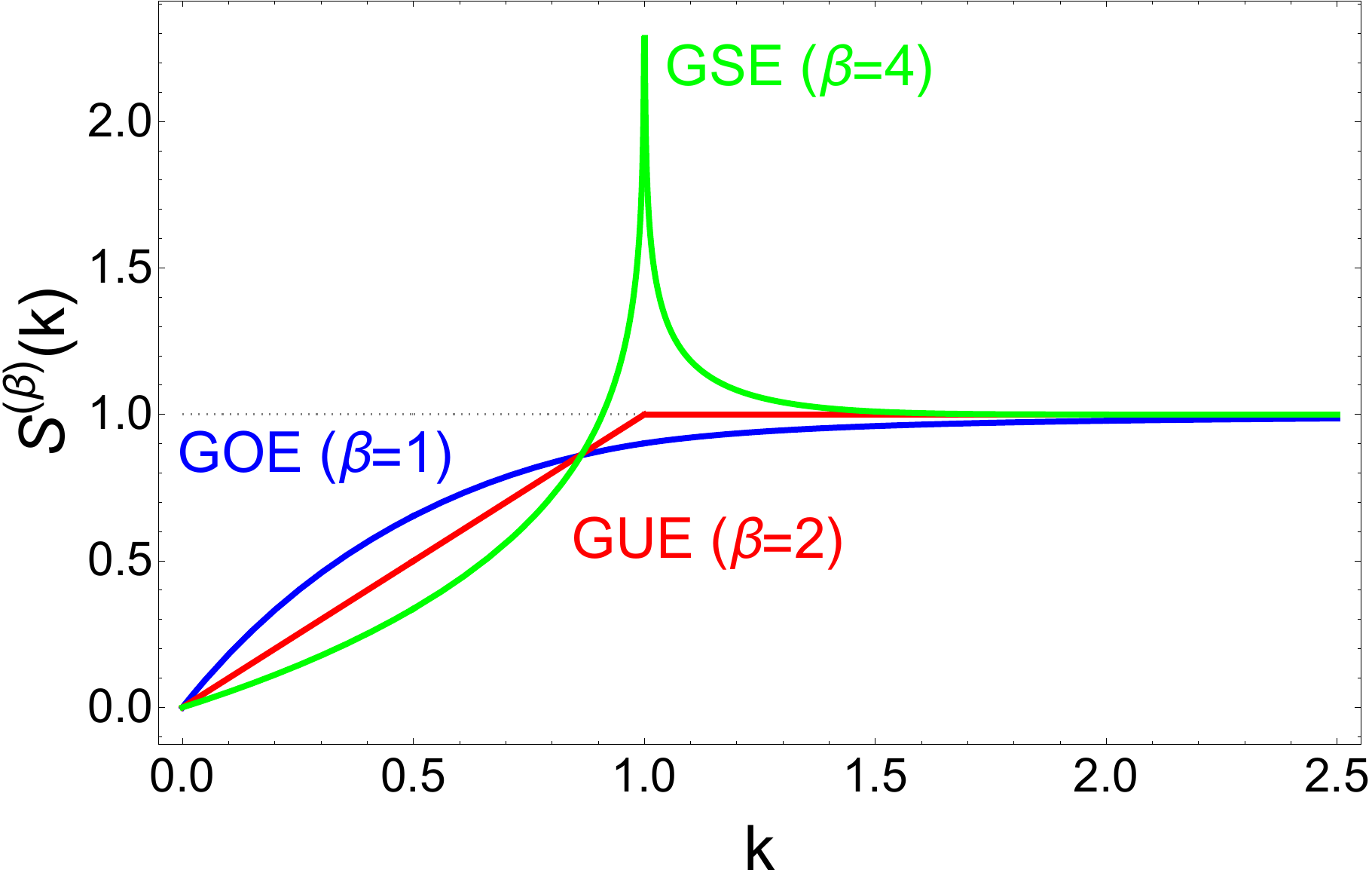}}
\caption{{\bf Left:} sine-kernel result~\eqref{chap1:sine-kernels} (meaning inside the bulk of the eigenvalue spectrum) for the three Dyson classes of the two-point correlation function. {\bf Right:} spectral   form factor of these two-point correlation functions which are essentially their Fourier transforms. The blue curve is the GOE bulk statistics ($\beta=1$), the red is the GUE bulk statistics ($\beta=2$) and the green curve is the GSE bulk statistics ($\beta=4$). The grey dotted curve at the value $1$ is the Poisson $2$-point correlation function (in the bulk) and its spectral form factor in both plots.}
\label{chap1:fig6:bulk}
\end{figure}

The local spectral statistics at a point inside the bulk of eigenvalues is given by the three $k$-point correlation functions following the Dyson classes~\cite{log-gas} with the kernels, see~\eqref{chap1:det.proc} and~\eqref{chap1:Pf.proc},
\begin{equation}\label{chap1:sine-kernels}
\begin{aligned}
K^{\rm (1,bulk)}(\lambda_1,\lambda_2)=&K^{\rm (2,bulk)}(\lambda_1,\lambda_2)=\frac{\sin[\pi(\lambda_1-\lambda_2)]}{\pi(\lambda_1-\lambda_2)},\qquad & K^{\rm (4,bulk)}(\lambda_1,\lambda_2)=&K^{\rm (2,bulk)}(2\lambda_1,2\lambda_2),\\
D^{\rm (1, bulk)}(\lambda_1,\lambda_2)=&\frac{1}{2}(\partial_{\lambda_a}-\partial_{\lambda_b})K^{\rm (2,bulk)}(\lambda_1,\lambda_2),\qquad &J^{\rm (1, bulk)}(\lambda_1,\lambda_2)=&\int_0^{\lambda_a-\lambda_b}K^{\rm (2,bulk)}(E,0) dE-\frac{1}{2}{\rm sign}(\lambda_a-\lambda_b),\\
D^{\rm (4, bulk)}(\lambda_1,\lambda_2)=&\frac{1}{2}(\partial_{\lambda_a}-\partial_{\lambda_b})K^{\rm (2,bulk)}(2\lambda_1,2\lambda_2),\qquad &J^{\rm (4, bulk)}(\lambda_1,\lambda_2)=&\int_0^{\lambda_a-\lambda_b}K^{\rm (2,bulk)}(2E,0) dE.
\end{aligned}
\end{equation}
Their dependence on the sine function has given them the name of these statistics, namely the sine kernel. All three statistics look very similar and surprisingly show a high similarity to the picket fence kernel~\eqref{chap1:Pf-kernel}. As can be checked via plotting the $2$-point correlation functions, see left plot in Fig.~\ref{chap1:fig6:bulk}, as well as the microscopic level densities, which are all $\overline{R}_1^{(\beta)}(\lambda)=1$, these expressions are properly unfolded. They show the translation invariance $\lambda_j\to\lambda_j+\widetilde{\lambda}$ and correspond to the normalised  unit mean level spacing.

Starting from these results, we can compute various other benchmarks used in the literature. For instance, the $2$-point correlation function is effectively a function of a single variable, namely the distance $\Delta\lambda=\lambda_2-\lambda_1$  of the two eigenvalues, which is plotted in the left graph of Fig.~\ref{chap1:fig6:bulk}. Starting from this quantity, we may compute the number variance, given by~\eqref{chap1:covariance} with $G(\lambda)=F(\lambda)=\Theta[L^2-4\lambda^2]$, meaning
\begin{equation}\label{chap1:number-var}
\Sigma_\beta^2(L)=\int_{[-L/2,L/2]^2}\left[\overline{R}_2^{\rm (\beta,bulk)}(\lambda_1,\lambda_2)-\overline{R}_1^{\rm (\beta,bulk)}(\lambda_1)\overline{R}_1^{\rm (\beta,bulk)}(\lambda_2)\right]d\lambda_1d\lambda_2+\int_{-L/2}^{L/2} \overline{R}_1^{\rm (\beta,bulk)}(\lambda)d\lambda.
\end{equation}
 This quantity is given by~\cite{Dyson-Mehta-number-variance}
\begin{equation}\label{chap1:number-var.results}
\Sigma_1^2(L)\approx\frac{2}{\pi^2}\left({\rm ln}(2\pi L)+\gamma+1-\frac{\pi^2}{8}\right),\qquad \Sigma_2^2(L)\approx\frac{1}{\pi^2}\left({\rm ln}(2\pi L)+\gamma+1\right),\qquad \Sigma_4^2(L)\approx\frac{1}{2\pi^2}\left({\rm ln}(4\pi L)+\gamma+1+\frac{\pi^2}{8}\right)
\end{equation}
 for the three Dyson classes in the limit of large $L$. As comparison, the number variance for the Poisson statistics is simply the linear function $\Sigma_{\rm Poisson}^2(L)=L$.  The constant $\gamma\approx0.577$ is the Euler-Mascheroni constant. 
 
 Another quantity often employed is the spectral form factor~\cite{Berry-spectral-form-factor}. Following~\eqref{chap1:covariance} with $G(\lambda)=e^{2\pi i k\lambda}$ and $F(\lambda)=e^{-2\pi i k\lambda}$ and dividing by $N$ so that limit $N\to\infty$ exists, it is
 \begin{equation}
 S^{(\beta)}(k)=1+\int_{\mathbb{R}^2}e^{2\pi i k(\lambda_1-\lambda_2)}\left[\overline{R}_2^{\rm (\beta,bulk)}(\lambda_1,\lambda_2)-\overline{R}_1^{\rm (\beta,bulk)}(\lambda_1)\overline{R}_k^{\rm (\beta,bulk)}(\lambda_2)\right]d\lambda_1d\lambda_2.
 \end{equation}
It takes the simple forms~\cite{GMW-review}
\begin{equation}
 S^{(1)}(k)=\left\{\begin{array}{cl} \displaystyle |k|(2-{\rm ln}(2|k|+1), & k\leq1, \\ \displaystyle 2-|k|{\rm ln}\left(\frac{2|k|+1}{2|k|-1}\right), &k\geq1,\end{array}\right.
 \qquad
 S^{(2)}(k)=\left\{\begin{array}{cl} |k|, & k\leq1, \\ 1, &k\geq1,\end{array}\right.
 \qquad
 S^{(4)}(k)=\left\{\begin{array}{cl} \displaystyle \frac{|k|}{4}(2-{\rm ln}|\, |k|-1|), & k\leq2, \\ 1, &k\geq2.\end{array}\right.
\end{equation}
These results are shown in the right plot of Fig.~\ref{chap1:fig6:bulk}. The very different, almost dull behaviour beyond frequencies $|k|>1$, comes from the fact that the corresponding eigenvalue scale is below the mean level spacing where it is very unlikely to find two or more eigenvalues so that the unfolded $1$-point function, which is constant to unity, is the only remaining contribution. Actually, at finite matrix size as well as in physical operators there is also an oscillatory dip for very small  frequencies $|k|\ll1$, e.g., see~\cite{Jia-Verbaarschot-spectral-form,Okuyamaa-Sakai-spectral-form,KHH-spectral-form,SPAD-spectral-form}. Those result from the fact that the unfolded regime of the spectrum has been left and the mesoscopic, and even more the macroscopic spectral structures become visible. Those parts are not very universal and depend on the details of the considered eigenvalue spectrum.

\begin{figure}[t!]
\centering
\includegraphics[width=.9\textwidth]{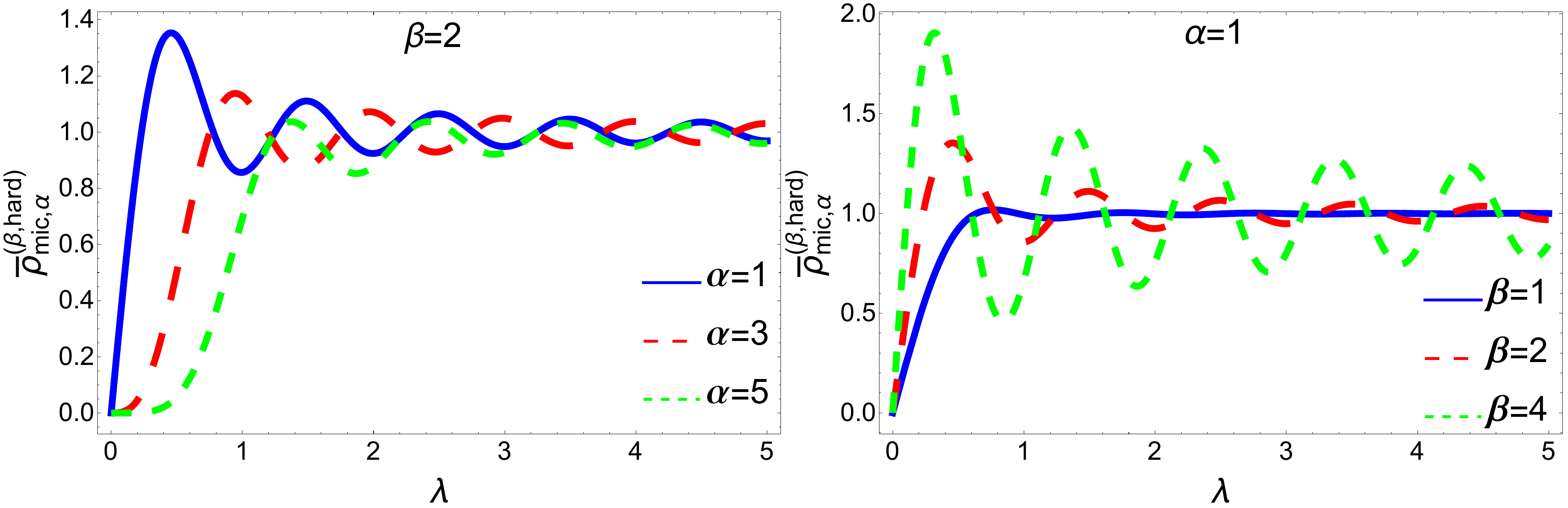}
\caption{The microscopic level densities~\eqref{chap1:hard-dens.U} and~\eqref{chap1:hard-dens.OS} at the hard edge of the seven Altland-Zirnbauer classes that follow the jpdf~\eqref{chap1:jpdf-non-Dyson}. In the left plot we fixed the Dyson index $\beta=2$ and varied the index $\alpha$ which is the source for the level repulsion from the origin. In the right panel we fixed $\alpha=1$ and varied the Dyson index $\beta$.}
\label{chap1:fig8:hard-mic}
\end{figure}

As aforementioned, the hard edge statistics are given when zooming into a point of the spectrum where the eigenvalues are prevented to go beyond a certain point. Those barriers follow from Linear Algebra considerations and are not probabilistic in nature. For instance the singular values of a matrix are always non-negative or the singular values of a truncated unitary matrix are always smaller or equal than $1$. When assuming no higher criticality the universal spectral statistics of the seven Altland-Zirnbauer classes~\cite{AZ-class} that are not in Dyson's threefold way, see Table~\ref{chap1:tab:RMT-Symmetry}, the kernels are
\begin{equation}\label{chap1:Bessel-GUE-kernels}
\begin{aligned}
K_{\alpha}^{\rm (2,hard)}(\lambda_1,\lambda_2)=&\pi^2\lambda_2\int_0^1 J_{(\alpha-1)/2}(\pi \lambda_1 t)J_{(\alpha-1)/2}(\pi \lambda_2 t)tdt=\pi \lambda_2\frac{\lambda_1J_{(\alpha+1)/2}(\pi \lambda_1)J_{(\alpha-1)/2}(\pi\lambda_2)- \lambda_2J_{(\alpha-1)/2}(\pi \lambda_1)J_{(\alpha+1)/2}(\pi\lambda_2)}{\lambda_1^2-\lambda_2^2}
\end{aligned}
\end{equation}
for $\beta=2$ and
\begin{equation}\label{chap1:Bessel-kernels}
\begin{aligned}
 K_{\alpha}^{\rm (1,hard)}(\lambda_1,\lambda_2)=&K_{2\alpha+1}^{\rm (2,hard)}(\lambda_1,\lambda_2)+\frac{\pi}{2}J_{\alpha}(\pi\lambda_1)\left(1-\int_0^{\pi\lambda_2}J_{\alpha}(\lambda')d\lambda'\right),\\
D_{\alpha}^{\rm (1, hard)}(\lambda_1,\lambda_2)=&\partial_{\lambda_2} K_{\alpha}^{\rm (1,hard)}(\lambda_1,\lambda_2),
\qquad J_{\alpha}^{\rm (1, hard)}(\lambda_1,\lambda_2)=\int_{\lambda_2}^{\lambda_1} K_{\alpha}^{\rm (1,hard)}(s,\lambda_2)ds-\frac{1}{2}{\rm sign}(\lambda_1-\lambda_2),\\
 K_{\alpha}^{\rm (4,hard)}(\lambda_1,\lambda_2)=&K_{\alpha-2}^{\rm (2,hard)}(2\lambda_1,2\lambda_2)-\frac{\pi}{2} J_{(\alpha-3)/2}(2\pi \lambda_1)\left(1-\int_{2\pi\lambda_2}^\infty J_{(\alpha-3)/2}(t)dt\right),\\
D_{\alpha}^{\rm (4, hard)}(\lambda_1,\lambda_2)=&\partial_{\lambda_2}  K_{\alpha}^{\rm (4,hard)}(\lambda_1,\lambda_2),
\qquad J_{\alpha}^{\rm (4, hard)}(\lambda_1,\lambda_2)=\int_{\lambda_2}^{\lambda_1}K_{\alpha}^{\rm (4,hard)}(s,\lambda_2)ds
\end{aligned}
\end{equation}
for $\beta=1$ and $\beta=4$, respectively, e.g., see~\cite{FNH-edges}. The function $J_\alpha(\lambda)$ is the modified Bessel function of the first kind. The index $\alpha$ corresponds to the jpdf~\eqref{chap1:jpdf-non-Dyson}  and its possible values for the seven classes can be found in Table~\ref{chap1:tab:RMT-param}. To see that these functions are properly unfolded one usually considers the microscopic level densities
			\begin{equation}\label{chap1:hard-dens.U}
			\bar{\rho}_{\rm mic,\alpha}^{\rm (2,hard)}(\lambda)=\frac{\pi^2}{2}\lambda\left(J_{(\alpha-1)/2}^2(\pi \lambda)-J_{(\alpha+1)/2}(\pi \lambda)J_{(\alpha-3)/2}(\pi \lambda)\right),
			\end{equation}
			and
			\begin{equation}\label{chap1:hard-dens.OS}
			\begin{split}
			\bar{\rho}_{\rm mic,\alpha}^{\rm (1, hard)}(\lambda)=
			\bar{\rho}_{\rm mic,2\alpha+1}^{\rm (2,hard)}(\lambda)+\frac{\pi}{2}J_{\alpha}(\pi \lambda)\left(1-\int_0^{\pi\lambda}J_{\alpha}(\lambda')d\lambda'\right),\qquad
			\bar{\rho}_{\rm mic,\alpha}^{\rm (4,hard)}(\lambda)=
			\bar{\rho}_{\rm mic,\alpha-2}^{\rm (2,hard)}(2\lambda)-\frac{\pi }{2}J_{(\alpha-3)/2}(2\pi \lambda)\left(1-\int_{2\pi\lambda}^\infty J_{(\alpha-3)/2}( \lambda')d\lambda'\right).
			\end{split}
			\end{equation}
Those densities are shown in Fig.~\ref{chap1:fig8:hard-mic}. In those plots it becomes apparent that $\alpha$ creates a level repulsion from the origin. In particular, the larger $\alpha$ becomes the further away the eigenvalues are pushed from the origin until a  spectral gap is created. Due to the relation of the dimensions of a rectangular matrix to the index $\alpha$, it is clear that the singular values exhibit a spectral gap for extremely rectangular matrices. We note that rectangular matrices are embedded in terms of chiral matrices in the classes AIII, B/DI and CII, see Table~\ref{chap1:tab:RMT-Symmetry}.

In contrast, the Dyson index $\beta$ creates strong oscillations when it becomes larger. This can be also observed in the microscopic level densities at the soft edges. Those edges are not enforced by Linear Algebra constraints but they originate from a probabilistic concentration onto a finite support in the limit of a large matrix dimension $N\to\infty$.

\begin{figure}[t!]
\centering
\includegraphics[width=.9\textwidth]{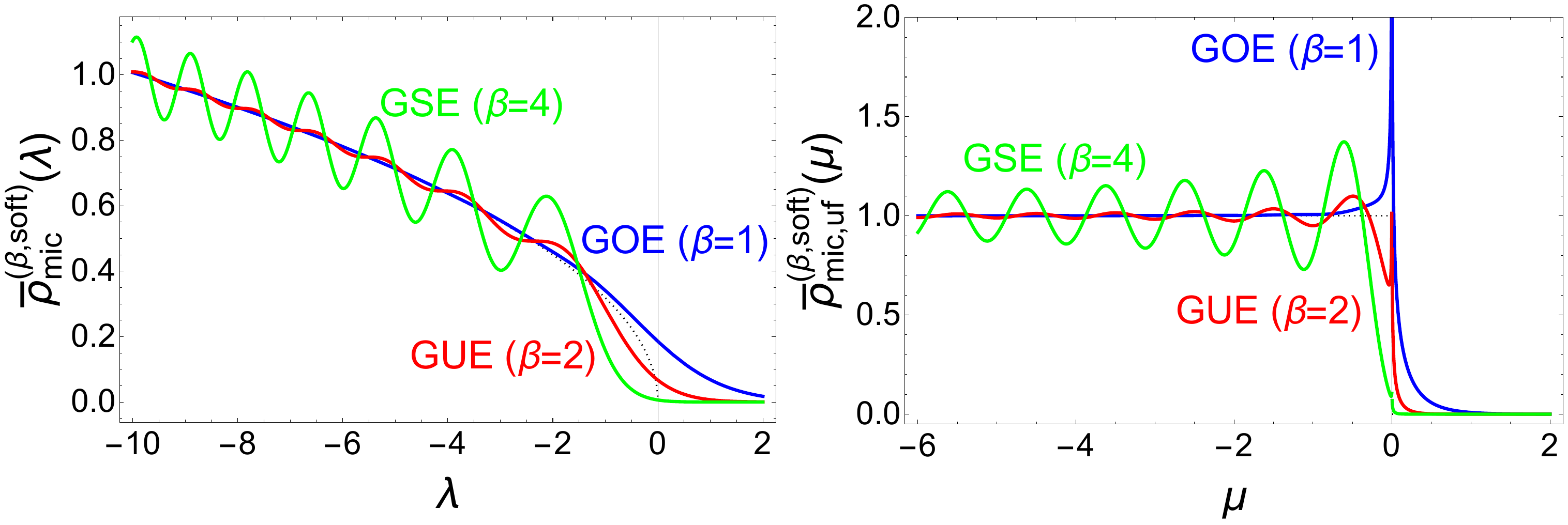}
\caption{Left plot: the microscopic level densities~\eqref{chap1:soft-dens.U} and~\eqref{chap1:soft-dens.OS} at the soft edge. The exponential damping to the right shows the end of the spectral support, while the oscillations hint to the bulk of eigenvalues and their repulsion. The growth like a square root of the level density is visible which also increases the frequency of the oscillations hinting of non-unfolded eigenvalue spectrum. In the right plot, we have unfolded the spectrum with the substitution $\lambda={\rm sign}(\mu)(3\pi |\mu|/2)^{2/3}$ which becomes visible in the asymptotic approach of a flat level density about $1$. The singularity at the origin is the prize one has to pay for such an unfolding.}
\label{chap1:fig9:soft-mic}
\end{figure}

Assuming no higher criticality the spectral statistics at the soft edges show the following kernels~\cite{FNH-edges}
\begin{equation}\label{chap1:Airy-GUE-kernels}
\begin{aligned}
K^{\rm (2,soft)}(\lambda_1,\lambda_2)=&\int_0^\infty {\rm Ai}(\lambda_1+t){\rm Ai}(\lambda_2+t)dt=\frac{ {\rm Ai}(\lambda_1){\rm Ai}'(\lambda_2)- {\rm Ai}'(\lambda_1){\rm Ai}(\lambda_2)}{\lambda_1-\lambda_2}
\end{aligned}
\end{equation}
for $\beta=2$ and
\begin{equation}\label{chap1:Airy-kernels}
\begin{aligned}
 K^{\rm (1,soft)}(\lambda_1,\lambda_2)=&K^{\rm (2,soft)}(\lambda_1,\lambda_2)+\frac{1}{2}{\rm Ai}(\lambda_1)\left(1-\int_{\lambda_2}^\infty{\rm Ai}(t)dt\right),\\
D^{\rm (1, soft)}(\lambda_1,\lambda_2)=&\partial_{\lambda_2} K^{\rm (1,soft)}(\lambda_1,\lambda_2),
\qquad J^{\rm (1, soft)}(\lambda_1,\lambda_2)=\int_{\lambda_2}^{\lambda_1} K^{\rm (1,soft)}(s,\lambda_2)ds-\frac{1}{2}{\rm sign}(\lambda_1-\lambda_2),\\
 K^{\rm (4,soft)}(\lambda_1,\lambda_2)=&2^{-1/3}K^{\rm (2,soft)}(2^{2/3}\lambda_1,2^{2/3}\lambda_2)-2^{-2/3}{\rm Ai}(2^{2/3}\lambda_1)\int_{\lambda_2}^\infty{\rm Ai}(2^{2/3}t)dt,\\
D^{\rm (4, soft)}(\lambda_1,\lambda_2)=&\partial_{\lambda_2} K^{\rm (4,soft)}(\lambda_1,\lambda_2),
\qquad J^{\rm (4, soft)}(\lambda_1,\lambda_2)=\int_{\lambda_2}^{\lambda_1}K^{\rm (4,soft)}(s,\lambda_2)ds
\end{aligned}
\end{equation}
for $\beta=1$ and $\beta=4$, respectively.
Here, we have employed the Airy function ${\rm Ai}(\lambda)=\frac{1}{\pi}\int_0^\infty\cos\left(\frac{t^3}{3}+\lambda t\right)dt$  which gives also the name for these kernels. The microscopic level densities, which are given by $K^{\rm (\beta,soft)}(\lambda,\lambda)$ for all $\beta$, are
\begin{equation}\label{chap1:soft-dens.U}
\overline{\rho}_{\rm mic}^{\rm (2,soft)}(\lambda)=[{\rm Ai}'(\lambda)]^2-\lambda[{\rm Ai}(\lambda)]^2
\end{equation}
and
\begin{equation}\label{chap1:soft-dens.OS}
\overline{\rho}_{\rm mic}^{\rm (1,soft)}(\lambda)=\overline{\rho}_{\rm mic}^{\rm (2,soft)}(\lambda)+\frac{1}{2}{\rm Ai}(\lambda)\left(1-\int_{\lambda}^\infty{\rm Ai}(t)dt\right),\qquad \overline{\rho}_{\rm mic}^{\rm (4,soft)}(\lambda)=2^{-1/3}\overline{\rho}_{\rm mic}^{\rm (2,soft)}(2^{2/3}\lambda)+2^{-2/3}{\rm Ai}(2^{2/3}\lambda)\int_{\lambda}^\infty{\rm Ai}(2^{2/3}t)dt.
\end{equation}
All three densities asymptotically behave like $\sqrt{|\lambda|}/\pi$ when going into the bulk $\lambda\to-\infty$. This underlines that they are actually not properly unfolded. This prevents a good comparison with other spectral statistics such as those at the hard edge, when studying soft-to-hard edge transitions as in~\cite{AGK-hard-soft,FNH-edges}, or with the picket fence statistics like it was found in~\cite{Johansson-Brownian} for Dyson-Brownian motion with initial conditions or in~\cite{ABK-product} for double scaling limits of a product of infinitely many matrices.  In~\cite{ABK-product,AGK-hard-soft}, we suggested to substitute $\lambda={\rm sign}(\mu)(3\pi |\mu|/2)^{2/3}$ which is not ideal at $\mu=0$ as one creates a singularity due to the Jacobian $\pi^{2/3} (3|\mu|/2)^{-1/3}/2$ which needs to be multiplied to the density. But already slightly away from the origin this unfolding is almost perfect, cf., Fig.~\ref{chap1:fig9:soft-mic} where the almost equidistant maxima highlight a mean level spacing of unity. This unfolding was extremely helpful in making visible how one spectral statistics transforms into another one.  For instance in~\cite{AGK-hard-soft}, we have seen that the bulk, hard edge and soft edge level spacing distribution are quite similar and only differ in about one percent. In contrast, one can see a significant difference in their microscopic level densities.

When going over to complex eigenvalue spectra, unfortunately only one universal bulk and edge kernel is explicitly known namely the one of the complex Ginibre ensemble which also exhibits a determinantal point process with the $k$-point correlation function~\cite{Bulk-3Gin,TaoVu-non-Herm,Edge-3Gin}
\begin{equation}\label{chap1:GinUE-uni-kernel}
\overline{R}_k^{\rm (GinUE,bulk)}(\lambda_1,\ldots,\lambda_k)=\det\left[e^{-|\lambda_a|^2+\lambda_a\lambda_b^*}\right]_{a.b=1,\ldots,k}\quad{\rm and}\quad \overline{R}_k^{\rm (GinUE, soft)}(\lambda_1,\ldots,\lambda_k)=\det\left[\frac{1}{2}e^{-|\lambda_a|^2+\lambda_a\lambda_b^*}{\rm erfc}\left(\frac{e^{-i\varphi_0}\lambda_a+e^{i\varphi_0}\lambda_b^*}{\sqrt{2}}\right)\right]_{a.b=1,\ldots,k}.
\end{equation}
where $e^{i\varphi_0}$ is the point where on the unit circle, which is the boundary of the support of the macroscopic level density, one zooms into. It was shown in~\cite{Bulk-3Gin,Edge-3Gin} that the real (GinOE) and quaternion (GinSE) Ginibre ensemble exhibit the very same spectral statistics in the spectral bulk and edge as long as one stays away from the symmetry axis of the spectrum which is the real axis in this case.  Also for their chiral counterparts~\cite{Akemann-chiral} it is now  confirmed that in the bulk and edge of the spectrum away from the symmetry axes (real axis for real and quaternion) and symmetry point (origin) the spectral statistics is the one of~\eqref{chap1:GinUE-uni-kernel}. The kernels and thus the whole spectral statistics around the real axis or the origin are also known~\cite{Akemann-complex-chiral-Gin,Osborn-complex-chiral-Gin,APS-real-chiral-Gin,Akemann-quaternion-chiral-Gin}.

Going beyond these results, to address the second of the two conjectures in~\cite{HKKU-univ} about that only three universality classes in the bulk statistics exist for the non-Hermitian symmetry classes, is  unfortunately currently out of reach. The reason is that it is very likely that most of the 38 classes do not exhibit the advantageous algebraic structures of determinantal and Pfaffian point processes even when choosing a Gaussian distribution. Even explicit expressions for the level density are not known for most of these classes so that other means have been used as explained in Subsection~\ref{chap1:secLagrangians}.

\subsubsection{Fredholm determinants and Pfaffians}

Coming back to the gap probabilities~\eqref{chap1:gap.prob}, we aim for expressing them in terms of the kernels. Once that is achieved one can plug in the limiting kernels and use the relations~\eqref{chap1:extreme.fredholm} and~\eqref{chap1:spacing.fredholm} to find expressions for the distributions of the smallest and largest eigenvalue as well as the level spacing distribution. This has been done in~\cite{Forrester-Fredholm} which led to the discovery that these quantities are related to Painl\'eve equation~\cite{TW-dist}.

For that purpose,  we consider the more general expectation value of
\begin{equation}\label{chap1:gap.compute}
\left\langle\prod_{j=1}^N[1-F(E_j)]\right\rangle=\int_{\mathbb{R}^N}\left(\prod_{j=1}^N[1-F(E_j)]\right)p(E) d[E]=\sum_{k=0}^N \frac{(-1)^k}{k!}\int \overline{R}_k(\lambda_1,\ldots,\lambda_k)\prod_{j=1}^kF(\lambda_j)d\lambda_j.
\end{equation}
with $F(\lambda)$ some suitably integrable test-function which is in terms of gap probabilities some indicator functions. When plugging in the formula of the $k$-point correlation function for the determinant point process~\eqref{chap1:det.proc}  we can expand the determinant. This would yield terms of the form
\begin{equation}
\int K(\lambda,\lambda)F(\lambda)d\lambda\qquad {\rm or}\qquad\int_{ \mathcal{I}} dx K(\lambda_1,\lambda)F(\lambda)K(\lambda,\lambda_2)F(\lambda_2)d\lambda.
\end{equation}
This gives rise to the definition of an integral operator $\mathcal{K}_{F}:\mathfrak{V}\to\mathfrak{V}$ given by
\begin{equation}
\mathcal{K}_{F}[G](\lambda)=\int K(\lambda,\lambda')F(\lambda') G(\lambda')d\lambda'.
\end{equation}
with $G(\lambda)$ inside the linear span $\mathfrak{V}={\rm span}\{f_1(\lambda),\ldots,f_N(\lambda)\}$ if the matrix dimension $N$ is finite dimensional or in the closed Banach space when $N\to\infty$, which is quite often the Hilbert space of square integrable functions on the restricted set which should be void of eigenvalues. We will stick at the moment with the finite dimensional version.

As $\mathcal{K}_{F}$ is an endomorphism and a trace class operator for suitable $F(\lambda)$, meaning the sum of its eigenvalues is finite, we can spectral decompose it. Let $\widetilde{f}_1(\lambda),\ldots,\widetilde{f}_N(\lambda)\in\mathfrak{V}$ be a basis of eigenfunctions of $\mathcal{K}_{F}$ to the eigenvalues $\alpha_1,\ldots,\alpha_N$, i.e., $\mathcal{K}_{F}[\widetilde{f}_j](\lambda)=\alpha_j\widetilde{f}_j(\lambda)$. Then, we can create a set of measures $d\nu_1(\lambda),\ldots,d\nu_N(\lambda)$ bi-orthonormal to this basis which means $\int \widetilde{f}_a(\lambda)d\nu_b(\lambda)=\delta_{ab}$. This creates a spectral decomposition of the kernel
\begin{equation}
\mathcal{K}_{F}[G](\lambda)=\sum_{j=1}^N\alpha_j \widetilde{f}_j(\lambda)\int G(\lambda')d\nu_j(\lambda').
\end{equation}
For the integral over the $k$-point correlation functions this means
\begin{equation}
\int \overline{R}_k(\lambda_1,\ldots,\lambda_k)\prod_{j=1}^kF(\lambda_j)d\lambda_j=\int \det\left[\sum_{j=1}^N\alpha_j\widetilde{f}_j(\lambda_a)d\nu_j(\lambda_b)\right]_{a,b=1,\ldots,k}=\frac{1}{k!}\sum_{j_1,\ldots,j_k=1}^N\int \det[\widetilde{f}_{j_b}(\lambda_a)]_{a,b=1,\ldots,k}\det[\alpha_{j_b}d\nu_{j_b}(\lambda_a)]_{a,b=1,\ldots,k},
\end{equation}
where we used the Cauchy-Binet formula for determinants of products of two rectangular matrices, which is the counterpart of the Andr\'eief identity~\cite{Andreief} for sums. Actually, the integrals can be carried out with the Andr\'eief integral identity for which we can employ the biorthonormality between the  eigenfunctions and the measures so that
\begin{equation}
\int \overline{R}_k(\lambda_1,\ldots,\lambda_k)\prod_{j=1}^kF(\lambda_j)d\lambda_j=\sum_{j_1,\ldots,j_k=1}^N\det\biggl[\alpha_{j_b}\overbrace{\int\widetilde{f}_{j_b}(\lambda)d\nu_{j_a}(\lambda)d\lambda}^{=\delta_{j_{a},j_{b}}}\biggl]_{a,b=1,\ldots,k}=k!\sum_{1\leq j_1<\ldots< j_k\leq N}\prod_{l=1}^k\alpha_{j_l}.
\end{equation}
We would like to mention that terms where two summing indices coincide must vanish due to the determinant as then at least two rows and columns must be the same. We plug this back into~\eqref{chap1:gap.compute} and arrive at the desired result for the expectation value
\begin{equation}\label{chap1:gap.result}
\left\langle\prod_{j=1}^N[1-F(E_j)]\right\rangle=\sum_{k=0}^N \sum_{1\leq j_1<\ldots< j_k\leq N}\prod_{l=1}^k(-\alpha_{j_l})=\prod_{j=1}^N(1-\alpha_{j})=\det[\mathbf{1}-\mathcal{K}_{F}].
\end{equation}
The determinant of the integral operator on the right is called Fredholm determinant.
For the level spacing distribution in the bulk of the GUE, see~\eqref{chap1:sine-kernels} for the kernel, we would have for example~\cite{TW-dist}
\begin{equation}
p_{\rm sp}^{\rm (GUE)}(s)=\partial_{s}^2\det[\mathbf{1}-\mathcal{K}_{s}]\qquad{\rm with}\quad \mathcal{K}_{s}[G](\lambda)=\int_{-s/2}^{s/2} \frac{\sin[\pi(\lambda-\lambda')]}{\pi(\lambda-\lambda')}G(\lambda')d\lambda',
\end{equation}
see~\eqref{chap1:spacing.fredholm}, where we omitted the dependence on $\lambda_0$ as it drops out in the bulk due to translation invariance. The vector space is the set of square integrable Lebesgue functions on the interval $[-s/2,s/2]$.

A similar expression can be derived for Pfaffian point processes~\eqref{chap1:Pf.proc} following the very same approach. The only difference is that one encounters a $2\times2$ matrix kernel.  The analogue of~\eqref{chap1:gap.result} is then a Fredholm-Pfaffian
\begin{equation}\label{chap1:gap.result-paff}
\left\langle\prod_{j=1}^N[1-F(E_j)]\right\rangle={\rm Pf}\left[\begin{bmatrix}0 & \mathbf{1}\\-\mathbf{1}&0\end{bmatrix}-\mathcal{K}_{F}\right]\qquad{\rm with}\quad \mathcal{K}_{F}\left[\begin{array}{c} G_1\\G_2\end{array}\right](\lambda)=\int\left[\begin{array}{cc} D(\lambda,\lambda') & K(\lambda,\lambda')F(\lambda') \\ -F(\lambda)K(\lambda',\lambda) & F(\lambda)J(\lambda,\lambda')F(\lambda') \end{array}\right]\left[\begin{array}{c} G_1(\lambda')\\G_2(\lambda')\end{array}\right]d\lambda'.
\end{equation}
The entry $G_2(\lambda)$ must be in ${\rm span}\{f_1(\lambda),\ldots,f_N(\lambda)\}$ while $G_1(\lambda)$ must be in the space of weight functions ${\rm span}\bigl\{\int F(\lambda)J(\lambda,\lambda')F(\lambda')f_1(\lambda')d\lambda',$ $\ldots,\int F(\lambda)J(\lambda,\lambda')F(\lambda')f_N(\lambda')d\lambda'\bigl\}$. Anew one can go over to the limit $N\to\infty$ to plug in directly the kernels~\eqref{chap1:sine-kernels} for the GOE and GSE bulk statistics.

Similar Fredholm determinants and Pfaffians have been used for the hard and soft edges~\cite{log-gas,Mehta,Forrester-Fredholm,TW-dist}. One needs only to use the kernels~\eqref{chap1:hard-dens.U},~\eqref{chap1:hard-dens.OS}, \eqref{chap1:soft-dens.U} and~\eqref{chap1:soft-dens.OS} instead of the sine kernels. Those have been used to derive expressions for the distribution of the smallest and largest eigenvalues in terms of Painlev\'e equations~\cite{Mehta,log-gas,TW-dist}. A famous result, which has been derived in this way, is the Tracy-Widom distribution~\cite{TW-dist} which gives the distribution of the largest (or smallest) eigenvalue at a soft edge.

There are various ways to evaluate Fredholm determinants and Pfaffians. One is to trace them back to Painlev\'e equations~\cite{log-gas} and solve those numerically. Another one is via quadratures~\cite{Bornemann-quadrature} such as the Gauss-Legendre quadratures. The latter method is extremely efficient as its convergence is exponential in the iteration when the kernels are analytic which is for the GUE and GSE the case, cf. Eq.~\eqref{chap1:sine-kernels}.

We would like to point out that these constructions of the Fredholm determinants and Pfaffians are not restricted to real eigenvalue spectra. It can be also applied to complex eigenvalue spectra, e.g., see~\cite{APS-gap}. Usually they can be avoided, however, as often isotropic ensembles are considered for which the bi-orthogonal functions are simply the monomials times the statistical weight.
 
\subsection{Effective Lagrangians}\label{chap1:secLagrangians}

\subsubsection{  Supersymmetry method}\label{chap1:sec:SUSY}

The $k$-point correlation functions can be generated from ratios of characteristic polynomials (also known as partition functions),
\begin{equation}\label{chap1:generating-function}
\mathcal{Z}_{k_{\rm B}|k_{\rm F}}(\kappa)=\left\langle\frac{\prod_{j=1}^{k_{\rm F}}\det(H-\kappa_{{\rm F},j}\mathbf{1})}{\prod_{j=1}^{k_{\rm B}}\det(H-\kappa_{{\rm B},j}\mathbf{1})}\right\rangle=\left\langle{\rm Sdet}^{-1}(H\otimes \mathbf{1}_{k_{\rm B}|k_{\rm F}}-\mathbf{1}\otimes \kappa)\right\rangle
\end{equation}
where $\kappa={\rm diag}(\kappa_{{\rm B},1},\ldots,\kappa_{{\rm F},k_{\rm B}};\kappa_{{\rm F},1},\ldots,\kappa_{{\rm F},k_{\rm F}})$ comprises the source variables which are arranged in terms of a diagonal  supermatrix. The source variables in the characteristic polynomials in the denominator must have an imaginary increment so that the integration of a Hermitian random matrix is well-defined, especially we assume $L_j\varepsilon={\rm Im}(\kappa_{{\rm B},j})\neq 0$ and its sign is $L_j={\rm sign}[{\rm Im}(\kappa_{{\rm B},j})]=\pm1$. The diagonal supermatrix $\kappa$ is sometimes replaced by a more general supermatrix $M$ called the mass matrix because it comprises also the masses of the dynamical fermions in QCD-like theories, e.g.,  see~\cite{NN-massive-1,NN-massive-2,AB-massive} as well as the book chapter~\cite{Oxford-QCD}.

The $k$-point correlation function can then be obtained by setting $k_{\rm B}=k_{\rm F}=k$, identifying $\kappa_{{\rm F},j}=\lambda_j-J_j+iL_j\varepsilon$ and $\kappa_{{\rm B},j}=\lambda_j+iL_j\varepsilon$, differentiating with respect to $J_j$ and taking the limit $J_j,\varepsilon\to0$ as follows
\begin{equation}
\lim_{J_1,\ldots,J_k,\varepsilon\to0}\sum_{L_1,\ldots, L_{k}=\pm1}\left(\prod_{j=1}^k\frac{L_j}{2\pi i N}\partial_{J_j}\right)\mathcal{Z}_{k_{\rm B}|k_{\rm F}}(\kappa)=\lim_{\varepsilon\to0}\left\langle\prod_{j=1}^k\left(\sum_{L_j=\pm1}\frac{L_j}{2\pi i N}{\rm tr}\left[H-(\lambda_j+iL_j\varepsilon)\mathbf{1}\right]^{-1}\right)\right\rangle=\left\langle\prod_{j=1}^k\rho(\lambda_j)\right\rangle=\widetilde{R}_k(\lambda_1,\ldots,\lambda_k)
\end{equation}
in terms of the empirical level density~\eqref{chap1:emp.density}. This is not exactly the very same as in~\eqref{chap1:k-point-int} but comprises self-interactions. The relation between the two are given by
\begin{equation}
\widetilde{R}_k(\lambda_1,\ldots,\lambda_k)=\frac{1}{N^k}\overline{R}_k(\lambda_1,\ldots,\lambda_k)+\frac{1}{N}\sum_{1\leq a<b\leq k}\delta(\lambda_a-\lambda_b) \frac{1}{N^{k-1}}\overline{R}_{k-1}(\{\lambda_j\}_{j=1,\ldots,k;\,j\neq b})+\ldots
\end{equation}
 in the highest order terms for $k\geq2$. Considering the fact that the $k$-point correlation functions $\overline{R}_j(\lambda_1,\ldots,\lambda_j)$ are normalised like $N!/(N-j)!\overset{N\gg1}{\approx} N^j$ for all fixed $j\in\mathbb{N}_0$, this difference becomes marginal (apart from the normalisation) in the limit of large matrix dimension $N\to\infty$ since the lower order terms are of order $N^{-1}$ or smaller. 

Before we go on, we would briefly recall the basics of superalgebra and superanalysis. For a deeper introduction we refer to the book by Berezin~\cite{Berezin-SUSY}. The definitions of the supertrace and superdeterminant are
\begin{equation}
{\rm Str}\, \sigma={\rm tr}\,\sigma_{\rm BB}-{\rm tr}\,\sigma_{\rm FF}\quad{\rm and}\quad {\rm Sdet}\,\sigma=\frac{\det (\sigma_{\rm BB}-\sigma_{\rm BF}\sigma_{\rm FF}^{-1}\sigma_{\rm FB})}{\det\sigma_{\rm FF}}=\frac{\det \sigma_{\rm BB}}{\det(\sigma_{\rm FF}-\sigma_{\rm FB}\sigma_{\rm BB}^{-1}\sigma_{\rm BF})}\quad{\rm with}\quad \sigma=\left[\begin{array}{cc} \sigma_{\rm BB} & \sigma_{\rm BF} \\ \sigma_{\rm FB} & \sigma_{\rm FF} \end{array}\right],
\end{equation}
in terms of the Boson-Boson and Fermion-Fermion blocks $\sigma_{\rm BB}$ and $\sigma_{\rm FF}$, which comprise only commuting matrix elements, and the Boson-Fermion and Fermion-Boson block $\sigma_{\rm BF}$ and $\sigma_{\rm FB}$, whose matrix entries are purely anti-commuting. The matrix $\sigma$ is called supermatrix of superdimension $(k_{\rm B}|k_{\rm F})\times(k_{\rm B}|k_{\rm F})$.

We also recall that in supersymmetric calculations we do not only have commuting real or complex variables but also anti-commuting Grassmann variables $\eta_j$, i.e., $\eta_j\eta_l=-\eta_l\eta_j$, which create together with the commuting variables the Grassmann algebra. Due to the anti-commutation relations, all Grassmann variables are nilpotent, meaning $\eta_j^2=0$, so that a function of these variables is always a finite Taylor series, e.g., $f(\eta_1\eta_2)=f(0)+f'(0)\eta_1\eta_2$. We can also introduce ad hoc complex conjugated Grassmann variables $\eta_j^*$ which represent nothing else than another set of independent Grassmann variables.

The integration of Grassmann variables is defined by the two rules
\begin{equation}
\int d\eta=0\qquad{\rm and}\qquad \int \eta d\eta=1,
\end{equation}
where the definition of an integral over higher moments is not necessary to  due to the nilpotence of $\eta$. The normalisation of the first moment is convention which may vary in the literature. The differentials are actually also anti-commuting variables and are equal to the derivative with respect to the Grassmann variables.
A simple but important example is the Gaussian integral over a complex supervector $v=(z_1,\ldots,z_{k_{\rm B}};\eta_1,\ldots,\eta_{k_{\rm F}})^T$ which is
\begin{equation}\label{chap1:Gaussian-int}
\frac{\int \exp[-v^\dagger \sigma v]d[v]}{\int \exp[-v^\dagger  v]d[v]}={\rm Sdet}\,\sigma^{-1}\qquad{\rm requiring}\quad \sigma_{\rm BB}+\sigma_{\rm BB}^\dagger>0\quad{\rm for\ convergence}.
\end{equation}
It is this integral which lies at the heart of the supersymmetry method in random matrix theory, condensed matter theory and many other fields in statistical physics  and quantum field theory. We refer to~\cite{Oxford-SUSY} for an in depth introduction into this technique.

The idea is now to use the identity~\eqref{chap1:Gaussian-int}  for the superdeterminant in the expectation value~\eqref{chap1:generating-function}. As we have $k_{\rm B}$ determinants in the denominator and $k_{\rm B}$ we need to replace the supervector $v$ in~\eqref{chap1:Gaussian-int} by an $N\times(k_{\rm B}|k_{\rm F})$ rectangular complex supermatrix $V$. Then, it is
\begin{equation}\label{chap1:part-Gaussian}
\mathcal{Z}_{k_{\rm B}|k_{\rm F}}(\kappa)={\rm Sdet}(iL)^N\frac{\int\left\langle \exp[-i{\rm Str}\, V^\dagger HVL]\right\rangle \exp[i{\rm Str}\, V^\dagger V L\kappa]d[V]}{\int \exp[-{\rm Str}\, V^\dagger V]d[V]}
\end{equation}
with the diagonal supermatrix $L={\rm diag}(L_1,\ldots,L_{k_{\rm B}};\mathbf{1}_{k_{\rm F}})$ of signs of the imaginary parts of $\kappa_{{\rm B},j}$ which needs to be introduced to guarantee the convergence of the integral in $V$. The expectation value
\begin{equation}
\mathcal{F}(VLV^\dagger)=\left\langle \exp[-i{\rm Str}\, V^\dagger HVL]\right\rangle=\left\langle \exp[-i\,{\rm tr}\, HVLV^\dagger ]\right\rangle
\end{equation}
is the characteristic function (Fourier transform) of the random matrix $H$. 

What happens now depends on the symmetric space $\mathfrak{p}$ we have started with, in particular which involutions it satisfies. Those symmetries are projective in combination with the trace. Say $H=-\iota(H)=H^T$ is symmetric, then it is
\begin{equation}\label{chap1:example1}
{\rm tr}\, HVLV^\dagger=\frac{1}{2}{\rm tr}(H+H^T)VLV^\dagger=\tfrac{1}{2}{\rm tr}\, H(VLV^\dagger+(VLV^\dagger)^T)={\rm tr}\, H\widehat{V}\widehat{L}\widehat{V}^\dagger\qquad{\rm with}\quad \widehat{V}=\tfrac{1}{\sqrt{2}}(V,V^*)
\end{equation}
and $\widehat{L}={\rm diag}(L_1,\ldots,L_{k_{\rm B}};\mathbf{1}_{k_{\rm F}};L_1,\ldots,L_{k_{\rm B}};-\mathbf{1}_{k_{\rm F}})$. The minus sign in  the last block of $\widehat{L}$ originates from the anti-commutativity of the Grassmann variables in the Fermion-Fermion block. In contrast, when  $H=\iota(H)=-\gamma_5 H\gamma_5$, with $\gamma_5={\rm diag}(\mathbf{1}_p,-\mathbf{1}_{N-p})$, is chiral we obtain
\begin{equation}\label{chap1:example2}
{\rm tr}\, HVLV^\dagger=\frac{1}{2}{\rm tr}(H-\gamma_5H\gamma_5)VLV^\dagger
={\rm tr}\, H\widehat{V}\widehat{L}\widehat{V}^\dagger\quad{\rm with}\quad \widehat{V}=\left[\begin{array}{cc} V_{\rm l}& 0\\ 0 & V_{\rm r}\end{array}\right] \quad{\rm and}\quad \widehat{L}=\left[\begin{array}{cc} 0 & \mathbf{1}_{k_{\rm B}|k_{\rm F}}\\ \mathbf{1}_{k_{\rm B}|k_{\rm F}} & 0\end{array}\right] 
\end{equation}
where $V^\dagger=(V_{\rm l}^\dagger,V_{\rm r}^\dagger)$ with $V_{\rm l}$ and $V_{\rm r}$ of sizes $p\times (k_{\rm B}|k_{\rm F})$ and $(N-p)\times (k_{\rm B}|k_{\rm F})$, respectively.
This procedure, here exercised for two examples,  works for all involutions we have encountered in the Altland-Zirnbauer classification which can be reduced to the chirality symmetry and the symmetry with respect to the transposition for Hermitian matrices. In the simplest case, we do not have any additional symmetry apart from $H$ being Hermitian. Then, there is no symmetrisation needed. Regardless, what the symmetry class is one can always write
${\rm tr}\, HVLV^\dagger={\rm tr}\, H\widehat{V}\widehat{L}\widehat{V}^\dagger$
so that $\widehat{V}\widehat{L}\widehat{V}^\dagger$ has the same symmetries under transposition and chirality operation as $H$.

For the next step, we assume that the probability measure is group invariant under the respective group $\mathcal{U}$ corresponding to the exponential image $\exp[\mathfrak{h}]\subset \mathcal{U}$ where $\mathfrak{h}$ is the Cartan partner of the symmetric space $\mathfrak{p}$, i.e., $P(H)=P(E)$ only depends on the eigenvalues $E$ of $H$. This carries over to the characteristic function, so that
\begin{equation}\label{chap1:duality}
\mathcal{F}(VLV^\dagger)=\mathcal{F}(\widehat{V}\widehat{L}\widehat{V}^\dagger)=\widetilde{\mathcal{F}}(\widehat{V}^\dagger\widehat{V}\widehat{L})
\end{equation}
is only a function of the supermatrix $\widehat{V}^\dagger\widehat{V}$. At this point there are two approaches to replace $\widehat{V}^\dagger\widehat{V}$ by an integral over a supermatrix $\sigma$. The first method is the superbosonisation formula~\cite{LSZ-superbosonisation,Sommers-superbosonisation}. For the symmetric space of Hermitian matrices it yields for the partition function the expression
\begin{equation}\label{chap1:superbosonisation}
\mathcal{Z}_{k_{\rm B}|k_{\rm F}}(\kappa)=\frac{\int\mathcal{F}(\sigma{L})\exp[i{\rm Str}\,\sigma L{\kappa}]{\rm Sdet}^N(iL\sigma) d\widetilde{\nu}(\sigma)}{\int\exp[-{\rm Str}\,\sigma]{\rm Sdet}^N\sigma d\widetilde{\nu}(\sigma)},
\end{equation}
where the Fermion-Fermion block $\sigma_{\rm FF}$ is drawn from the unitary group and the Boson-Boson block $\sigma_{\rm BB}$ is a positive definite Hermitian matrix. The measure $d\widetilde{\nu}(\sigma)$ is group invariant under the supergroup ${\rm U}(k_{\rm B}|k_{\rm F})$ which fixes it uniquely up to normalisation. Also for the other symmetric matrix spaces a superbosonisation formula exists~\cite{LSZ-superbosonisation}. These  superbosonisation formulas have in common that the Fermion-Fermion block is
drawn from a compact symmetric space which is a subset of the unitary group, such as  the circular orthogonal or symplectic ensembles, and  the Boson-Boson block is some positive definite matrix drawn from a subset of the Hermitian matrix space. This method can be understood as a generalisation of the residue theorem in Complex Analysis as the expansion in Grassmann variables is equal to an application of a complicated and high order differential operator on the integrand which then can be expressed in terms of contour integrals.

In the second method, one transforms the expression back via a Fourier-transformation in superspace with the help of a generalisation of the Hubbard-Stratonovich transformation~\cite{Guhr-HS-trafo,GGK-HS-trafo}. This method is easier to understand why it works, though there are analytical problems to be overcome due to issues in the convergence of the integrals when $L_{\rm BB}\neq\pm1$. Fyodorov and Strahov~\cite{FS-hyperbolic,Fyodorov-hyperbolic} solved this problem by deforming the integration domain for the Boson-Boson block into a hyperbolic hypersurface. We want to highlight that both approaches are equivalent as shown in~\cite{SUSY-comparison}.

There is actually a third method to short-cut the calculation of Fourier transforms which is called supersymmetric projection formula~\cite{Kieburg-Phdthesis,KKG-projection}. In this approach, the ordinary matrix space is extended to a superspace via Wegner integration theorems~\cite[Chapter 15]{Wegner}, which is a generalisation of the Cauchy theorem. In the case of a unitarily invariant, suitably differentiable probability density $P(H)$ on the Hermitian matrices ${\rm Herm}(N)$ this looks like~\cite[Chapter 11]{Kieburg-Phdthesis}
\begin{equation}
\frac{\int \widehat{P}\left(\left[\begin{array}{cc} H & W^\dagger \\ W & \widetilde{\sigma} \end{array}\right]\right)d[W]d[\widetilde{\sigma}]}{\int R\left(\left[\begin{array}{cc} H & W^\dagger \\ W & \widetilde{\sigma} \end{array}\right]\right)/R\left(\left[\begin{array}{cc} H & 0 \\ 0 & 0 \end{array}\right]\right)d[W]d[\widetilde{\sigma}]}= \widehat{P}\left(\left[\begin{array}{cc} H & 0 \\ 0 & 0 \end{array}\right]\right)=P(H)\qquad{\rm with}\quad R\left(\left[\begin{array}{cc} H & W^\dagger \\ W & \sigma \end{array}\right]\right)=\exp[-{\rm tr}H^2-{\rm Str}(\widetilde{\sigma}^2+2WW^\dagger)].
\end{equation}
where $W$ is a $(k|k)\times N$ complex rectangular supermatrix and $\widetilde{\sigma}$ is $(k|k)\times (k|k)$ Hermitian supermatrix.
The normalisation with the help of the Gaussian is chosen out of convenience but has not much of significance. The factor can be also replaced by  a general supergroup invariant probability function $R$. Then, the partition function~\eqref{chap1:generating-function} for $k_{\rm B}=k_{\rm F}=k$ is given by~\cite[Chapter 11]{Kieburg-Phdthesis}
\begin{equation}\label{chap1:SUSY.rel}
\mathcal{Z}_{k_{\rm B}|k_{\rm F}}(\kappa)=\int{\rm Sdet}^{-N}( \widetilde{\sigma}- \kappa)Q(\widetilde{\sigma})d[\widetilde{\sigma}]\qquad{\rm with}\quad Q(\widetilde{\sigma})=\frac{\int \widehat{P}\left(\left[\begin{array}{cc} H & W^\dagger \\ W & \widetilde{\sigma} \end{array}\right]\right)d[W]d[H]}{\int \exp[-{\rm Str}(\widetilde{\sigma}^2+2WW^\dagger)]d[W]d[\widetilde{\sigma}]}.
\end{equation}
Similar formulas were also found for the other two Dyson classes~\cite[Chapter 11]{Kieburg-Phdthesis} and the three chiral counterparts~\cite{KKG-projection}. It is to be expected that also for the remaining four Altland-Zirnbauer classes such supersymmetric projection formulas exist.

The relation~\eqref{chap1:SUSY.rel} highlights the duality between the integral over the oridinary matrix $H$ and the supermatrix $\sigma$. This duality is also present in the other two supersymmetry approaches where the original matrix dimension $N$ becomes the number of characteristic polynomials, while the number of characteristic polynomials becomes the new dimension. This duality has its origin in representation theory and is known as Howe duality~\cite{CFZ-Howe}. It essentially says that representations of two different groups may agree with each other which is indirectly exploited, here. Interestingly, this duality can be even exploited when mild symmetry breaking terms are involved in the random matrix model, e.g., see~\cite{Guhr-scattering,Oxford-SUSY}.

\subsubsection{Fermionic partition functions and effective Lagrangians}

The expressions~\eqref{chap1:superbosonisation} and~\eqref{chap1:SUSY.rel} underline the advantage of the supersymmetry method. It maps the integral over an ordinary matrix whose dimension tends to infinity to an integral over a supermatrix with a fixed matrix dimension. The supersymmetric expression invites for a saddle point approximation which leads to the effective Lagrangians.

Our aim is to figure out what the universal kernels at the hard edge and in the bulk, see Subsection~\ref{chap1:sec:universal}, will look like and avoid the discussion about higher criticality. Therefore, we can use the Gaussian ensembles, see~\eqref{chap1:Gaussian.Wigner}. To simplify the discussion, we restrict ourselves to the fermionic sector, only, meaning only averages of characteristic polynomials in the numerator but none in the denominator are considered ($k_{\rm B}=0$ in~\eqref{chap1:generating-function}). Such averages play an important role in QCD, e.g., see~\cite{NN-massive-1,NN-massive-2,AB-massive,Oxford-QCD}, and in number theory about the Riemann-zeta function and its generalisation to L-functions~\cite{Oxford-numbers}. Other applications can be found in random landscapes, e.g, see~\cite{Fyodorov-landscape,WW-landscape}

Any of the two expressions~\eqref{chap1:superbosonisation} and~\eqref{chap1:SUSY.rel} lead to the same limit when zooming into the point $\lambda_0$ and setting $\kappa_{{\rm F},j}=\lambda_0+\overline{s}(\lambda_0)\mu_j$ as well as dividing by $\mathcal{Z}_{k_{\rm B}|k_{\rm F}}(\lambda_0)$. It is explicitly given by~\cite{Oxford-QCD,Zirnbauer-superclasses}
\begin{equation}\label{chap1:Lagrangian}
\lim_{N\to\infty}\frac{\mathcal{Z}_{0|k_{\rm F}}(\kappa)}{\mathcal{Z}_{0|k_{\rm F}}(\lambda_0)}=\int_{\mathcal{G}/\mathcal{H}}\exp[{\rm tr}\,\widehat{M} U ]\det \widetilde{U}^\nu d\widetilde{\mu}(U)\quad{\rm with}\quad \widehat{M}\quad {\rm defined\ by}\quad {\rm tr}\, {\rm diag}(M_1,\ldots,M_{k_{\rm B}})V^\dagger  V L={\rm tr}\,\widehat{M}\widehat{V}^\dagger \widehat{V}\widehat{L}
\end{equation}
where  $d\widetilde{\mu}(U)$ is the Haar measure of the coset $\mathcal{G}/\mathcal{H}$ induced by the Haar measure of the Lie group $\mathcal{G}$. The cosets can be read off in Table~\ref{chap1:tab:Symmetry-breaking}. The term $\det \widetilde{U}^\nu$ only appears for symmetry classes with non-trivial index $\nu$, see Table~\ref{chap1:tab:RMT-param}. The matrix $\widetilde{U}$ is related to  $U$ depending on which ordinary symmetric matrix space one started with and reminiscent of how the measure $d[V]$ of the rectangular supermatrix $V$ (see Subsection~\ref{chap1:sec:SUSY}) transformed under the group action $\mathcal{G}$. Thus, $\det \widetilde{U}^\nu$ can be seen more as part of the measure instead of the exponential as it was done in the superbosonisation formula~\cite{LSZ-superbosonisation}.

\begin{table}[t!]
\centering
\rotatebox{0}{\small
\begin{tabular}{|c|c|c|c|}
  \hline
 Local Kernels at the Origin
                 & \hspace*{-0.2cm}$\begin{array}{c} \text{Abbreviation} \\ \text{Gaussian RMT}\end{array}$ & $\begin{array}{c} \text{Cartan} \\ \text{Class}\end{array}$ & $\begin{array}{c} {\rm Symmetry\ Breaking\  Pattern} \\ \mathcal{G}\rightarrow\mathcal{H} \end{array}$   \\
  \hline\hline
  $\begin{array}{c} \text{sine kernel}\ \beta=2 \end{array}$		&	GUE$(N)$	& A &	${\rm U}(2k)\rightarrow{\rm U}(k)\times{\rm U}(k)$  \\ \hline
  $\begin{array}{c} \text{sine kernel} \ \beta=1 \end{array}$		&	GOE$(N)$	& AI &	${\rm USp}(4k)\rightarrow{\rm USp}(2k)\times{\rm USp}(2k)$  \\ \hline
   $\begin{array}{c} \text{sine kernel}\ \beta=4 \end{array}$		&	GSE$(N)$	&	AII & ${\rm O}(2k)\rightarrow{\rm O}(k)\times{\rm O}(k)$  \\ \hline\hline
  $\begin{array}{c} \text{Bessel kernel}\ \beta=2\ {\rm and}\ \alpha=2\nu \end{array}$	&	GAOE$_\nu(N)$	& B/D &	${\rm O}(2k)\rightarrow{\rm U}(k)$  \\ \hline
  $\begin{array}{c} \text{Bessel kernel} \ \beta=2\ {\rm and}\ \alpha=2 \end{array}$	&	GASE$(N)$	&	C &		${\rm USp}(2k)\rightarrow{\rm U}(k)$  \\ \hline
  $\begin{array}{c} \text{Bessel kernel} \ \beta=2\ {\rm and}\ \alpha=2\nu+1 \end{array}$ &	$\chi$GUE$_\nu(N)$	& AIII &	${\rm U}(k)\times{\rm U}(k)\rightarrow{\rm U}(k)$  \\ \hline
  $\begin{array}{c} \text{Bessel kernel} \ \beta=1\ {\rm and}\ \alpha=\nu \end{array}$ &	$\chi$GOE$_\nu(N)$ & B/DI & 	${\rm U}(2k)\rightarrow{\rm USp}(2k)$  \\ \hline
   $\begin{array}{c} \text{Bessel kernel} \ \beta=4\ {\rm and}\ \alpha=4\nu+3 \end{array}$ &	$\chi$GSE$_\nu(N)$	 & CII &	${\rm U}(k)\rightarrow{\rm O}(k)$  \\ \hline
  $\begin{array}{c} \text{Bessel kernel} \ \beta=1\ {\rm and}\ \alpha=1 \end{array}$ &	GBOE$(N)$	& CI &	${\rm USp}(2k)\times{\rm USp}(2k)\rightarrow{\rm USp}(2k)$  \\ \hline
   	$\begin{array}{c} \text{Bessel kernel} \ \beta=4\ {\rm and}\ \alpha=4\nu+1 \end{array}$ &	GBSE$_\nu(N)$	&	B/DIII & ${\rm O}(k)\times{\rm O}(k)\rightarrow{\rm O}(k)$	  \\ \hline
\end{tabular}}
\caption{
The ten Hermitian symmetric (flat) matrix spaces (second column is the Gaussian realisation) and their microscopic statistics at the origin (first column). The Dyson index $\beta$ and the index $\alpha$ correspond to the jpdf~\eqref{chap1:jpdf-non-Dyson} and can be read off in Table~\ref{chap1:tab:RMT-param}. The symmetry breaking pattern $\mathcal{G}\rightarrow\mathcal{H}$ for the corresponding non-linear $\sigma$-model is stated in the third column. The co-set $\mathcal{G}/\mathcal{H}$ is the integration manifold of the Goldstone modes in the effective field theory and after carrying out the integrals we would find the microscopic spectral statistics mentioned in the first column.
}\label{chap1:tab:Symmetry-breaking}
\end{table}

Surprisingly, the result~\eqref{chap1:Lagrangian} holds true for all ten symmetry classes of the Altland-Zirnbauer scheme. Only the explicit form of $\widehat{\mu}$, $U$ and $\widetilde{U}$ changes. For instance, for the $\beta=2$ sine kernel scaling limit and $k_{\rm F}=2k$, it is $\widehat{M}={\rm diag}(m_1,\ldots,m_{2k})$ and $U=V{\rm diag}(\mathbf{1}_{k}),-\mathbf{1}_{k})V^{-1}\in {\rm U}(2k)/[{\rm U}(k)\times{\rm U}(k)]$ with $V\in{\rm U}(2k)$ Haar distributed and the term $\det \widetilde{U}^\nu$ is not there. For the example~\eqref{chap1:example1} with real symmetric matrices which corresponds to the $\beta=1$ sine kernel and also $k_{\rm F}=2k$, we have $\widehat{M}={\rm diag}(m_1\mathbf{1}_{2},\ldots,m_{2k}\mathbf{1}_{2})$ and $U=V{\rm diag}(\mathbf{1}_{2k},-\mathbf{1}_{2k})V^{-1}\in {\rm USp}(4k)/[{\rm USp}(2k)\times{\rm USp}(2k)]$ with $V\in{\rm USp}(4k)$ also Haar distributed and no $\det \widetilde{U}^\nu$ term. For the example~\eqref{chap1:example2} with complex Hermitian chiral matrices we have $\widehat{M}={\rm diag}(m_1,\ldots,m_{k_{\rm F}},m_1,\ldots,m_{k_{\rm F}})$ and $U={\rm diag}(V_1,V_2)\tau_1\otimes \mathbf{1}_{k_{\rm F}} {\rm diag}(V_1^{-1},V_2^{-1})={\rm diag}(V_1V_2^{-1},V_2V_1^{-1})\in [{\rm U}(k_{\rm F})\times{\rm U}(k_{\rm F})]/{\rm U}(k_{\rm F})$ with $V_1,V_2\in{\rm U}(k_{\rm F})$ and $\det \widetilde{U}^\nu=\det (V_1V_2^{-1})^\nu$, which holds true for arbitrary $k_{\rm F}\in\mathbb{N}$. The cases of odd $k_{\rm F}$ for the sine kernels involve a union of two disconnected integration manifolds which is reflected by the discrete group $\mathbb{Z}_2$ and is reflected in the parametrisation $U=V{\rm diag}(\mathbf{1}_{k},-\mathbf{1}_{k+1})V^{-1}$ and $U=V{\rm diag}(\mathbf{1}_{k+1},-\mathbf{1}_{k})V^{-1}$ for the $\beta=2$ case and similar for $\beta=1$.

Let us briefly outline where this result is coming from. The original random matrix $H$ may satisfy several symmetries with respect to involutions, say $\iota_1,\ldots,\iota_{I}$. These involuting symmetries carry over from $H$ to the dyadic nilpotent matrix $VLV^\dagger$ because of the trace ${\rm tr}\, HVLV^\dagger$ so that we arrive at the new combination $\widehat{V}\widehat{L}\widehat{V}^\dagger$. After switching to $\widehat{V}^\dagger\widehat{V}\widehat{L}$ when we employ the duality~\eqref{chap1:duality} we can again ask whether the new matrix satisfies any symmetries with respect to involutions. This is actually the case. However, often these involutions are not exactly the same as the one of the matrix $H$ but dual to those. For example, when $H=H^T$ was symmetric under transposition the matrix $\widehat{V}^\dagger\widehat{V}=\widehat{\tau}_2(\widehat{V}^\dagger\widehat{V})^T\widehat{\tau}_2$ will be self-dual; in this case $\widehat{L}=\mathbf{1}_{2k}$ is trivial. The invariance group of these dual symmetric matrix spaces are denoted by $\mathcal{G}$. The measure $d[V]$ is however not necessarily invariant under $\mathcal{G}$ and gives rise to an anomalous symmetry breaking which will be reflected in the term $\det \widetilde{U}^\nu$. When taking the limit $N\to\infty$ and performing the saddle point analysis, the invariance group $\mathcal{G}$ will be broken by the saddle point solution which is only invariant under the subgroup $\mathcal{H}\subset\mathcal{G}$. The remaining saddle point manifold one needs to integrate over is given by the coset $\mathcal{G}/\mathcal{H}$. The directions of integrations which are orthogonal to the saddle point manifold are expanded up to second order yielding Gaussians which are integrated out.

This mechanism is a spontaneous symmetry breaking and the saddle point manifold is the Goldstone manifold in effective field theory~\cite{Oxford-QCD}. The Lagrangian is in the present case fairly simple
\begin{equation}\label{chap1:Lagrangian-real}
\mathcal{L}(U)={\rm tr}\,\widehat{M}U,
\end{equation}
where we consider $\det \widetilde{U}^\nu d\widetilde{\mu}(U)$ as the reference measure encoding the anomalous symmetry breaking.
Higher order expansions in terms of $1/N$ would reveal more complicated but systematic corrections of this Lagrangian.
The Lagrangian~\eqref{chap1:Lagrangian-real} should agree with the potential of non-linear $\sigma$-models as it does for QCD. In this field the approach described above is also known as chiral perturbation theory because of obvious reasons. The matrix dimension plays then the role of the space-time volume, e.g., see~\cite{Oxford-QCD} for deeper discussion. The higher order corrections in the volume/matrix dimension and thus also the influence of the kinetic part of the non-linear models can be accessed with random matrix theory as it was shown in~\cite{AI-pion} where the influence of the pion decay constant was modelled.

At the soft-edge the Lagrangian must be expanded up to third and not second order in the integration directions which are orthogonal to the saddle point manifold. This gives the characteristic result in terms of Airy functions, cf., Eqs.~\eqref{chap1:soft-dens.U} and~\eqref{chap1:soft-dens.OS}. Surely, one can extend this approach to higher multicritical points such as the Pearcy kernel statistics~\cite{TW-Pearcy} where two disjoint supports of the level density start to merge. Then, the expansion in the integration variables orthogonal to the saddle point manifold must involve even higher terms.

What happens when we go over to general ratios of characteristic polynomials? Zirnbauer has actually classified and identified all ten supersymmetric coset spaces for the Altland-Zirnbauer tenfold way in~\cite{Zirnbauer-superclasses}. Implicitly, he has also shown that the supersymmetric version  of the Lagrangian~\eqref{chap1:Lagrangian-real} holds true for all these classes. There is, however, one subtlety in the spontaneous symmetry breaking pattern due to the interplay of the non-compact Boson-Boson block and the Boson-Boson block of the constant supermatrix $\widehat{L}$. The signature in $\widehat{L}$ selects the saddle points which is reflected in the pattern as well as in the hyperbolic group structure. This was already observed in a very early work~\cite{VWZ-nuclear} by Verbaarschot, Weidenm\"uller and Zirnbauer.

\subsubsection{  Effective Lagrangians for non-Hermitian matrix spaces and the first conjecture in~\cite{HKKU-univ}}

The Lagrangian approach, especially the average of products of characteristic polynomials have also helped in settling the first conjecture in~\cite{HKKU-univ}, namely that the bulk statistics are different for the classes of the complex Ginibre ensemble (class A) and the Gaussian ensembles of the complex symmetric matrices (class AI$^\dagger$) and the complex self-dual matrices (class AII$^\dagger$), see Fig.~\ref{chap1:fig3:edge-difference}. In~\cite{AAKP-determinants,LZ-duality,KKR-spacing,Forrester-determinants}, these averages were studied by different groups with the help of various methods. Two of these approaches~\cite{AAKP-determinants,KKR-spacing} involved a result in terms of effective Lagrangians. 

In the case of non-Hermitian matrices one needs to consider two types of characteristic polynomials so that the partition function takes the form
\begin{equation}\label{chap1:generating-function-complex}
\mathcal{Z}_{k,k_*}(z,w^*)=\left\langle\left(\prod_{j=1}^{k}\det(X-z_{j}\mathbf{1})\right)\left(\prod_{j=1}^{k_*}\det(X^\dagger-w_{j}^*\mathbf{1})\right)\right\rangle.
\end{equation}
Those partition functions can be dealt with the supersymmetry method in the very same way as in the Hermitian case as long as one sticks with determinants in the numerator, only. For determinants in the denominator, as they are need to compute $k$-point correlation functions, one needs to apply Girko's Hermitisation trick~\cite{Girko}.

Our aim is to zoom into a point $\lambda_0\in\mathbb{C}$ inside the bulk of the complex eigenvalue spectrum, meaning we substitute $z_j=\lambda_0+\overline{s}(\lambda_0)\delta z_j$ and $w_j=\lambda_0+\overline{s}(\lambda_0)\delta w_j$. For simplicity, we set $k=k_*$. Unfortunately, the limit of the normalised partition function $\mathcal{Z}_{k,k}(z,w^*)/\mathcal{Z}_{k,k}(\lambda_0\mathbf{1}_k,\lambda_0^*\mathbf{1}_k)$ does not exist unlike in the Hermitian statistics. The reason is that inside the bulk the local mean level spacing $\overline{s}(\lambda_0)$ is of order $1/\sqrt{N}$ and not of order $1/N$ which leads to mixing terms of the integration variables orthogonal to the saddle point manifold and those of the manifold. the exponentially diverging terms are reminiscent of this mixture. To cure this we paired up the variables $\{z_j,w_j^*)$ in~\cite{AAKP-determinants} and considered the limit
\begin{equation}\label{chap1:generating-function-complex}
\lim_{N\to\infty}\left(\prod_{j=1}^k\frac{\mathcal{Z}_{1,1}(\lambda_0,\lambda_0^*)}{\mathcal{Z}_{1,1}(z_j,w^*_j)}\right)\frac{\mathcal{Z}_{k,k}(z,w^*)}{\mathcal{Z}_{k,k}(\lambda_0\mathbf{1}_k,\lambda_0^*\mathbf{1}_k)}=\int_{\mathcal{G}/\mathcal{H}}\exp\left[\frac{1}{2}{\rm tr} [\delta z\otimes\mathbf{1}_2]U^\dagger [\delta w^*\otimes\mathbf{1}_2] U-{\rm tr}\,\delta z\delta w^*\right]d\widetilde{\mu}(U)
\end{equation}
with $\delta z={\rm diag}(\delta z_1,\ldots,\delta z_k)$ and $\delta w^*={\rm diag}(\delta w^*_1,\ldots,\delta w^*_k)$. This limiting structure is again true for all three bulk statistics where only the coset changes which are $[{\rm U}(k)\times{\rm U}(k)]/{\rm U}(k)\simeq{\rm U}(k)$ for the class A (complex Ginibre), $[{\rm USp}(2k)\times{\rm USp}(2k)]/{\rm USp}(2k)\simeq{\rm USp}(2k)$ for the class AI$^\dagger$ (complex symmetric matrices) and  $[{\rm O}(2k)\times{\rm O}(2k)]/{\rm O}(2k)\simeq{\rm O}(2k)$ for the class AII$^\dagger$ (complex self-dual matrices). Thence, the measure $d\widetilde{\mu}(U)$ is simply the Haar measure of the corresponding groups.

The Lagrangian
\begin{equation}
\mathcal{L}(U)=\tfrac{1}{2}{\rm tr} [\delta z\otimes\mathbf{1}_2]U^\dagger [\delta w^*\otimes\mathbf{1}_2] U-{\rm tr}\,\delta z\delta w^*
\end{equation}
 is in this case quadratic in the source variables and the integration variable $U$. This could have been anticipated from the scaling $\overline{s}(\lambda_0)=\mathcal{O}(1/\sqrt{N})$ of the mean level spacing. But from the ${\rm U}(1)$-symmetry of the random matrix $X\to e^{i\varphi}X$ one could have deduced that $\delta z$ and $\delta w^*$ can only appear in combinations of pairs.

Comparing this Lagrangian with the local spectral statistics of Hermitian symmetric matrix spaces, one may ask whether this Lagrangian may also be the same one on symmetry axes such as the real axis and symmetry points like the origin. The three classes A, AI$^\dagger$ and AII$^\dagger$ do not exhibit such axes and points so that it could not be observed. Yet, the remaining 35 classes have such points and a similar analysis may shed light on this open problem.

In~\cite{AAKP-determinants}, the Lagrangian for the soft edge statistics of the three non-Hermitian classes have been calculated, too. Those lead to Lagrangians that are quartic in the integration variables orthogonal to the saddle-point manifold. Interestingly, those can be carried out in the case of the complex Ginibre ensemble leading to the error function kernel~\eqref{chap1:GinUE-uni-kernel}. The reason for this is that they involve the Harish-Chandra--Itzykson--Zuber integral~\cite{HC-int,IZ-int} which is known. Sadly, this was not possible, yet, for the other two classes since for those one requires the knowledge of the Itzykson--Zuber integrals for the orthogonal and the unitary symplectic groups.

\section{Conclusion}

We have summarised the aspects of the local spectral statistics which are one of the main objects when studying chaotic quantum systems. Particularly, random matrix results of such statistics serve as benchmarks to judge how chaotic a system is.

For instance, we outlined the relation between the classification of symmetric matrix spaces and the universal eigenvalue statistics. While this classification is settled in the case of Hermitian matrix spaces, which is important for closed quantum systems, it is still a mathematical challenge to rigorously prove the inequivalence of the 38 classes of non-Hermitian symmetric matrix spaces identified in~\cite{KSUS-class}. The second conjecture of~\cite{HKKU-univ}, which says that there are only 3 universal bulk statistics of complex eigenvalue spectra, is even more challenging. It will be a tedious task to go through all 38 classes and check their spectral statistics without a unifying method. Even more so when considering that the $k$-point correlation functions still remain to be unknown for two of the classes. The approach via effective Lagrangians have, however, proven quite successful in circumventing this obstacle and may be a way to show that this conjecture also holds true.

Especially, the Lagrangians in the neighbourhood of symmetry axes and points in the spectrum which are very distinct for most symmetric matrix spaces will be easier accessible than the computation of the local spectral statistics by other means. First numerical studies have been done in~\cite{XSK-hard-nonHerm}. Nonetheless, this does not mean that one may give up at all on those. The supersymmetry method at least shows a path to get expressions in terms of coset integrals. Due to the Boson-Boson blocks they might be still analytically challenging as the convergence of integrals pose serious challenges. One would need to go over to hyperbolic deformations of the integration manifolds as already suggested for the real spectral analysis in~\cite{FS-hyperbolic,Fyodorov-hyperbolic}.

We have also described the unfolding procedure which is an important preparation of eigenvalue spectra. Without a proper unfolding of spectra the statistics do not become comparable and very likely skewed. This may prevent a clean comparison of empirical data with RMT benchmarks, e.g., see Fig.~\ref{chap1:fig5:unfolding-mic}.
Inside a bulk of eigenvalues away of any edge, symmetry points or axes and multicritical points, such an unfolding can be circumvented with spacing ratios~\cite{OH-spacing,ABGVV,B-spacing,SRP-spacing}. It will, however, break down when one does not zoom into the bulk of eigenvalue spectra as this procedure needs the translation invariance of the local spectral statistics to work properly. This is the reason why we discussed another way of unfolding at edges such as the hard and soft edge. This type of unfolding is based on the limiting mesoscopic level density. It is still not ideal as it creates artificial singularities right at the edge where the mesoscopic level density vanishes, cf., Fig.~\ref{chap1:fig8:hard-mic}. Even so, the result already slightly away from the edge is astoundingly good as the local mean level spacing becomes unity after this type of unfolding. What is still open is to find a theory of unfolding such that even these singularities have vanished. Ideally this theory would be diffeomorphism invariant so that local spectral statistics should be independent of the level density. It should only take into account what type of criticality the spectral statistics exhibits. Furthermore, this method should be also easily accessible for empirical spectra. The future will show whether this wish list is satisfiable.

When going over to complex eigenvalue spectra, a proper unfolding of 2-dimensional point are still a serious challenge if one does not zoom into a bulk of eigenvalues. In potential theoretic approach the conformal mapping method, nicely explained in~\cite[Appendix A]{Byun-conformal}, may hint at that only conformal maps should be applied to spectra to unfold them. From the experimental and numerical point this remains a challenge and an art in itself, nevertheless.

Due to limited space we did not discuss the developments of the eigenvector statistics which has its own long history and very recent developments. Especially two developments should briefly mentioned here. Firstly, the eigenvector statistics of Hermitian Hamiltonians experience a revival in recent years due to the computation of quantum informational benchmarks, for instance in the investigation of many-body systems~\cite{Z-localisation}. See the review~\cite{Qinfo-review} and book chapter~\cite{LZ-QC-QI}. Following the ideas of Lubkin~\cite{Lubkin}, Lloyd and Pagels~\cite{LP-qinfo} and Page~\cite{Page} of considering a uniformly distributed pure state the corresponding reduced density matrix becomes a random matrix called the fixed trace ensemble. In recent years, these ideas were generalised to more structured and physical Hilbert spaces, e.g., see~\cite{Qinfo-review,LSZ-boson,AHHJK,PHFR-nonAbel,CHK-nonAbel}. The BGS conjecture~\cite{BGS-conj} serves as an explanation to describe the eigenvectors of a quantum chaotic one-body Hamiltonian by a uniform  measure over the respective manifold. Open questions are, for instance, how to generalise those to systems with a non-Abelian group invariance, bosonic systems, which are inherently infinite dimensional, as well as how to deal with multi-partite systems. The first steps in this direction have been done in very recent approaches~\cite{LSZ-boson,AHHJK,PHFR-nonAbel,CHK-nonAbel}.

The second development lies in the statistics of eigenvectors of non-Hermitian Hamiltonians as they appear in open quantum chaotic systems. Those eigenvectors are usually not orthogonal to each other. Hence, Chalker and Mehlig~\cite{CM-eigenvect} introduced the concept of the overlap. In a series of works, the overlaps and overlap distributions, even under conditioning the eigenvalues, have been computed. for instance see~\cite{NT-eigenvect,AFK-eigenvect,Dubach-eigenvect,Noda-eigenvect,CFW-eigenvect,AFS-eigenvectors}, which is only a tiny fraction of the true burst works in recent years about this topic.

When considering these very recent developments a trend becomes clear in RMT but also in physics. Increasingly more open systems are studied where non-Hermitian Hamiltonians and random matrices play a predominant role. This is the reason why we have outlined here and there what the non-Hermitian counterpart looks like and what is still posing a challenge. Some problems to be overcome are more serious and mathematically challenging than others. What is, with out a doubt, clear is that those new developments have already exhibited interesting and sometimes surprising new results which expand our knowledge about eigenvalue spectra of operators.

\paragraph{\bf Acknowledgements}

The author has been  supported by the Australian Research Council via Discovery Project grant DP25010255. I want to thank Barbara Dietz for explaining to me the Pad\'e approximation of the level spacing distribution in~\cite{HaakeDietz} and Peter Forrester for a careful reading of the first draft.

\providecommand{\href}[2]{#2}\begingroup\raggedright\endgroup

\end{document}